\documentclass[useAMS,usenatbib,babel,times,superscriptaddress]{mnras}

\usepackage[english,english]{babel}
\usepackage{amsmath}
\usepackage{amssymb,amsfonts,textcomp}
\usepackage{array}
\usepackage{supertabular}
\usepackage{hhline}
\usepackage{hyperref}
\usepackage[usenames]{color}
\usepackage{algpseudocode}
\algrenewcommand\textproc{}

\def\gtrsim{\lower.5ex\hbox{$\; \buildrel > \over \sim \;$}}
\usepackage{graphicx}

\def \prd {{\it Phys. Rev. D}}

\def \apj {{\it ApJ}}
\def \apjl {{\it ApJ Let.}}
\def \apjs {{\it ApJ Sup.}}

\def \aa {{\it A\&A}}
\def \aap {{\it A\&A}}
\def \aj {{\it AJ}}
\def \mnras {{\it MNRAS}}
\def \nat {{\it Nature}}

\def \jcap {{\it JCAP} }

\definecolor{grey}{rgb}{0.75,0.75,0.75}
\definecolor{Orange}{rgb}{1.0,0.5,0.15}
\definecolor{brown}{rgb}{0.7,0.25,0.0}
\definecolor{pink}{rgb}{1.0,0.5,0.5}
\definecolor{darkerred}{rgb}{0.8,0,0}
\definecolor{darkerblue}{rgb}{0,0,0.8}
\definecolor{Blue}{rgb}{0,0.08,0.65}
\definecolor{Red}{rgb}{0.65,0.08,0.05}
\definecolor{Green}{rgb}{0.15,0.45,0.25}

\def\green{}

\begin{document}

\author[S.~ Codis, C.~Pichon and D.~Pogosyan]{
\parbox[t]{\textwidth}{
Sandrine Codis$^{1}$\thanks{E-mail: codis@iap.fr}, Christophe Pichon$^{1,2}$ and Dmitry Pogosyan$^{3}$}
\vspace*{6pt}\\
\noindent$^{1}$ Institut d'Astrophysique de Paris, CNRS \& UPMC, UMR 7095, 98 bis Boulevard Arago, 75014, Paris, France\\
$^{2}$  {Institute of Astronomy, University of Cambridge, Madingley Road, Cambridge, CB3 0HA, United Kingdom}\\
$^{3}$ Department of Physics, University of Alberta, 11322-89 Avenue, Edmonton, Alberta, T6G 2G7, Canada\\
}

\title[Anisotropic tidal torque theory]{
Spin alignments within the cosmic web:\\   a   theory of  constrained tidal torques near filaments
}

\maketitle

\begin{abstract}
{
The geometry of the cosmic web drives in  part the spin acquisition of galaxies. 
This can be explained in a Lagrangian framework,  by identifying  the specific long-wavelength correlations  within  the primordial Gaussian 
random field which  are relevant to spin acquisition. 
Tidal Torque Theory is revisited in the context of such anisotropic environments, biased by the presence of a filament within a wall.  
The  point process of filament-type saddles represents it most efficiently.
The constrained misalignment between the tidal  and the inertia tensors in the vicinity of  filament-type saddles
simply  explains  the distribution of spin directions.
This  misalignment implies in particular an {\sl azimuthal} orientation for the spins of more massive galaxies
and a spin {\sl alignment} with the filament for less massive galaxies.
This prediction is found to be in  qualitative agreement with measurements in Gaussian random fields and 
N-body simulations. It  relates the transition mass to  the 
geometry of the saddle, and  accordingly predicts  its  measured scaling with the mass of non-linearity.
 Implications for galaxy formation and weak lensing are briefly discussed, as is the dual theory of spin alignments in 
walls.
}
\end{abstract}



\section{Introduction}

 Modern simulations based on  a well-established paradigm of cosmological structure formation predict a significant connection between the geometry and dynamics of the large-scale structure on the one hand, and the evolution of the physical properties of forming galaxies on the other. Key questions formulated decades ago are nevertheless not fully answered. What are the main processes which determine the morphology of galaxies? What is the role played by angular momentum in shaping  them? 

\cite{pichonetal11} have suggested that the large-scale coherence of the inflow, inherited 
from the  low-density cosmic web, explains why cold flows are so efficient at producing thin high-redshift discs from the inside out \citep[see also][]{stewartetal2013,laigle2014,prietoetal2014}. 
  On the scale of a given gravitational patch, gas is expelled from adjacent voids, towards sheets and filaments forming at their boundaries. Within these sheets/filaments, the gas shocks and radiatively loses its energy before streaming towards the nodal points of the cosmic  network. 
In the process, it advects angular momentum, hereby seemingly driving the morphology of galaxies (bulge or disc). The evolution of the Hubble sequence in such a scenario is  therefore at least in part initially driven   by the geometry of the cosmic web. As a consequence, the distribution of the properties of galaxies measured relative to their cosmic web environment should reflect such a process. In particular, the spin distribution of 
galaxies should display a preferred mass-dependent orientation relative to the cosmic web. 

Both numerical \citep[e.g.][]{calvoetal07,hahnetal07,sousbie08, pazetal08,zhangetal09,codisetal12,Libeskind13a,Calvo13,dubois14}, and observational evidence (e.g. Tempel et al. 2013) have recently supported this scenario. 
In parallel, much {\green analytical \citep[e.g.][]{Cat++01,H+S04}, numerical \citep[e.g.][]{HRH00,C+M00,S+B10,sfc12,Joa++13b,codis14,tenneti14} and observational \citep[e.g.][]{Bro++02,L+P02,B+N02,Hey++04,Hir++04,H+S04,Man++06,Hir++07,Man++11,Joa++11,Joa++13a}} efforts have been invested to control the level of intrinsic alignments of galaxies as a potential source of systematic errors in weak gravitational lensing measurements. 
Such alignments are believed to be a worrisome  source of systematics of the future generation of lensing surveys like Euclid or LSST.
It is therefore of interest to understand from first principles why such intrinsic alignments arise,
so as to possibly temper their effects.  

Hence we should try and refine a theoretical framework to study the dynamical influence of filaments on galactic scales, via an extension of the peak theory to the truly three-dimensional anisotropic geometry of the circum-galactic medium, and amend the standard galaxy formation model  to account for this anisotropy. Toward this end, we will develop here a filament version of an anisotropic ``peak-background-split'' formalism, i.e. make use of the fact that walls and filaments are the interference patterns of primordial fluctuations on large scales, and induce a corresponding anisotropic boost in over-density. 
 Indeed, filaments feeding galaxies with cold gas are themselves embedded in larger scale walls imprinting their global geometry {\citep{danovichetal11,Dubois2012}}. 
 On top of these modes, constructive interferences of high frequency modes produce peaks which thus get a boost in density that allows them to pass the 
critical threshold necessary to decouple from the overall expansion of the Universe, as envisioned in the spherical collapse model \citep{Gunn1972}. 
This well-known biased clustering effect has been invoked to justify the clustering of galaxies around the nodes of the cosmic web \citep{WhiteTullyDavis88}.
It also  explains why galaxies form in filaments: {in walls alone, the actual density boost is typically not sufficiently large to trigger galaxy formation.  The main nodes of the cosmic web are where galaxies migrate,
not where they form. They thus inherit the anisotropy of their birth place
as spin orientation. During migration, they may collide with other galaxies/haloes and erase part of their birth heritage when converting orbital momentum into spin via merger \citep[e.g.][]{codisetal12}.
Tidal torque theory should therefore  be re-visited  to account  for  the   anisotropy of this filamentary environment on various scales in order to model
 primordial {\sl and} secondary spin acquisition.  

 In this paper, we will quantify and model the intrinsically 3D geometry of galactic spins  while accounting for the geometry of  ÒsaddleÓ points of the density field.
Indeed, saddle points  define an anisotropic {\sl point process} which accounts for the presence of 
 filaments embedded in walls  \citep{Pogosyanetal1998}, two critical ingredient in shaping the spins of galaxies.
Taking them into account will in particular allow us to predict the biased  geometry of the  tidal field
in the vicinity of saddle points. This can be formalized using the two-point joint probability of the gravitational potential field and its first to fourth derivatives and imposing a  saddle point constraint. For Gaussian (or quasi-Gaussian) fields these two-point functions are within reach from first principle \citep{BBKS}. 
 A proper account of the anisotropy of the  environment in this context will allow us to demonstrate why the spin of the forming galaxies field are first aligned with the filament's  direction.
  We will also show that  massive galaxies will have their spin preferentially along 
 the azimuthal direction. While relying on a straightforward extension of Press Schechter's theory, 
 we will predict the corresponding   transition mass' scaling
 with the (redshift dependent) mass of non-linearity, while relying on the so-called cloud-in-cloud problem
 applied to the  filament-background split.

The paper is organised as follows.  Section~\ref{sec:principe} qualitatively presents the 
basis of  the physical process at work in aligning the spin of dark haloes relative to the 
cosmic web.
Section~\ref{sec:2D} then presents the expected Lagrangian spin distribution near filaments, assuming cylindrical symmetry, and 
explains the observed mass transition while carrying a multi-scale analysis of the fate of 
collapsing haloes in the vicinity of 2D saddle points. 
Section~\ref{sec:3D}  revisits this distribution in three dimensions
for realistic typical 3D saddle points. 
Section~\ref{sec:statistics} investigates the predictions of the theory using Gaussian random field and N-body simulations,
while 
we finally conclude in Section~\ref{sec:conclusion}.
Appendix~\ref{sec:A1} discusses possible limitations and extensions of this work.
Appendix~\ref{sec:voids} presents the dual theory for spin alignment near wall saddles.
Finally, Appendix~\ref{sec:technical} gathers some technical complements.
  
\section{Tidal torquing  near a  saddle}
\label{sec:principe}

Before presenting  analytical estimates for the expected spins near filament in two and three dimensions and their transition mass, let us discuss qualitatively what 
underpins the corresponding theory.

\subsection{Spin acquisition by tidal torquing}
\label{sec:TTT}
\begin{figure}
\begin{center}
\includegraphics[width=\columnwidth]{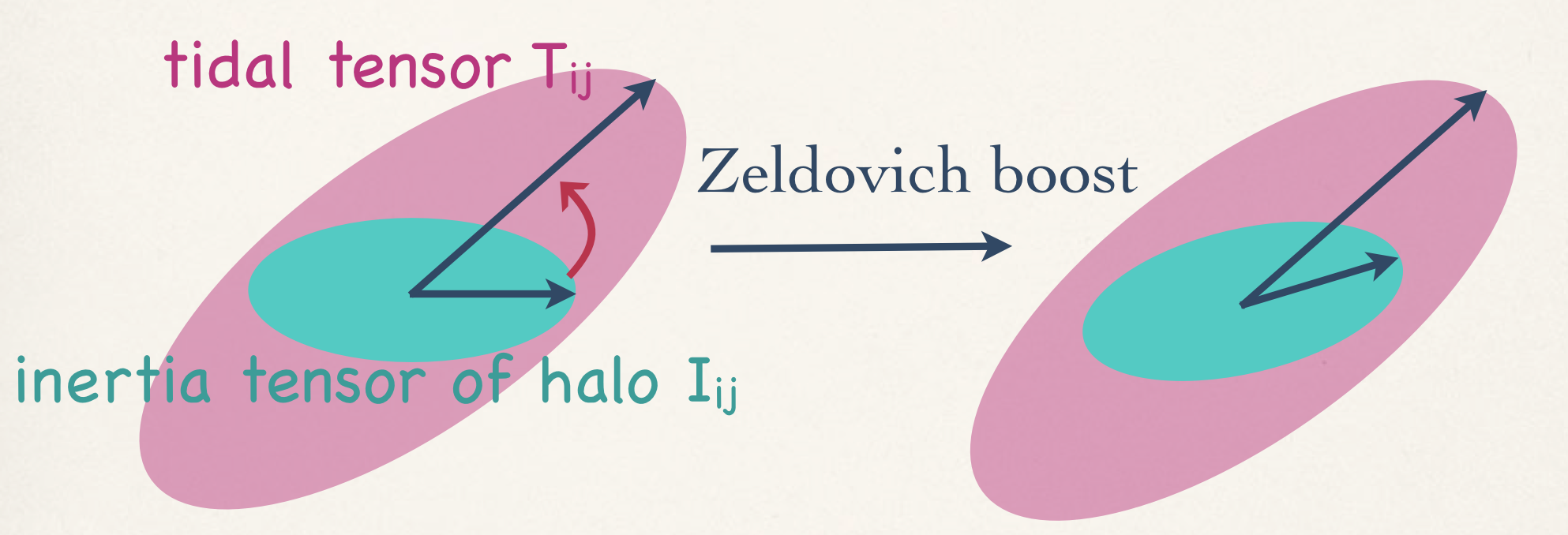}
\caption{Spin
acquisition by tidal torquing. At linear order, the misalignment between the inertia tensor of the proto-object and the surrounding tidal tensor induces an inhomogeneous Zel'dovich boost which corresponds to the acquisition of a net intrinsic angular momentum in Eulerian space.
\label{fig:TTT}}
\end{center}
\end{figure}
In the standard paradigm of galaxy formation, protogalaxies acquire their spin\footnote{Note that in this paper we will call interchangeably ``spin'' or ``angular momentum'' the intrinsic angular momentum of (proto-) haloes.
} by tidal torquing coming from the surrounding matter distribution \citep{hoyle49,peebles69,doroshkevich70,white84,catelan96,Cri++01}. At linear order, this spin is acquired gradually until the time of maximal extension (before collapse) and is proportional to the misalignment between the inertia tensor of the protogalaxy and the surrounding tidal tensor \citep[see][for a review]{schaefer09}
\begin{equation}
\label{eq:TTT}
L_{i}=\sum_{j,k,l}a^{2}(t)\dot D_{+}(t)\epsilon_{ijk} I_{jl}T_{lk}\,,
\end{equation}
where $a(t)$ is the scale factor, $D_{+}$ the growth factor, $T_{ij}$ the tidal tensor (detraced Hessian of the gravitational potential), $I_{ij}$ the protogalactic inertia tensor (only its traceless part, $\overline I_{ij}$ contributes to
the spin). 
As this work focuses on the spin direction, the factor $a^{2}(t)D_{+}(t)$ will
henceforth be dropped for brievity.
This process of spin acquisition by tidal torquing is illustrated on Figure~\ref{fig:TTT}.

In the Lagrangian picture, $I_{ij}$ is the moment of inertia of a uniform
mass distribution within the Lagrangian image of the halo, while $T_{ij}$ is
the tidal tensor averaged within the same image. Thus, to rigorously determine
the spin of a halo, one must know the area from which  matter is assembled,
beyond the spherical approximation. While this can be determined in numerical
experiments, theoretically we do not have the knowledge of the exact boundary
of a protohalo. As such, one inevitably 
has to introduce an approximate proxy for 
the moment of inertia (and an approximation for how  the tidal field is averaged over that region).

The most natural approach is to consider that protohaloes form around an elliptical
peak in the initial density and approximate its Lagrangian boundary with
the elliptical surface where the over-density drops to zero. 
This leads to the following 
approximation for the traceless part of the inertia tensor 
\citep[e.g.][see also equations~(\ref{Iij})-(\ref{eq:inertia})]{Schafer2012}
\begin{equation}
\overline{I}_{ij} = \frac{2}{5} \nu \sigma_2 M \overline{H}_{ij}^{-1}
= \frac{2}{5} \nu \sigma_2 \frac{M}{\det{\mathbf{H}}} \widetilde{H}_{ij}\,,
\label{eq:Hinvproxy}
\end{equation}
where $\overline{H}_{ij}^{-1}$ is the traceless part of the
inverse Hessian of the density field, 
$H_{ij}=\partial_{i}\partial_{j}\delta$, $\nu$ is the overdensity at the peak, 
and $M$ is the mass of the protohalo. In the second form we explicitly
presented the inverse Hessian via the (detraced) matrix of the Hessian minors,
$\widetilde{H}_{ij}$. While $\widetilde{H}_{ij}$ is a simple polynomial in
second derivatives of the density, ${M}/{\det{\mathbf{H}}} $ is not, which is   the source
of  most technical difficulties when statistical studies of the spin are
attempted.

Let us point at the following considerations  to bypass these difficulties. First, 
 the supplementary condition for the halo to be at a peak of the density
yields an extra $\det{\mathbf{H}}$ factor in all statistical measures \citep[see e.g.][]{BBKS}. 
This factor exactly cancels the determinant  in the denominator. Secondly, 
all  quantities in equation~(\ref{eq:Hinvproxy})
are computed after the density field is smoothed at a particular
scale $R_h$ which sets the corresponding mass scale. Therefore, it is more
appropriate to apply equation~(\ref{eq:Hinvproxy}) to  haloes  at {\sl fixed}
mass $M$, determined by that smoothing.
Hence we could argue for the proxy
$\overline{I}_{ij} \propto \nu \widetilde{H}_{ij}$
for the moment of inertia for haloes of a given fixed mass, where the  change in mass
is reflected in the corresponding change in the smoothing scale.
In two dimensions, we show in Appendix~\ref{sec:A1} that  this  multi-scale approximation
gives qualitatively the  same statistical results as just using $H_{ij}$ as a proxy.
While this approximation
is relatively simple, since we 
are only concerned with the direction of the spin, we will now
go one step further and use throughout this paper the Hessian as a proxy for the inertia tensor, 
even in three dimensions. Indeed, $I_{ij}$, $\widetilde{H}_{ij}$ and $H_{ij}$
share the same eigen-directions \citep{catelan96,Schafer2012}, 
so we {\sl define} the spin for the rest of the paper as 
\begin{equation}
\label{eq:spin-ATTT}
s_{i}\equiv  \sum_{j,k,l}\epsilon_{ijk} H_{jl}T_{lk}\,.
\end{equation}
The vector field $s_{i}$ is then quadratic in the successive derivatives of the potential: its (possibly constrained) expectation can therefore be computed for Gaussian random fields.
This approximation  is further discussed in Appendix~\ref{sec:A1}.
Note that equation~(\ref{eq:spin-ATTT}) improves upon simple parametrisations of the mean misalignment between inertia and tidal tensors \citep[see, e.g.][]{lee&pen00,Cri++01} by {\sl ab initio} explicitly taking into account  the  correlations between both tensors.

\subsection{Geometry of the cosmic web}
Galaxies are not forming everywhere but preferentially in filaments and nodes  which define the so-called cosmic web \citep{ks93,bkp96}.
The origin of these structures lies in the asymmetries of the initial Gaussian random field describing the primordial universe, amplified by
 gravitational collapse \citep{zeldovich70}. The presence of such large-scale structure (walls, filaments, nodes) induces local preferred directions for both the tidal tensor and the inertia tensor of forming objects which will eventually turn into preferred alignments of the spin w.r.t the cosmic web. It is therefore of  interest to understand what is the expected spin direction predicted by equation~(\ref{eq:spin-ATTT}) given the presence of a typical filament nearby. As a filament is typically the field line that joins two maxima of the density field through a filament-type saddle point (where the gradient is null and the density Hessian has two negative eigenvalues), we choose to study in this paper the expected spin direction of proto-objects \textit{in the vicinity of a filament-type saddle point} with a given geometry (which imposes the direction of the filament and the wall).  Figure~\ref{fig:cosmicweb} illustrates the geometry of  filaments near peaks and saddles in 
 a 2D Gaussian field.

\begin{figure}
\begin{center}
\includegraphics[width=0.9\columnwidth]{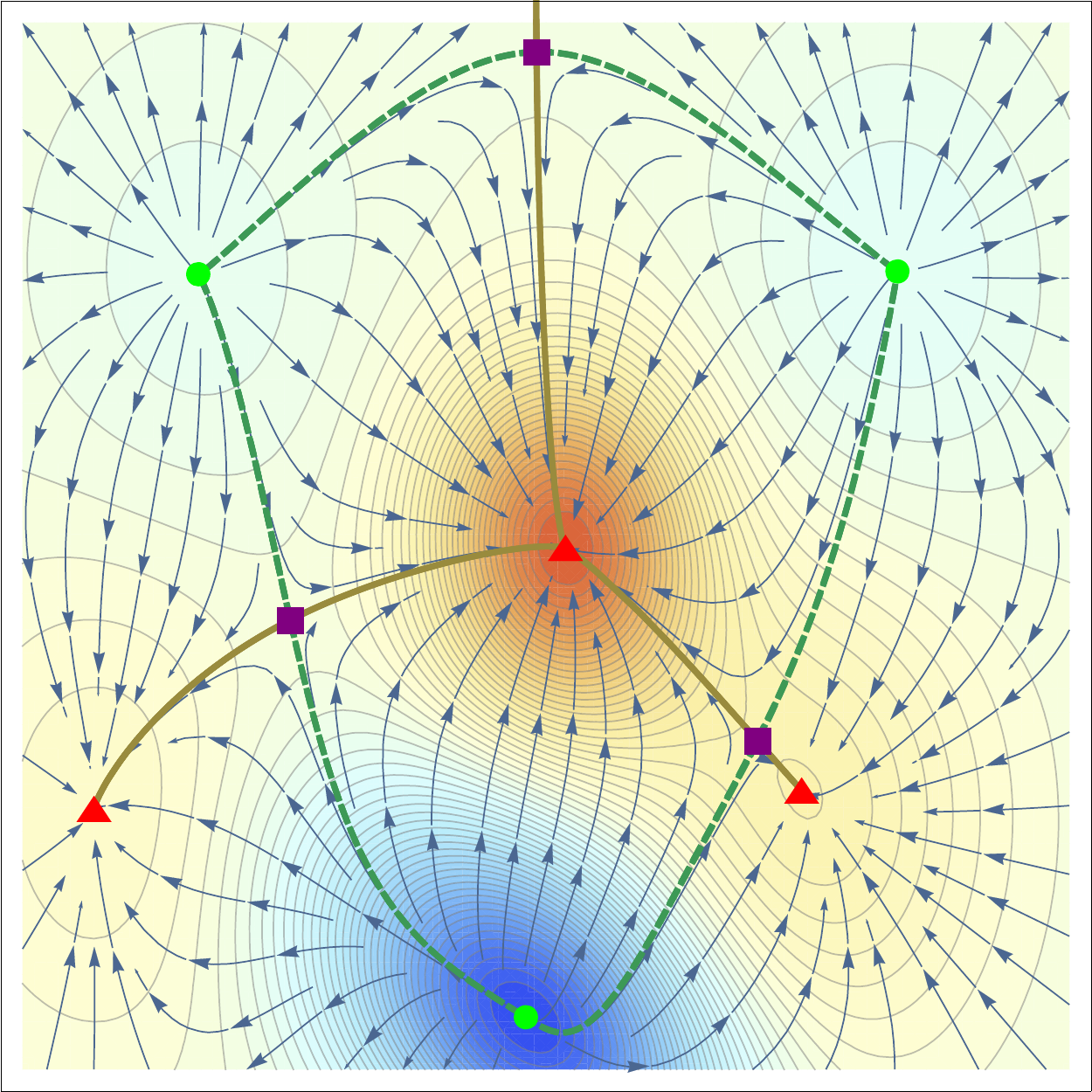}
\caption{ 
On top of the density contours (from dark blue to dark red),
the three ({\green{red triangle}}) maxima (resp. the three ({\green{green points}}) minima) are connected by the crest lines (in {\green{solid}} gold,  resp. the through-lines in {\green{dashed}} green)
 which intersect through saddles points ({\green{purple squares}}).  The blue arrows represent stream lines of the gradient flow.  
 Throughout this paper, we will assume that the geometry of the regions of intermediate densities are  set by the shape of the (purple) saddles.
 \label{fig:cosmicweb}}
\end{center}
\end{figure}

\subsection{Constrained tidal torque theory in a nutshell}
\subsubsection{spin alignments and flips}
It has been shown in simulations \citep[among others]{bailin&steinmetz05,calvoetal07, pazetal08,zhangetal09,codisetal12,Libeskind13a,Forero-Romero2014} that the spin of dark haloes is correlated to the direction of the filaments of the cosmic web in a mass-dependent way. The alignment between the spin and the closest filament increases with mass until a mass of maximum alignment \citep{laigle2014} that we call here critical mass. As mass increases, the direction of the spin becomes less aligned with the filament before becoming perpendicular to it \citep{codisetal12}. This transition -- from aligned to perpendicular -- occurs at a mass that we call here the transition mass. 

This paper will claim that the critical mass is directly related to the size of the quadrant of coherent angular momentum imposed 
by the tides of the saddle point (which are effectively the Lagrangian counter parts of the quadrant of vorticity found in \cite{laigle2014}). This mass
can be captured using a cylindrical model that would correspond to the plane perpendicular to the filament at the saddle point (which amounts to assuming an infinitely long filament). This 2D toy model (see Section~\ref{sec:2D} below) shows that near a 2D peak (i.e near an infinitely long 3D filament), the quadrupolar structure seen in simulation naturally arises in a Lagrangian framework. We investigate the size of that quadrants and shows that it qualitatively predicts the right critical mass.

The second stage of accretion, that flips the spin of more massive haloes from aligned to perpendicular to the filaments, requires a 3D analysis (see Section~\ref{sec:3D}). It is shown that indeed small-haloes that form close to the saddle point, acquire spin along the filaments while more massive haloes that form further from the saddle (i.e closer to the peaks/nodes) acquire a spin perpendicular to the filaments (while accreting 
smaller haloes). The transition mass will be predicted as a function of redshift and shown to agree with measurements in simulations.

\subsubsection{The premices of anisotropic tidal torque theory} \label{sec:premises}
Let us  present here an outline of the extension of TTT within the context of a peak (or saddle) background split.
Given the an-isotropically triaxial  saddle constraint, we will argue that the  misalignment between the tidal tensor and the hessian of the density field  
simply  explains  the transverse and  longitudinal antisymmetric geometry of angular momentum distribution in their vicinity.
It arises because the two tensors probe different scales: given their relative correlation lengths,
 the hessian probes more directly its closest neighbourhood, while the tidal field, 
somewhat larger scales. 

Within the plane of the saddle point perpendicular to the filament axis (the midplane hereafter), the dominant wall (corresponding to the longer axis of the cross section
of the saddle point) will  re-orient more the Hessian than the tidal tensor, which also feels the denser, but typically further away saddle point,
see Figure~\ref{principe}, top panels.
This net misalignment will induce  a spin perpendicular to that plane i.e along the filament. 
This effect will produce a quadrupolar, antisymmetric distribution of the {longitudinal} component of the angular momentum which will
 be strongest at some four points,  not far off axis. 
Beyond a couple of correlation lengths away from those  four points, the effect of the tidal field induced by the saddle point will subside, as both tensors become
more spherical.  

Conversely, in planes containing the filament,  e.g.  containing the main wall, a similar process will misalign both tensors.
This time, the two anisotropic features differentially pulling the tensors are the filament on the one hand, and the density gradient towards 
the peak on the other.  The net effect of the corresponding misalignment will be to also spin up haloes perpendicular to that plane,
 along the azimuthal direction, see Figure~\ref{principe}, bottom panels. By symmetry, the  anti-clockwise tidal spin will be generated on the other side of the saddle point.
 \begin{figure*}
\begin{center}
\includegraphics[width=1.6\columnwidth]{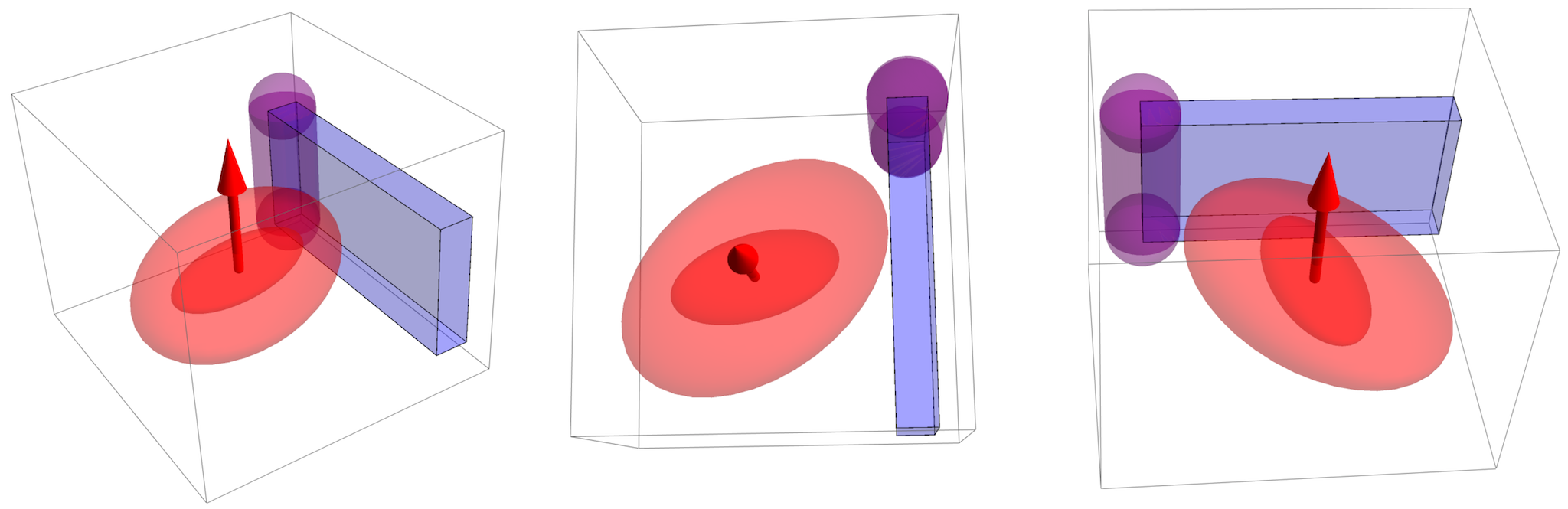}
\includegraphics[width=1.6\columnwidth]{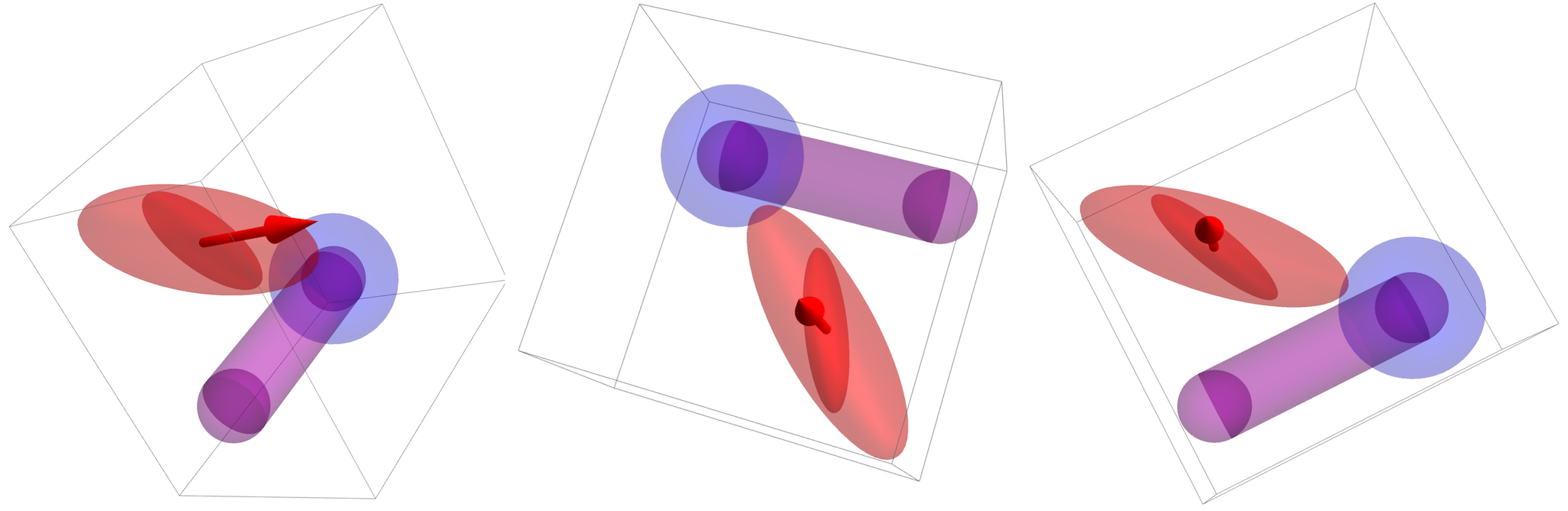}
\caption{Sketch of main differential alignment between halo shapes and tidal tensor responsible for $\mathbf e_z$ and $\mathbf e_\phi$
component of momentum.
{\sl Top:} the two tensors in light and dark red, end up being misaligned 
as they feel differently the neighbouring wall (blue) and filament (purple), inducing a spin
parallel to the filament (red arrow). Three projections are shown for clarity.
 {\sl Bottom:} correspondingly, the differential pull from the filament 
(purple)
and the density gradient towards the peak (blue) generates a spin (red arrow) along the azimuthal direction. By symmetry, the other peak(s) on the other side of the saddle point
will spin up massive haloes in the opposite direction.
\label{principe}}
\end{center}
\end{figure*}

Hence the geometry of angular momentum near filament-saddle points is the following: it is aligned with the filament
in the median plane (within four antisymmetric quadrants), and (anti-)aligned with the azimuthal direction away from that plane. 
The stronger the  triaxiality the stronger the amplitude.  Conversely, if the saddle point becomes degenerate in one or two directions,
the component of the angular momentum in the corresponding direction will vanish. For instance, a saddle point in the middle of 
a very long filament will only display alignment with that filament axis, with no azimuthal component.
For a typical triaxial configuration, two pairs of four points define the loci of maximal {longitudinal} and azimuthal spin.

\subsubsection{Geometry of spin flip}

Figure~\ref{geometry} gives a more quantitative account of the geometry of 
the tidal field around a given saddle point embedded in a given dominant wall.
\begin{figure}
\begin{center}
\includegraphics[width=0.95 \columnwidth]{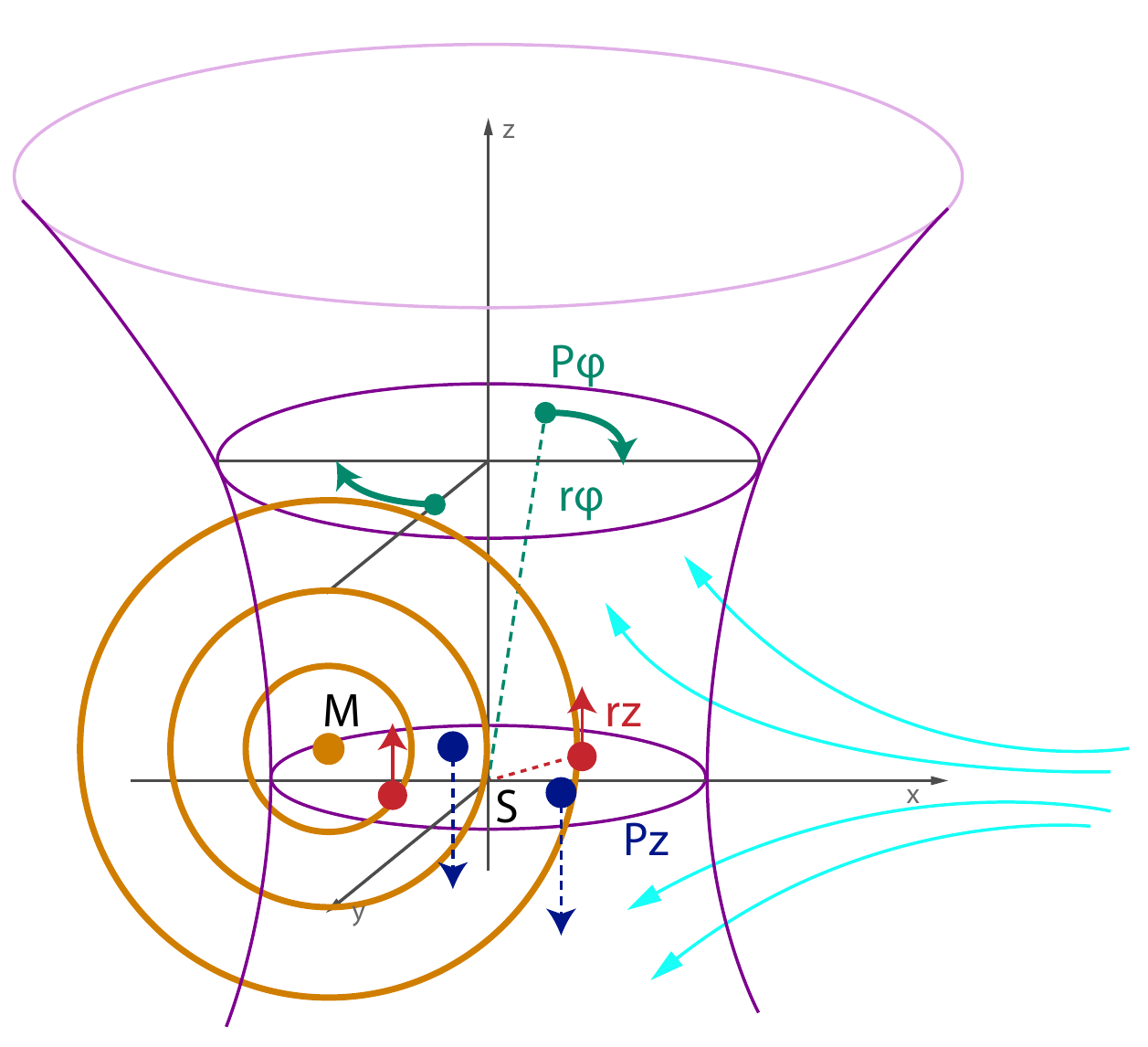}
\caption{\small{Qualitative geometry of the angular momentum distribution near a  
elliptical saddle point $S$ (see also Figure~\ref{fig-momentum3D}). 
The shape of a given tri-axial iso-density is shown in purple,
together with two cross sections, resp. in the $Sxy$ mid plane and  in a  plane containing 
the maxima of transverse angular momentum. The velocity flow in the 
$Sxz$ plane is shown in cyan.  The locus of the Lagrangian extent of 
haloes  is shown in orange concentric spheres centered on $M$. In the $Sxy$ plane, the four points, $P_z$ correspond
 to the maxima of the modulus of angular momentum.
 They point respectively  along $\mathbf e_z$ in the the first and third quadrants (in red), and along $-\mathbf e_z$ in the second and fourth quadrant (in blue).
 Conversely, the four points, $P_\phi$, correspond to the maxima  of the 
 alignment of $\mathbf s$ along $\mathbf e_\phi$. Only two (above the mid plane) are shown.
 As the orange sphere capture more than one quadrant, the $z$ component of $\mathbf L$
 subside, as it encompasses more of the neighbourhood of $P_\phi$, and
 its $\mathbf e_\phi$ component increases. In this Lagrangian framework, the spin flip
 as a function of mass is a direct
 consequence of the geometry of the tidal field imposed by the saddle point.  
 }}
\label{geometry}
\end{center}
\end{figure}
%
We consider here the  angular momentum distribution near a  
filament-saddle point, $S$. It is assumed that the three eigenvalues of the density are such that the filament going through this saddle point 
is along the vertical axis and that the other two eigenvalues are different, reflecting the presence 
of a dominant wall, in the $Sxz$ plane, in which the filament is embedded. 
 The shape of a given triaxial iso-density is shown in purple,
together with two cross sections, resp. in the $Sxy$ mid plane and  in a  plane containing 
the maxima of transverse angular momentum. 
As we will demonstrate later, the spin is mostly confined in the neighbourhood of the 
$Sz$ axis, up to a couple of correlation length of the density. It would in fact vanish,
 should the saddle become isotropic.
 In the $Sxy$ plane, we identify four quadrants corresponding to regions in which 
 the spin is parallel to the filament. Within theses quadrant, the spin
  point respectively  along $\mathbf e_z$ in the the first and third quadrants, and along $-\mathbf e_z$ in the second and fourth quadrant.
By symmetry, the spin has to vanish along $Sx$ and $Sy$.

\subsubsection{Towards a transition mass?}

The  twisted geometry of the spin near the saddle point
  also allows us to identify   the  Lagrangian transition mass corresponding to  the  alignment of dark matter haloes' spin relative to the direction of  their neighbouring filament.  Let us first consider the Lagrangian counterpart of a low-mass halo and assume it lies near the median plane.
  it will typically fall into one of the quadrant corresponding to an orientation of the spin parallel to the filament axis. 
  Now consider a halo of larger Lagrangian extent. 
  As long as its size is smaller   than the typical size of a quadrant (which will be defined
  more precisely below)
   the alignment increases, until it  over-extends the quadrant. 
  As it does, two things  happen i) it will start capturing tides from the next quadrant, 
  which would anti-align it; as the Lagrangian patch radius increases more, it reaches a size comparable to the whole tidal region of influence of the  saddle point.
It then encompasses both the clock-wise and anti-clockwise azimuthal regions, and add up  to a net momentum  of null amplitude.
 ii) it will start capturing the effect of the azimuthal tide, hence 
  inducing a spin flip. Depending on the ratio of the eigenvalues of the Hessian, the two might be concurrent or not.
  In parallel, as the radius increases, the patch collects the mean potential gradient which defines the Zel'dovich boost 
  which will drive it away from the neighbourhood of the  saddle point.
  The above description clearly accounts for the influence of only one saddle point. As we consider regions further away from that 
  saddle, we should account for the influence of other critical points, as discussed in Section~\ref{sec:statistics}.

  We have up to now considered a patch centered near the midplane close to the saddle point. Indeed, typically, in the peak background split framework,
  such patches will collapse preferentially where the density is boosted,  that is within the wall containing the filament, close to the filament.
The rarer (more massive) haloes will form in turn in the denser regions, away from the saddle point, along the filament, 
while the more common lighter haloes will form everywhere and in particular
 near  the saddle point. 
The former will have a spin perpendicular to the filament.
The latter will have a spin parallel to the filament.
The relative number of light to small haloes will depend on 
 curvilinear coordinate along the filament because consumption is important: object above the transition mass have 
swallowed their lighter parents. At a given redshift, the left overs will decide what matters.
This effect is the anisotropic  version of the well-known  cloud-in-cloud problem.

\subsubsection{Lagrangian dynamics  of spin flip}

In order to understand how a {\sl given} halo flips, let us split the original Lagrangian patch in two concentric shells.
The inner shell will correspond to the Lagrangian extent  of the halo as it initially forms, while the outer shell 
will correspond to secondary infall.
The  reasoning presented in Section~\ref{sec:premises} can be applied independently to both the inner and outer shell, and we would typically conclude that 
the outer shell would be more likely to have its spin perpendicular to the filament axis. It follows that, as far as this halo is concerned,
it will undergo a spin flip as it moves towards the core of the filament and away from the saddle.
This process will also correspond to an acquired net {\sl helicity} for the secondary infall, which will last as long as the 
transverse anisotropy of the saddle  point correlates the local tidal field. 
In effect, this consistent helicity will build up the spin of the forming
galaxy via secondary infall as it drifts, up to the point where mergers will re-orient the direction of its spin.
Hence this constructive build up of disc should only last so long as the galaxy drifts within the high-helicity region.
Note that the transverse motion will correspond to the halo entering the vortex rich caustic corresponding to the multi-flow region 
near the filament, so that this Lagrangian description remains fully consistent with the Eulerian discussion given in \cite{laigle2014}. 
We can anticipate that the longitudinal motion generates azimuthal vortices as well.

The scenario described in this section can be formalized  at two levels. 
First, within the framework of constrained random fields, one can compute 
the expected geometry of the  spin configuration near a given saddle. 
This will yield a map of the mean alignment between spin and filament in the vicinity of the saddle point.
 We will then marginalize over the expected distribution of such saddles,
 and model correspondingly the evolution of the expected mass of dark haloes around the filament.
 This will allow us to recover the numerically measured mass transition for spin flip. 
We may also test the 
mass-dependent alignment w.r.t. $\mathbf e_\phi$ in Gaussian random fields and N-body simulations.
For the sake of clarity, we will proceed in two steps: first, while assuming cylindrical symmetry
we will compute the expected spin distribution within the most likely cross section of a filament of 
infinite extend (Section~\ref{sec:2D});  then we will compute this expectation  around the most likely  3D saddle point (Section~\ref{sec:3D}).

\section{ Spin along infinite filament}
\label{sec:2D}

Let us first start while assuming that the filament is of infinite extent, so that we can restrict ourselves to cylindrical symmetry in two dimensions.
This is of interest as the angular momentum is then along the filament axis by symmetry and its derivation in the context of TTT is much simpler.
It captures already in part the mass transition, in as much as 
we can define the mean extension of a given quadrant of momentum with a given 
polarity. 
In this context, it is of interest to study the spin geometry in the median plane i.e in the vicinity of a 2D peak. This 2D spin is  along the filament,
 and will be denoted $s_{z}$ in what follows.

\subsection{Shape of the spin distribution near filaments}
\label{sec:shape2D}

Under the assumption that the {\sl  direction} of  the spin 
 along the $z$ direction is well represented by the 
 fully anti-symmetric (Levi Civita) contraction 
of the tidal tensor and  density hessian given by equation~(\ref{eq:spin-ATTT}) \citep[e.g.][]{Schafer2012},
it becomes a quadratic function of the second and fourth derivatives of the potential.
As such, it becomes possible to compute expectations of it subject to its relative position
to a peak with a given geometry (which would correspond to the cross section of the filament in the midplane).
Note that, as mentioned  in Section~\ref{sec:TTT}, standard TTT  relies, more correctly, on the  inertia tensor in place of the Hessian.
Even though they have inverse curvature of each other, their set of eigen-directions are locally the same, 
 so we expect the induced spin {\sl direction}-- which is the focus of this paper, to be the same,
 so long as the inertia tensor is well described by its local Taylor expansion.

\subsubsection{ Constrained joint PDF near peak}
\label{sec:AM2Dmap}
\label{sec:2DS}

Any matrix of second derivatives $f_{ij}$ -- rescaled so that $\left\langle (\Delta f)^{2} \right\rangle=1$-- can be decomposed into its trace $\Delta f$, and its detraced components in the frame of the separation 
\begin{equation}
f^{+}=(f_{11}-f_{22})/2 \,,\,\,\,\,
 f^{\times}=f_{12}.
 \end{equation}
  Then all the correlations between two such matrices, $f_{ij}$ and $g_{ij}$ can be decomposed irreducibly as follows.
Let us call $\xi_{f g}^{\Delta\Delta}$ ,
$\xi_{f g}^{\Delta+}$ and $\xi_{f g}^{\times\times}$ 
the correlation functions in the frame of the separation (which is the first coordinate here) between the second derivatives of the field $f$ and $g$ separated by a distance $r$:
\begin{align}
\label{eq:xi}
\xi_{f g}^{\Delta\Delta}(r)\!&=\!\left\langle\Delta f \Delta g\right\rangle\!,\nonumber \\
\xi_{f g}^{\Delta +}(r)\!&=\!\left\langle\Delta f g^{+}\! \right\rangle\!,\nonumber\\
\xi_{f g}^{\times\times}(r)\!&=\!\left\langle f^{\times}\!g^{\times}\!\right\rangle\!. 
\end{align}
All other correlations are trivially expressed in terms of the above as 
\begin{align}
\left\langle f^{\times} \Delta g\right\rangle&=0,\,\,
\left\langle f^{+}\! g^{\times} \right\rangle=0,\,\, \nonumber \\
\left\langle f^{+}\! g^{+} \right\rangle&=\frac 1 4 \xi_{f g}^{\Delta\Delta}(r)-\xi_{f g}^{\times\times}(r)\,.
\end{align}
Here, we consider two such fields, namely the gravitational potential $\Phi$ and the density $\delta$. In the following these two fields and their first and second derivatives are assumed to be rescaled by their variance $\sigma_{0}^{2}=\left\langle \Phi^{2}\right\rangle$, $\sigma_{1}^{2}=\left\langle  (\nabla\Phi)^{2}\right\rangle$, $\sigma_{2}^{2}=\left\langle  (\delta=\Delta\Phi)^{2}\right\rangle$, $\sigma_{3}^{2}=\left\langle  (\nabla\delta)^{2}\right\rangle$ and $\sigma_{4}^{2}=\left\langle  (\Delta\delta)^{2}\right\rangle$. The shape parameter of the density field is defined as 
\begin{equation}
\gamma=\sigma_{3}^{2}/(\sigma_{2}\sigma_{4}).
\end{equation}
 The rescaled potential and density will be denoted by $\phi$ and $x$ and the rescaled first and second derivatives by $\phi_{i}$, $x_{i}$ and $\phi_{ij}$, $x_{ij}$.

Let us gather the first and second derivatives of the gravitational field and the first and second derivatives of the density in a vector denoted by $\mathbf{X}$ spatially located in ${\bf r}_{X}$ and $\mathbf{Y}$ located in ${\bf r}_{Y}$. 
The Gaussian joint PDF of  $\mathbf{X}$ and  $\mathbf{Y}$ at the two given locations (${\bf r}_{X}$ and ${\bf r}_{Y}$ separated by a distance $r=|{\bf r}_{X}-{\bf r}_{Y}|$) 
obeys 
\begin{equation}
{\cal P}(\mathbf{X},\mathbf{Y})=\! \frac{1}{\sqrt{{\rm det}| 2\pi \mathbf{C}|}  }
\exp\!\left(\!\!-\frac{1}{2}\left[\!\begin{array}{c}\label{eq:defPDFMT}
 \mathbf{X} 
\\
\mathbf{Y} 
 \\
\end{array} \!\right]^{\rm T}
 \!\!\!\!\cdot \mathbf{C}^{-1} \!\!\!\!\!\cdot \left[\!\begin{array}{c}
 \mathbf{X} 
\\
\mathbf{Y} 
 \\
\end{array}\! \right] \right)\,,
\end{equation} 
where $\mathbf{C}_{0}\equiv \langle  \mathbf{X}\cdot \mathbf{X}^{\rm T} \rangle$,
  $\mathbf{C}_{\gamma}\equiv \langle  \mathbf{X}\cdot \mathbf{Y}^{\rm T} \rangle$
and 
\[
\quad \mathbf{C}=\left[
\begin{array}{cc}
\mathbf{C}_{0} &\mathbf{C}_\gamma
\\
\mathbf{C}_\gamma^{\rm T}  &\mathbf{C}_{0} 
 \\
\end{array}
\right]\,. 
\]
All these quantities depend on the separation vector $\textbf{r}$ only because of statistical homogeneity. 
This PDF is sufficient to compute the expectation of any quantity involving  derivatives of
 the potential  and the density up to second order. All the coefficients can easily be computed from the power spectrum of the potential 
  \begin{multline}
\left\langle{\partial_{1}^{i_{1}} \partial_{2}^{i_{2}}\phi} ,\partial_{1}^{j_{1}} \partial_{2}^{j_{2}}\phi\right\rangle=
 \int_{0}^\infty \!\!\! \int_0^{2\pi}  \!\!\! \!\!\!  {\rm d} \theta\; {\rm d} k \; P_k(k) \exp(\imath k\, r\cos \theta )\\
 \imath^{i_{1}+i_{2}} (-\imath)^{j_{1}+j_{2}}(\cos\! \theta)^{i_{1}+j_{1}}  (\sin \! \theta)^{i_{2}+j_{2}} \frac{  k^{i_{1}+i_{2}+j_{1}+j_{2}+1}}{\sigma_{i_{1}+i_{2}} \sigma_{j_{1}+j_{2}}}
 \,,
\label{eq:defpowespectra}
\end{multline}
and
\begin{equation}
\sigma_{n}^{2}=
 \int_{0}^\infty \!\!\! \int_0^{2\pi}  \!\!\! \!\!\!  {\rm d} \theta\; {\rm d} k \; P_k(k) 
  k^{2n+1}
 \,,
 \nonumber
\end{equation}
{\green{where the power spectrum of the potential $P_k(k) $ can include a filter function on a given scale. In this work,  we use a Gaussian filter defined in Fourier space by
\begin{equation}
    W_{G}(\mathbf{k},R)=
    \frac{1}{(2\pi)^{3/2}}\exp\left(\frac{-k^{2}R^{2}}{2}\right)
    \,.
    \end{equation}
}}
 For instance, the one-point covariance matrix for $(\Delta\phi,\phi^{+}\!,\phi^{\times}\!,\Delta x, x^{+}\!,x^{\times}\!)$
 at  a given point simply reads
 \begin{equation}
\mathbf{C}_{02}=
 \left(
\begin{array}{cccccc}
1&0&0&-\gamma&0&0\\
0&1/8&0&0&-\gamma/8&0\\
0&0&1/8&0&0&-\gamma/8\\
-\gamma&0&0&1&0&0\\
0&-\gamma/8&0&0&1/8&0\\
0&0&-\gamma/8&0&0&1/8\\
\end{array}
\right)\,, \nonumber
 \end{equation}
 where $\gamma=\sqrt{(n+2)/(n+4)}$ for a scale-invariant density power-spectrum with spectral index $n$ (i.e $n-4$ for the potential).
 Note that the first derivatives of the density and the potential fields are decorrelated from the second derivatives meaning that
  \begin{equation}
\mathbf{C}_{0}=
 \left(
\begin{array}{cc}
\mathbf{C}_{01}&0\\
0&\mathbf{C}_{02}\\
\end{array}
\right)\,, \nonumber
 \end{equation}
 where $\mathbf{C}_{01}$ is the one-pt covariance matrix of the gradients of the potential and the density fields $(\phi_{1},\phi_{2},x_{1},x_{2})$ as a function of $\gamma'=\sigma_{2}^{2}/\sigma_{1}/\sigma_{3}$
  \begin{equation}
\mathbf{C}_{01}=
 \left(
\begin{array}{cccccc}
1/2&0&-\gamma'/2&0\\
0&1/2&0&-\gamma'/2\\
-\gamma'/2&0&1/2&0\\
0&-\gamma'/2&0&1/2\\
\end{array}
\right)\,. \nonumber
 \end{equation}
 
The two-point covariance matrix, $ \mathbf{C}_\gamma$  can be similarly derived. In particular, its restriction to the second derivatives of the density and the potential fields can be written as a function of the nine $\xi$ functions defined in equation~(\ref{eq:xi}) (for $fg=\phi\phi,\phi x,xx$) (see Figure~\ref{fig-xi-2D}
and Appendix~\ref{sec:xi}):
 \begin{equation}
\left(
\begin{array}{cccccc}
\xi_{\phi\phi}^{\Delta\Delta} & \xi_{\phi\phi}^{\Delta+} &0& \xi_{\phi x}^{\Delta\Delta} & \xi_{\phi x}^{\Delta+} &0\\
\xi_{\phi\phi}^{\Delta+} & \frac{\xi_{\phi\phi}^{\Delta\Delta} }{4}-\xi_{\phi\phi}^{\times\times}  &0& \xi_{\phi x}^{\Delta+} & \frac{\xi_{\phi x}^{\Delta\Delta} }4-\xi_{\phi x}^{\times\times}  &0\\
0&0 &\xi_{\phi\phi}^{\times\times} &0 &0&\xi_{\phi x}^{\times\times} \\
\xi_{\phi x}^{\Delta\Delta} & \xi_{\phi x}^{\Delta+} &0&\xi_{x x}^{\Delta\Delta} & \xi_{x x}^{\Delta+} &0\\
 \xi_{\phi x}^{\Delta+} & \frac{\xi_{\phi x}^{\Delta\Delta} }4-\xi_{\phi x}^{\times\times}  &0& \xi_{x x}^{\Delta+} &\frac{ \xi_{x x}^{\Delta\Delta} }4-\xi_{x x}^{\times\times}  &0\\
0 &0&\xi_{\phi x}^{\times\times} & 0&0 &\xi_{x x}^{\times\times}\\
\end{array}
\right) \nonumber
 \end{equation}

\begin{figure}
\begin{center}
\includegraphics[width=0.95\columnwidth]{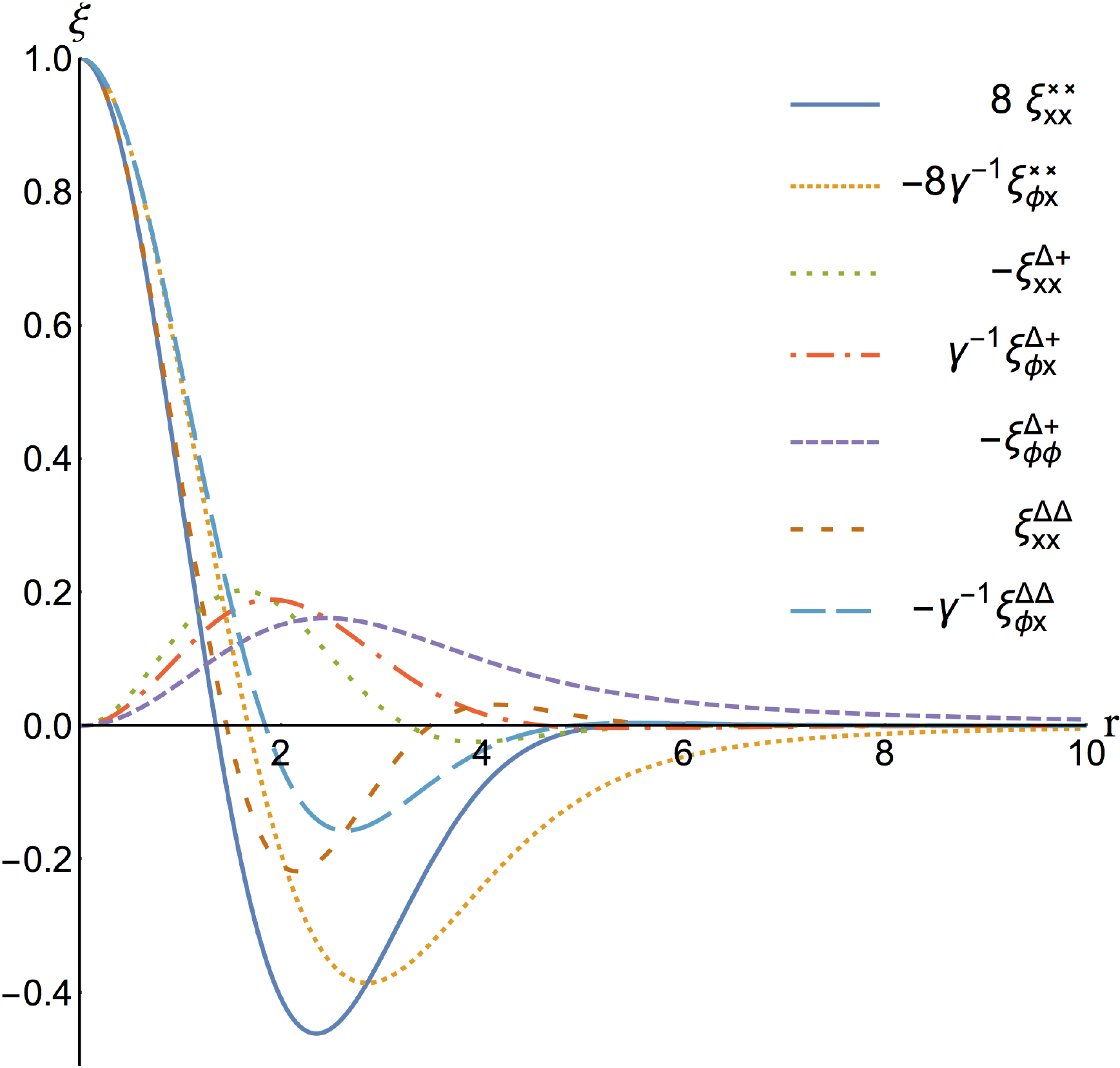}
\caption{Two-point correlation functions as a function of the separation $r$ in units of the smoothing length for a power-law 2D power spectrum with spectral index $n=-1/2$ i.e $\gamma=\sqrt{3/7}$. 
For aesthetic purpose, these functions have been rescaled by their value in $r=0$ as explicitly written in the legend.}
\label{fig-xi-2D}
\end{center}
\end{figure}

Once the joint PDF given by Equation~(\ref{eq:defPDFMT}) is known, it is straightforward to compute conditional PDFs (in particular subject to a critical point constraint ${\cal C}({\rm crit})=|\det\! \left(x_{ij}\right)|\delta_\mathrm{ D}(x_{i})$).
Given the conditionals, simple 
algebra then yield the conditional density and spin.
More specifically, relying on Bayes theorem, the conditional can be expressed in terms of the joint PDF -- equation~(\ref{eq:defPDFMT})-- as
\begin{equation}
{\cal P}(\mathbf{X}| \mathbf{Y},{\rm pk})=\! \frac{{\cal P}(\mathbf{X} ,  \mathbf{Y}, {\rm pk}) }{{\cal P}(\mathbf{Y}, {\rm pk}) }\,,
 \nonumber
\end{equation}
where
\begin{equation}
{\cal P}(\mathbf{Y}, {\rm pk})=\int {\rm d}  \mathbf{Y} {\cal P}(\mathbf{X} ,  \mathbf{Y}) {\cal C}({\rm pk})
 \nonumber
\end{equation}
is the marginal distribution describing the likelihood of a given peak, $\rm pk$ (the transverse cross-section of an infinite filament)
with a given geometry. Once the conditional, ${\cal P}(\mathbf{X}| \mathbf{Y},{\rm pk})$ is known, it is straightforward\footnote{see \url{http://tinyurl.com/mmbse3z} which describes an implementation in {\sc mathematica} of the conditional probability. }
 to compute
the expectation of any function, $f(\mathbf{X})$ as 
\begin{equation}
\langle f(\mathbf{X}) | {\rm pk} \rangle= \int {\rm d}  \mathbf{X}\, {\cal P}(\mathbf{X} |  {\rm pk})  f(\mathbf{X})\,,
\label{eq:defexpectation}
\end{equation}
which, when $ f(\mathbf{X})$ is multinomial  in  the components of $\mathbf{X}$ can be carried out analytically.
In the following, we will consider in turn functions which are indeed algebraic function of  $\mathbf{X}$.
\subsubsection{Constrained density maps}
\label{sec:dens2Dmap}
\label{sec:2DS}

\begin{figure*}
\begin{center}
\includegraphics[width=.95\columnwidth]{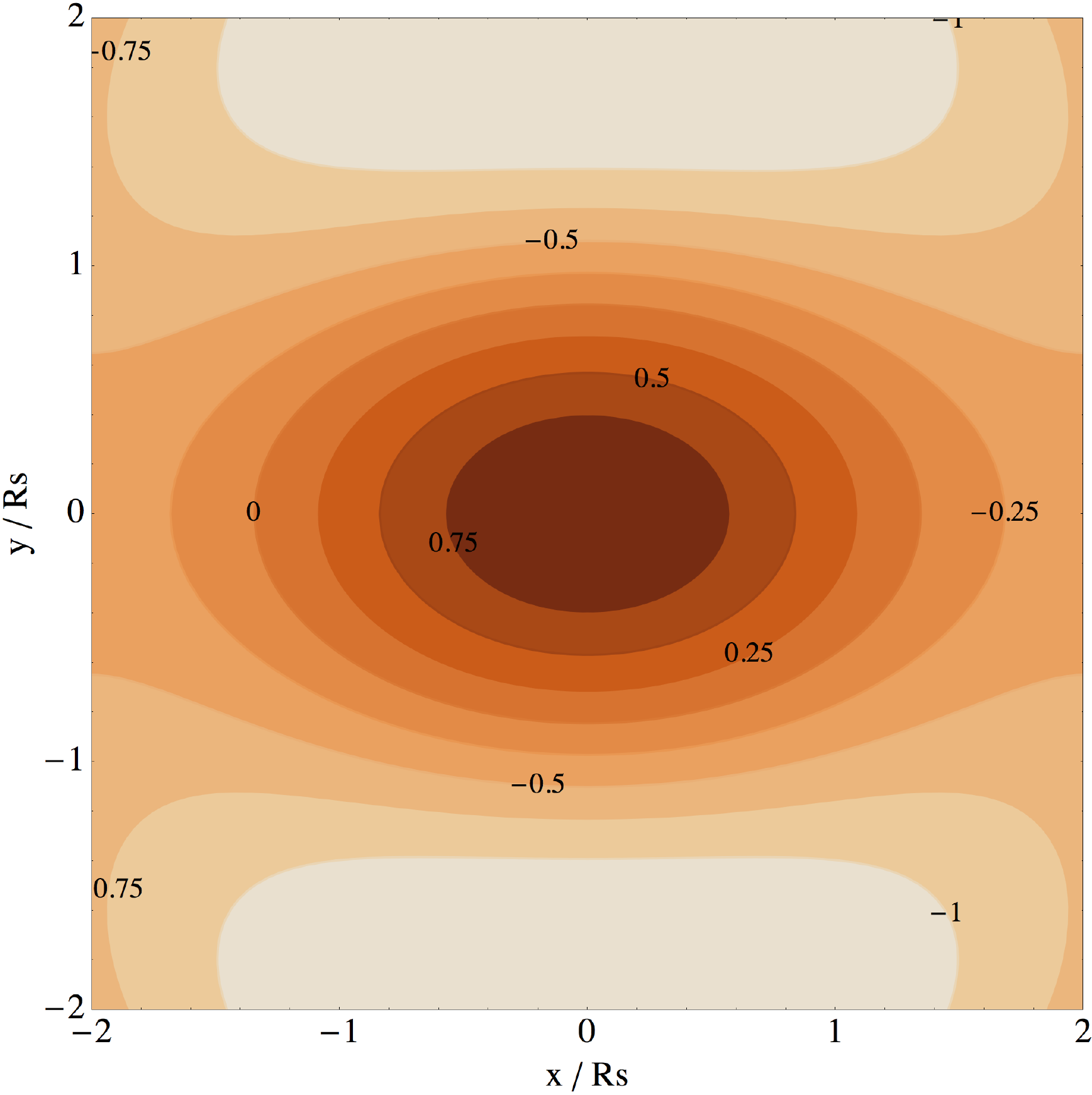}\hspace{1cm}
\includegraphics[width=.95\columnwidth]{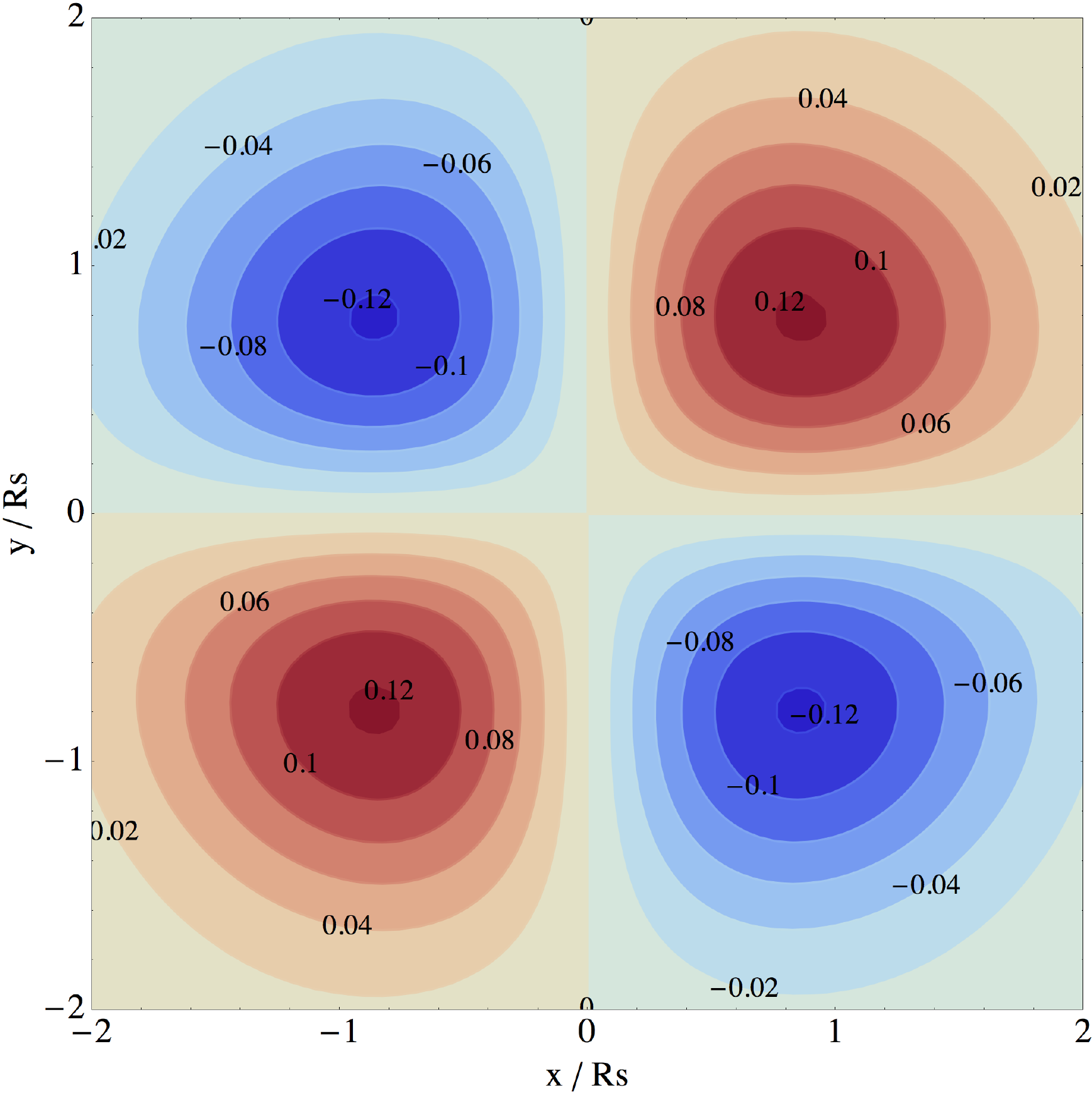}
\caption{\small{Left: mean density (contrast) field near a 2D peak of height $\nu=1$, $\lambda_{1}=-1$ and $\lambda_{2}=-2$ for a power spectrum with index $n=1/2$ computed from
Equation~(\ref{eq:defDensity2D}). Contours are displayed from $\delta=-1$ to 1 by step of 1/4 as labeled. The $x$ and $y$ axes are in units of the smoothing length.
Right: corresponding mean spin colour coded from blue (negative) to red (positive) computed from
Equation~(\ref{eq:defL2D}). The flattening of the filament's cross section induces a clear quadrupolar
spin distribution in its vicinity. 
 }}
\label{fig:2Dspin}
\end{center}
\end{figure*}

From Equation~(\ref{eq:defPDFMT}), given a contrast $\nu$ and  a geometry for the saddle (or any critical point) defined by $\kappa=\lambda_{1}-\lambda_{2},I_{1}=\lambda_{1}+\lambda_{2}$ (where $\lambda_{1}>\lambda_{2}$ are the two eigenvalues of the Hessian of the density field $\mathbf H$ -- both negative for a peak),
 the mean density contrast, $\delta_{\rm ext}=\langle \delta  | {\rm ext} \rangle$,
 (in units of $\sigma_{2}$) around the corresponding critical point
can be analytically computed
\begin{align}
\delta(\mathbf{r}| {\rm ext},\kappa,I_{1},\nu)=&\frac{I_{1} (\xi_{\phi x}^{\Delta\Delta} +\gamma \xi_{\phi\phi}^{\Delta\Delta} )+
\nu  (\xi_{\phi \phi}^{\Delta\Delta}+\gamma \xi_{\phi x}^{\Delta\Delta})}{1-\gamma ^2} \nonumber
\\&
+4\left(\mathbf{\hat r}^{\rm T} \!\!\cdot \overline{\mathbf{ H}}\cdot\mathbf{\hat r}\right) \xi_{\phi x}^{\Delta +} \,,
\label{eq:defDensity2D}
\end{align}
where $\overline{\mathbf{ H}}$ is the detraced Hessian of the density and
$\mathbf{\hat r}={\mathbf{r} }/{r}$
so that 
\begin{equation}
\mathbf{\hat r}^{\rm T}\!\! \cdot \overline{\mathbf{ H}}\cdot\mathbf{\hat r}= \kappa\frac{\cos (2 \theta )}{2}\,,
\end{equation}
$r$ being the distance to the critical point and $\theta$ the angle from the eigen-direction corresponding to the first eigenvalue $\lambda_{1}$ of the critical point. When $r$ goes to zero,  given the properties of the $\xi$ functions (see Figure~\ref{fig-xi-2D}),
the density trivially converges to the constraint $\nu$.

\subsubsection{  Constrained 2D spin maps}
\label{sec:dens2Dmap}
\label{sec:2DS}

In two dimensions, the spin is a  scalar  given by
\begin{equation}
{s}_z (\mathbf r) = \sum_{i,j,k} \epsilon_{ij3}  \phi_{ik}   x_{jk}\,, \label{eq:defL2D}
\end{equation}
where   $\epsilon=\epsilon_{ij3}$ is built upon the totally anti-symmetric rank 3 Levi-Civita tensor $\epsilon_{ijk}$.
Since equation~(\ref{eq:defL2D}) is quadratic in the fields $x$ and $\phi$, 
equation~(\ref{eq:defexpectation}) can be readily applied to compute analytically its conditional expectation.
The angular momentum generated by TTT as a function of the polar
position $(r,\theta)$  subject to the same critical point constraint at the origin with 
contrast
$\nu$, and principal curvatures $(\lambda_{1},\lambda_{2})$
is given by the sum of a quadrupole ($\propto \sin 2\theta$) and an octupole ($\propto \sin 4\theta$)
\begin{align}
\langle {s}_z  | {\rm ext}\rangle&=   s_z(\mathbf{r} | {\rm ext},\kappa,I_{1},\nu) \,,  \label{eq:L2D}\\
 &=-16(\mathbf{\hat r}^{\rm T}\!\!\cdot \epsilon \cdot \overline{\mathbf{ H}}\cdot\mathbf{\hat r})\,
 (s_{z}^{(1)}+2 (\mathbf{\hat r}^{\rm T}\!\! \cdot \overline{\mathbf{ H}}\cdot\mathbf{\hat r}\nonumber
) s_{z}^{(2)})
\,,
\end{align}
 where  the octupolar coefficient $s_{z}^{(2)}$ can be written as
  \begin{equation}
  s_{z}^{(2)}(r)=  \left(\xi_{\phi x}^{\Delta\Delta}\xi_{x x}^{\times\times}-\xi_{\phi x}^{\times\times}\xi_{x x}^{\Delta\Delta}
  \right)\,,
   \nonumber
     \end{equation}
     and the quadrupolar coefficient
     $s_{z}^{(1)}$ reads
     \begin{align}	
  s_{z}^{(1)}(r)=& 
  \frac{\nu}{1-\gamma^{2}}
   \left[
  (\xi_{\phi \phi}^{\Delta +}+\gamma \xi_{\phi x}^{\Delta +})\xi_{x x}^{\times\times}
  -(\xi_{\phi x}^{\Delta +}+\gamma \xi_{x x}^{\Delta +})\xi_{\phi x}^{\times\times}
  \right]\nonumber
  \\
 +&\frac{ I_{1}}{1-\gamma^{2}} \left[
 (\xi_{\phi x}^{\Delta +}+\gamma \xi_{\phi \phi}^{\Delta +})\xi_{x x}^{\times\times}
-
  (\xi_{x x}^{\Delta +}+\gamma \xi_{\phi x}^{\Delta +})\xi_{\phi x}^{\times\times}
    \right]\,,
   \nonumber
     \end{align}
     while 
     \begin{equation}
\mathbf{\hat r}^{\rm T}\!\!\cdot \epsilon \cdot \overline{\mathbf{ H}}\cdot\mathbf{\hat r}= -\kappa\frac{\sin (2 \theta )}{2}\,.
\end{equation} 
Equation~(\ref{eq:L2D}) is remarkably simple.
  As expected, the spin, $s_z$, is identically null if the filament is axially symmetric  ($\kappa=0$).
It is zero along the principal axis of the Hessian (where $\theta=0 \mod \pi/2$ for which $\mathbf{\hat r}^{\rm T}\cdot \epsilon \cdot \mathbf{\hat r}=0$).
Near the peak, the anti-symmetric, $\sin(2\theta)$,  component dominates, and the spin distribution is quadrupolar. 
For scale-invariant density power spectra with index $n$ ($n-4$ for the {\sl potential}), $s_z$ 
can be computed explicitly.
At small separation, $s_z$  behaves like
 \begin{equation}
s_z \propto   \kappa(  (n+2)\nu +\sqrt{(n+2)(n+4)}I_{1})r^{2}\sin(2 \theta ), \label{eq:Lz-asymptote}
 \end{equation}
 which shows explicitly  that the quadrupolar term dominates. 
 Figure~\ref{fig:2Dspin} displays the mean density and spin map for a power-law power spectrum with index $n=1/2$ around a 2D peak of the density field with geometry $\nu=1$, $\lambda_{1}=-1$ and $\lambda_{2}=-2$. 
 
 At this stage it is interesting to understand how much angular momentum is contained into spheres of increasing radius that would feed the forming object at different stages of its evolution.
 For instance let us assume there is a small-scale overdensity at (one of the four) location of maximum angular momentum (denoted $r_{\star}$ hereafter) and let us filter the spin field with a top-hat window function centered on $r_{\star}$ and of radius $R_{\rm TH}$. The resulting amount of angular momentum as a function of this top-hat scale is displayed in Figure~\ref{fig:spinTH}. During the first stage of evolution, the central object will acquire spin constructively until it reaches a Lagrangian size of radius $R_{\rm TH}=r_{\star}$ and feels the 2 neighbouring quadrants of opposite spin direction. The spin amplitude then decreases and becomes even negative before it is fed by the last quadrant of positive spin. The minimum is reached for radius around $2.4 r_{\star}$. This result does not change much with the contrast and the geometry of the peak constraint.
 %
 \begin{figure}
\begin{center}
\includegraphics[width=0.95\columnwidth]{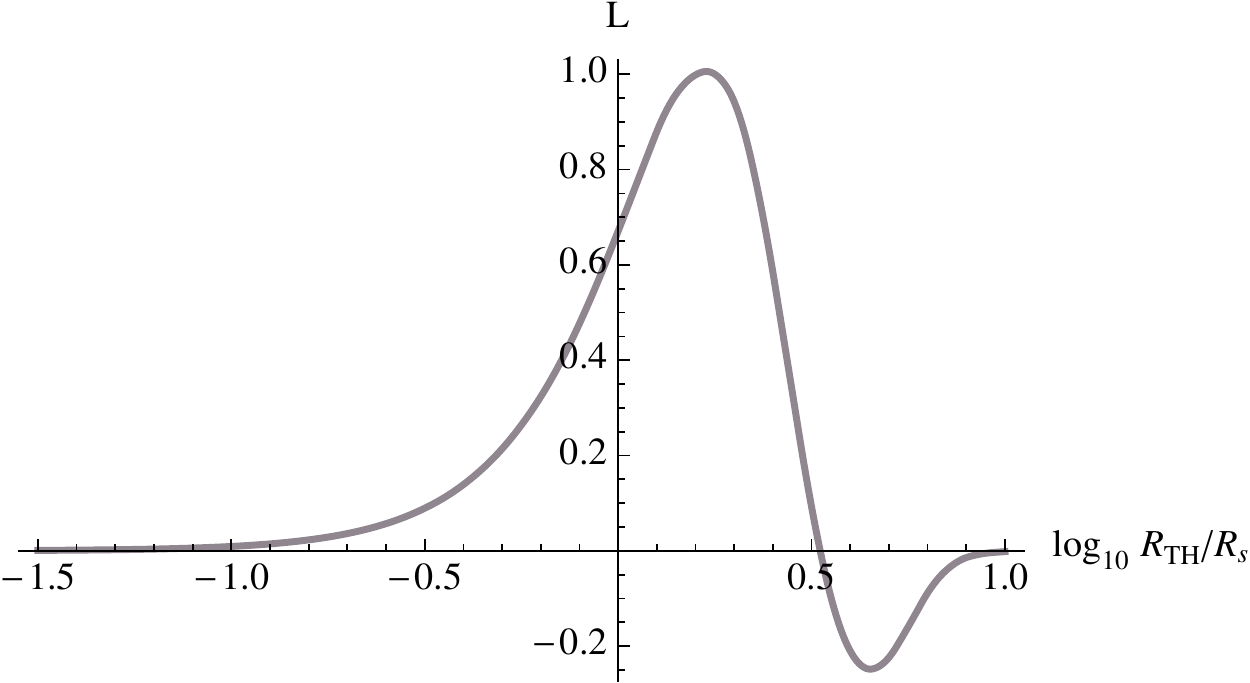}
\caption{Evolution of the amount of algebraic angular momentum in sphere of radius $R_{\rm TH}$ centered on $r_{\star}$ The density power spectrum index is $n=-3/2$, the height of the peak in $(0,0)$ is $\nu=1$ and principal curvatures $\lambda_{1}=-1,\lambda_{2}=-2$. The amplitude of the spin is normalised by its maximum value around $R_{\rm TH}=r_{\star}$
}
\label{fig:spinTH}
\end{center}
\end{figure}
%
Figure \ref{fig:spinTH} is the Lagrangian counterpart of Figure~4 of \cite{laigle2014} (or Figure~7 of \cite{pichonbernardeau1999}) which displays the quadrant of vorticity in the vicinity of filaments.
\subsubsection{Cosmic variance on spin }
\label{sec:variance2D}
On top of the mean spin, one can also compute the dispersion of the spin described by
\begin{equation}
\label{eq:sigma2D}
\sigma(\mathbf r)=\sqrt{\left\langle s_{z}^{2}(\mathbf r)\right\rangle-\left\langle s_{z}(\mathbf r)\right\rangle^{2}}\,.
\end{equation}
A map of this spin dispersion is shown on Figure~\ref{fig:sigma2D}.
 Comparing Figure~\ref{fig:sigma2D} to Figure~\ref{fig:2Dspin}, we see that  spin direction fluctuates  along the 
major axis of the filament cross section, and best defined along its minor axis. 
As the conditional statistics is Gaussian, 
the whole spin statistics (third moments, ...)
 can in principle be similarly computed.

\begin{figure}
\begin{center}
\includegraphics[width=0.95\columnwidth]{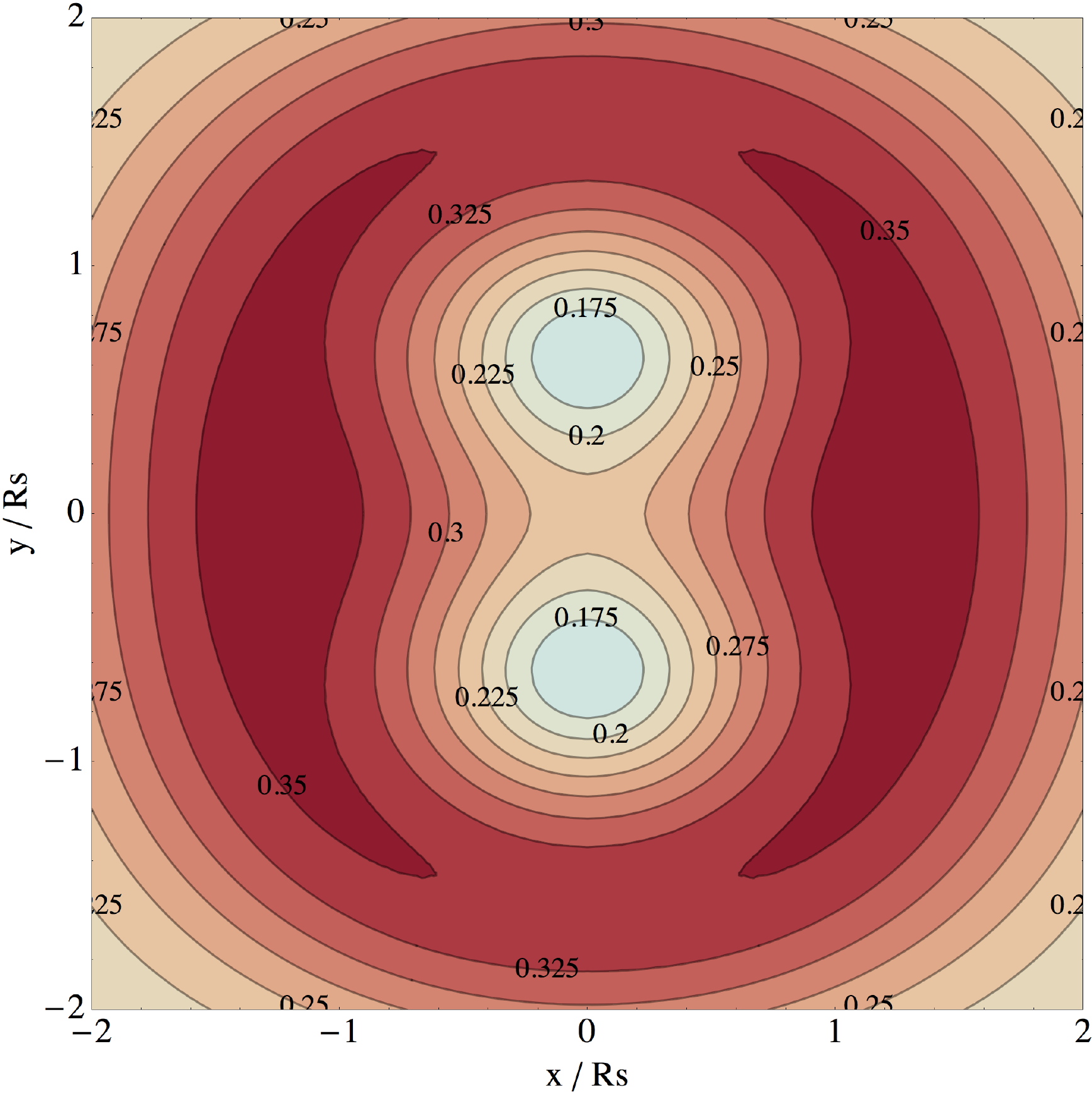}
\caption{\small{ 2D spin dispersion (defined in Equation~(\ref{eq:sigma2D})) near a 2D peak of height $\nu=1$ and curvatures $\lambda_{1}=-1$ and $\lambda_{2}=-2$ for a power spectrum with index $n=1/2$}}
\label{fig:sigma2D}
\end{center}
\end{figure}

\subsubsection{  Zel'dovich mapping  of the Spin }
\begin{figure*}
\begin{center}
\includegraphics[width=0.95\columnwidth]{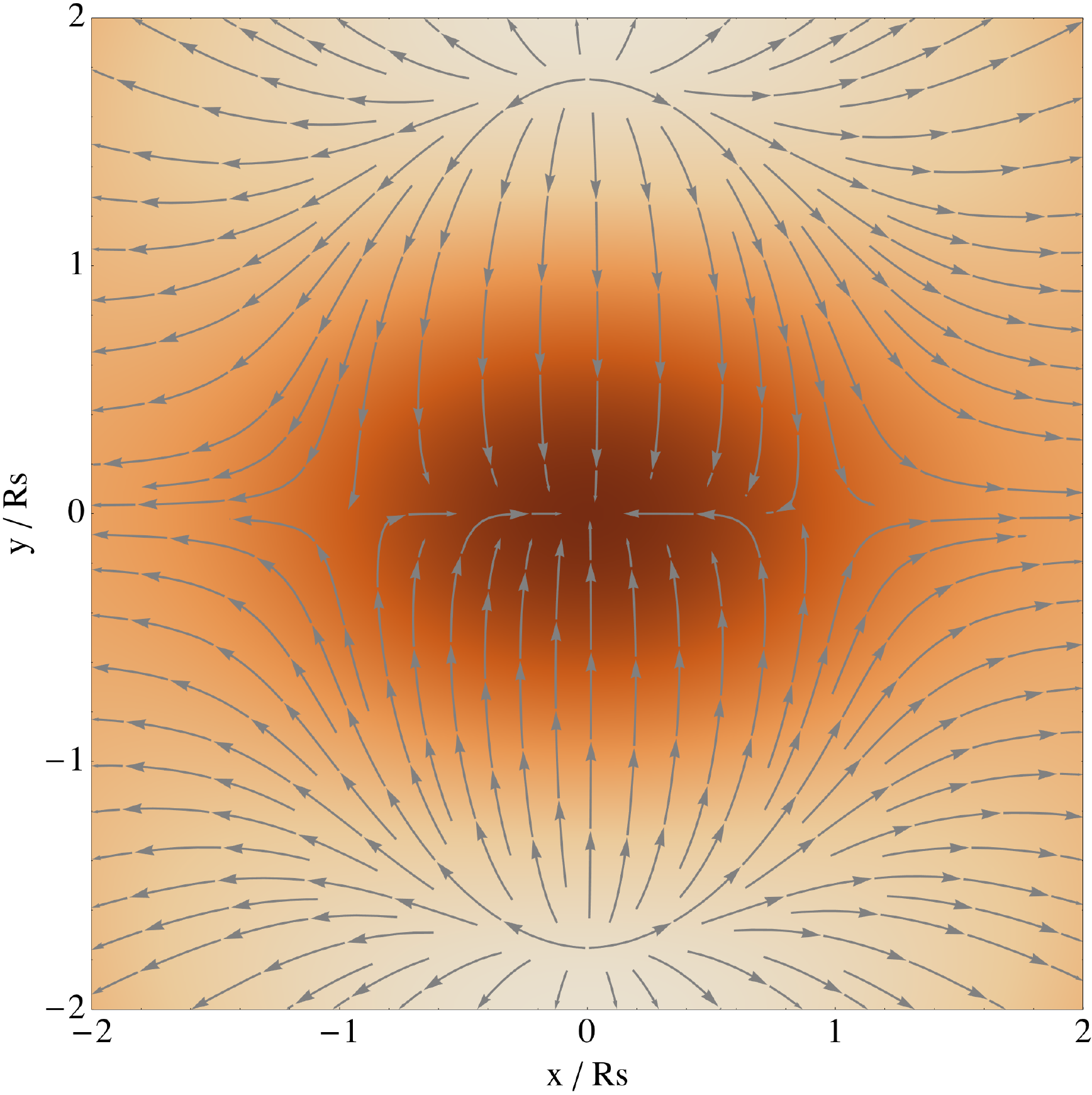}\hspace{1cm}
\includegraphics[width=0.95\columnwidth]{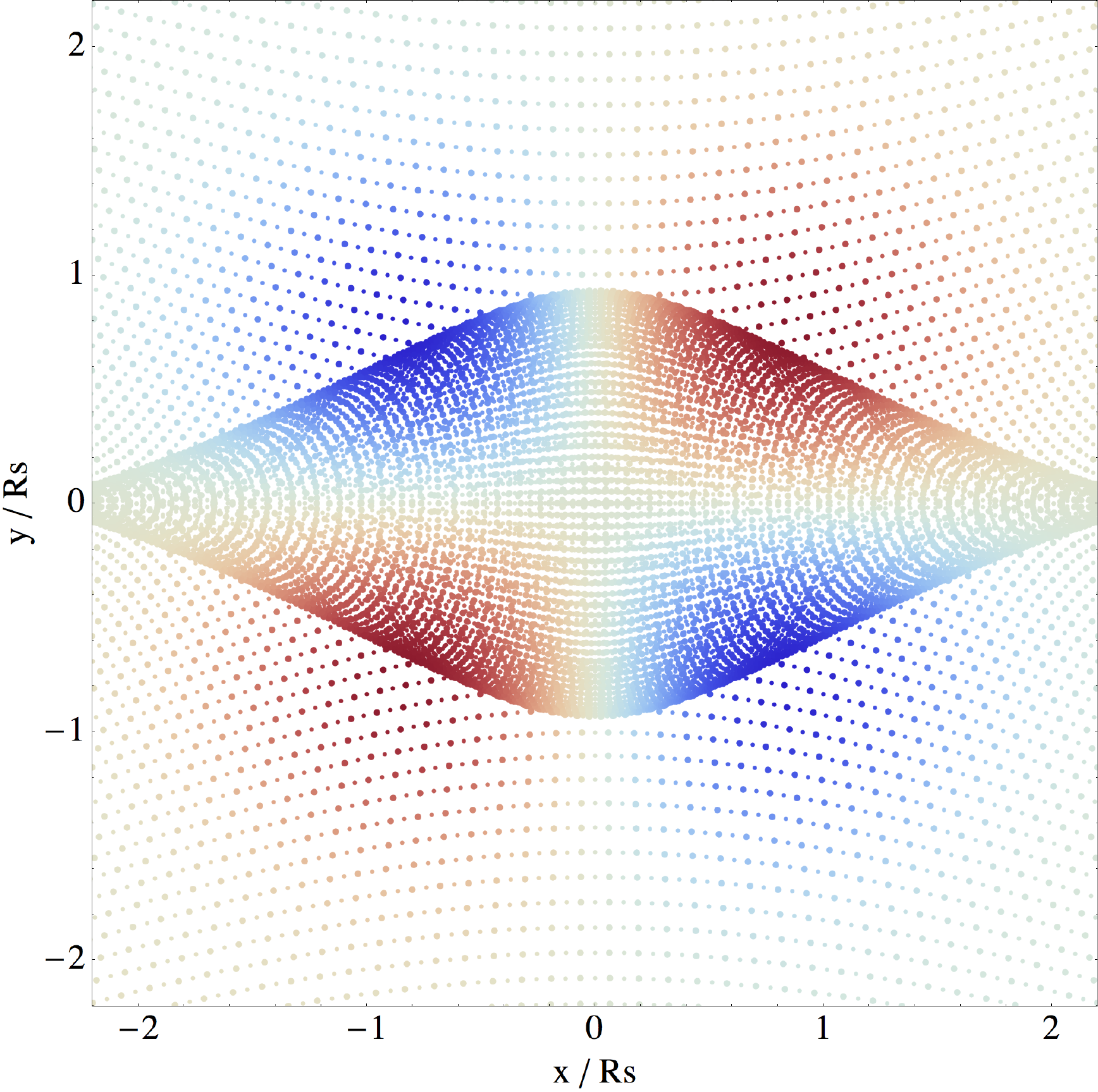}
\caption{\small{{\sl Left:} Stream lines of the 2D velocity field (defined as the potential gradient) near a 2D peak of height $\nu=1$ and curvatures $\lambda_{1}=-1$ and $\lambda_{2}=-2$ for a power spectrum with index $n=1/2$. {\sl Right}: Zel'dovich mapping of the spin distribution.
There  is a good qualitative agreement between the vorticity section presented in \protect\cite{laigle2014} and this spin map.}
}
\label{fig:zeldomap}
\end{center}
\end{figure*}

Figure~\ref{fig:zeldomap} displays the image of the initial density field  (resp. initial spin field) translated  by a Zel'dovich displacement. The displacement is proportional to $(\phi_{1},\phi_{2})$ here and its expectation given a central peak is trivially computed from the conditional PDFs. The resulting quadrupolar caustics is qualitatively similar to the quadrupolar geometry of the vorticity field measured in numerical simulations \citep{laigle2014}.
Indeed, as discussed in that paper, there is a dual relationship between such Eulerian vorticity maps and  
the geometry of the spin distribution within the neighbouring patch of a 3D saddle point.

\subsection{Transition mass for long filaments }
\label{sec:Transition}

Up to know we assumed  that the geometry of the critical point was given. Let us now build the joint statistics of the spin and the mass  near 2D peaks.

\subsubsection{ Geometry of the most likely cross section}

Let us now study what should be the typical geometry of a peak.
 Following \cite{pogo09}, it is straightforward to derive the PDF for a point to have height $\nu$ and geometry $\kappa,I_{1}$ as in their notation $J_{2}=\kappa^{2}$ so that
 \begin{equation}
 {\cal P}(\nu,\kappa,I_{1})=\frac{\kappa}{\pi\sqrt{1-\gamma^{2}}}\exp \left(\!\!-\frac 1 2 \!\left(\frac{\nu+\gamma I_{1}}{\sqrt{1-\gamma^{2}}}\right)^{\!2}\!\!\!\!-\frac 1 2 I_{1}^{2}\!\!-\kappa^{2}\!\!\right)\,. \nonumber 
 \end{equation}
 Now the PDF for a {\it peak} to have height $\nu$ and geometry $\kappa,I_{1}$ becomes:
  \begin{multline}
 {\cal P}(\nu,\kappa,I_{1}|pk)=\frac{\sqrt{3}\kappa|(I_{1}-\kappa)(I_{1}+\kappa)|}{2\pi\sqrt{1-\gamma^{2}}}\Theta(-\kappa-I_{1})\times
 \\
 \exp \left(\!\!-\frac 1 2 \!\left(\frac{\nu+\gamma I_{1}}{\sqrt{1-\gamma^{2}}}\right)^{\!2}\!\!\!\!-\frac 1 2 I_{1}^{2}\!\!-\kappa^{2}\!\!\right)
\,.\label{eq:2Dgeom}
 \end{multline}
The maximum of this PDF is trivially reached for $\bar \nu=\sqrt{7/3}\,\gamma$, $\bar \kappa=\sqrt{1/3}$ and $\bar {I_{1}}=-\sqrt{7/3}$.

\subsubsection{ The size and area of constant polarity quadrants}

   From equation~(\ref{eq:L2D}), it appears clearly that the extension of the region of influence of the critical point is limited, and peaks within each quadrant at some 
   specific $(r_\star,\theta_\star)$ position. 
   Moreover, for small enough $\kappa$,
   the quadrupole dominates, and the extremum is along $\theta=\pi/4$.
   It is therefore possible to use $r_\star$
   to define  an area in which the spin is significantly non zero within each quadrant.
Let us compute $r_\star$,   as the radius for which $s_z(\theta=\pi/4)$
is maximal as a function of $r$\footnote{Setting $\theta=\pi/4$ effectively neglect
the octupolar part of $s_z$.}. 
The area of a typical quadrant, in which the spin has the same orientation, can then simply be expressed as  
\begin{equation}
{\cal A}= \pi  r_{\star}^{2  }\,, \label{eq:defarea}
\end{equation}
where 
 $r_{\star}=r_{\star}(\nu,\kappa)$ is the position of a maximum of angular momentum from the peak.
 {
Because of the quadrupolar antisymmetric geometry of the angular momentum distribution near the saddle point,
it is typically twice as small (in units of the smoothing length) as one would naively expect.}

For power-law density power spectrum with spectral index in the range 
$n\in ]-2,2]$, a good fit to its scaling  is  given by

\begin{equation}
\frac{r_\star}{R_{s}} \approx \frac{3}{250} \left(n-5\right)^2+\frac{13}{10}\,,
\label{eq:rstarOfn}
\end{equation}
where $r_{\star}$ was computed for the mean geometry given by $\bar \nu=\sqrt{7/3}\,\gamma$, $\bar \kappa=\sqrt{1/3}$ and $\bar {I_{1}}=-\sqrt{7/3}$.

\subsubsection{Critical mass scaling}
The critical mass is the mass of maximum spin alignment. In simulation, 
it has been shown by \cite{laigle2014} to be $M_{\rm crit}\approx 10^{12} M_{\sun}$ at redshift 0. The authors claimed that the critical mass is related to the mass contained in a typical quadrant of vorticity.
In this work, we have computed in Lagrangian space the typical area of a quadrant (see Equation~\ref{eq:defarea}).
This area is a function of the smoothing scale. In order to compute it, we need to define a scale. It is reasonable  that the maximum spin alignment should be reached for filament that has just collapsed at redshift 0. Indeed, for larger scale filaments, part of the haloes do not lie inside the filament but in the nearby wall which will therefore decrease the mean spin-filament alignment. In previous sections, we focused on $\nu=0.9$ filaments. The model of the cylindrical collapse then say that those filaments have just collapsed at redshift 0 for a top-hat initial smoothing scale $\sigma(R_{\textrm{ TH}})=1.6$ which corresponds to a smoothing length $R_{\textrm{ TH}}=2.2 $Mpc$/h$. We can therefore compute the corresponding $r_{\star}$ which is $r_{\star}\approx 1.6 R_{s}\approx 0.7 R_{\textrm {TH}}=1.5 $Mpc$/h$ and corresponds to the mass $M_{\textrm {crit}}= \frac 4 3 \pi r_{\star}^{3}\rho_{c}\Omega_{\rm m}\approx 1.5\, 10^{12} M_{\sun}$, in good agreement with the value measured in simulations. Its redshift evolution is also predicted by the formalism through the cylindrical collapse and could be compared to simulation in future works.

Note that this line of reasoning  could be made more rigorous by adding  new ingredients in the formalism: a peak constraint at the location of the spin with a smoothing length $R_{h}<R_{s}$ so that one can vary $R_{h}$ (without any assumption on the additivity property of the spin) and see how the spin changes. This formalism can be implemented in two dimensions (see Appendix~\ref{sec:2D-Rh} and ~\ref{sec:2D-pkpk}) and leads to the same order of magnitude for $r_{\star}$.

\section{3D spin near and along filaments }
\label{sec:3D}

Let us now turn to the truly three dimensional theory of tidal torques 
in the vicinity of a typical filament saddle point.
Beyond the obvious increased realism,
the main motivation is that 
the 3D saddle theory fully captures the mass transition.

In three dimensions, we must consider two competing processes.
If we vary the  radius corresponding to the Lagrangian patch centered on the 
running point,
 we have a spin-up (along $\mathbf e_z$) arising from  the running to wall-running to saddle tidal misalignment and 
 a second spin-up  (along $\mathbf e_\phi$) arising from running to filament- running peak tidal misalignment.
To each position in the vicinity of the central saddle point,  we can assign $M(\mathbf{r})$ together  with 
   $\cos \mu_z (\mathbf{r})$ and $\cos \mu_\phi(\mathbf{r})$, the cosines of the angle between 
   the spin of the patch and the $\mathbf e_z$ and $\mathbf e_\phi$ direction respectively.
 Eliminating $\mathbf{r}$ yields $\cos \mu_z (M)$ and $\cos \mu_\phi(M)$ and therefore yields an estimate of the transition mass.

\subsection{Spin distribution along and near filaments}
\label{sec:3DS}

\begin{figure}
\begin{center}
\includegraphics[width=0.95\columnwidth]{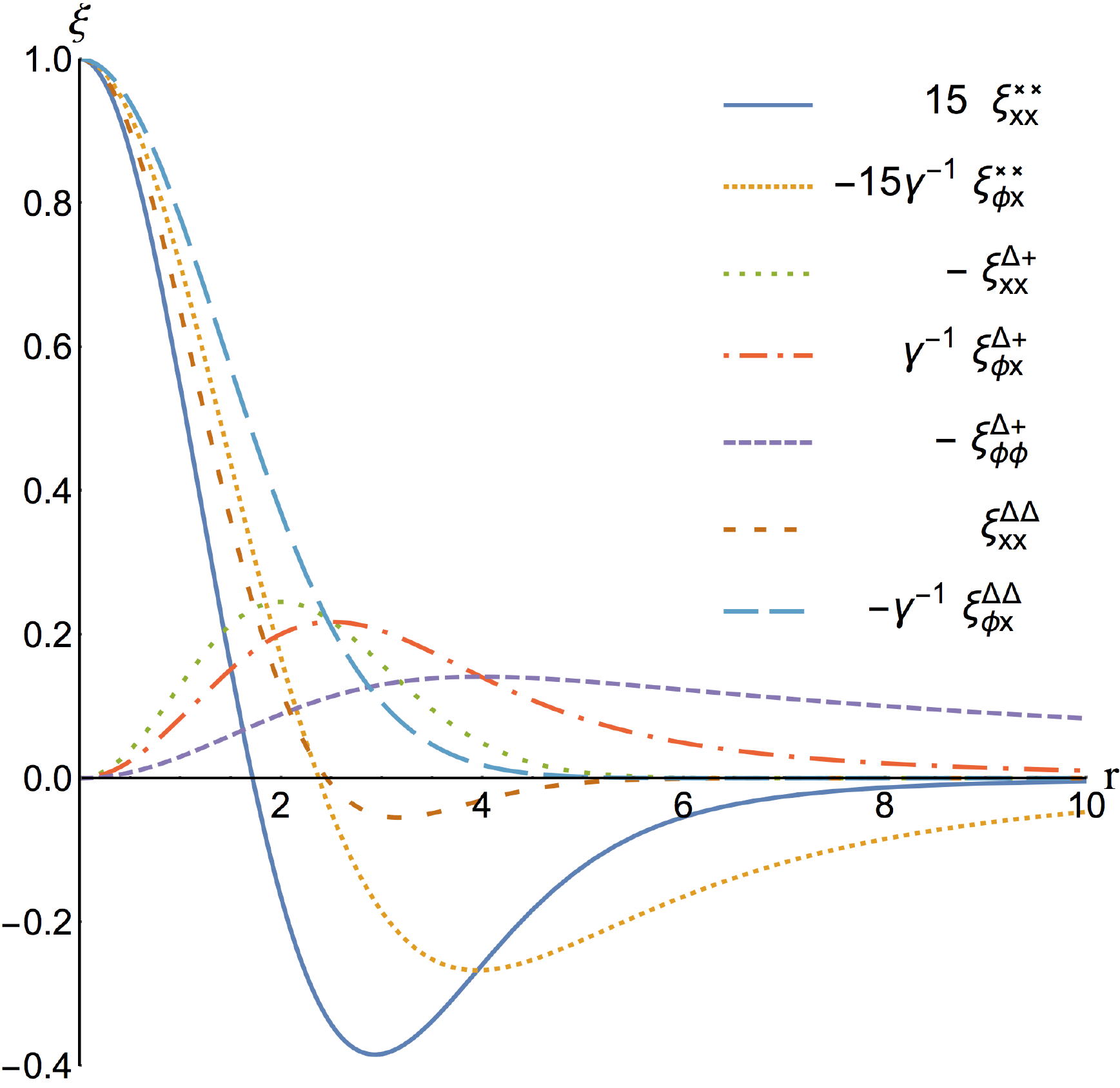}
\caption{Two-point correlation functions as a function of the separation $r$ in units of the smoothing length for a power-law 3D power spectrum with spectral index $n=-2$ i.e $\gamma=\sqrt{3}/3$. As in two dimensions, the 
correlations are rescaled at the origin.
See also Figure~\ref{fig-xi-LCDM}}
\label{fig-xi}
\end{center}
\end{figure}

 The formalism developed in Section~\ref{sec:2D} can easily be extended to three dimensions. A critical (saddle) point constraint is now imposed. This critical point is defined by its geometry, namely its height $\nu$ and eigenvalues $\lambda_{1}\geq\lambda_{2}\geq\lambda_{3}$. Note that such a critical point is a filament-type saddle point  if $\lambda_{1}\geq0\geq\lambda_{2}\geq\lambda_{3}$. In what follows, we decouple the trace from the detraced part of the density Hessian and therefore define the three curvature parameters $I_{1}=\lambda_{1}+\lambda_{2}+\lambda_{3}$, $\kappa_{1}=\lambda_{1}-\lambda_{2}$ and $\kappa_{2}=\lambda_{2}-\lambda_{3}$.
 
 \subsubsection{Mean density field around a critical point}

 The resulting mean density (contrast) field subject to that critical point constraint becomes
 (in units of $\sigma_{2}$) :
\begin{multline}
\delta(\mathbf{r}| {\rm crit},I_{1},\kappa_{1},\kappa_{2},\nu)=
\frac{I_{1} (\xi_{\phi x}^{\Delta\Delta} +\gamma \xi_{\phi\phi}^{\Delta\Delta} )}{1-\gamma ^2}+
\frac{\nu  (\xi_{\phi \phi}^{\Delta\Delta}+\gamma \xi_{\phi x}^{\Delta\Delta})}{1-\gamma ^2}\\
+\frac {15}{2}\left(\mathbf{\hat r}^{\rm T} \cdot \overline{\mathbf{ H}}\cdot\mathbf{\hat r}\right) \xi_{\phi x}^{\Delta +} \,,
\end{multline}
where again $\overline{\mathbf{ H}}$ is the {\sl detraced} Hessian of the density and
$\mathbf{\hat r}={\mathbf{r} }/{r}$
and we define in 3D $\xi_{\phi x}^{\Delta +} $ as
\begin{equation}
\xi_{\phi x}^{\Delta +} =\left\langle\Delta x \,\phi^{+}\right \rangle,
\end{equation}
with 
$\phi^{+}=\phi_{11}-(\phi_{22}+\phi_{33})/2$. The other $\xi$ functions are defined in the same way as in two dimensions (see equations~\ref{eq:xi}) and displayed on Figure~\ref{fig-xi}. Note also that $  \hat {\mathbf r}^{\rm T}\cdot \overline{\mathbf{ H}}\cdot \hat {\mathbf{r} }$
is a  scalar defined explicitly as 
$   \sum_{ij} \hat { r}^{}_{i}\overline{{ H}}_{ij}\hat {{r} }_{j}$.
%
\begin{figure*}
\begin{center}
\includegraphics[width=1.05\columnwidth]{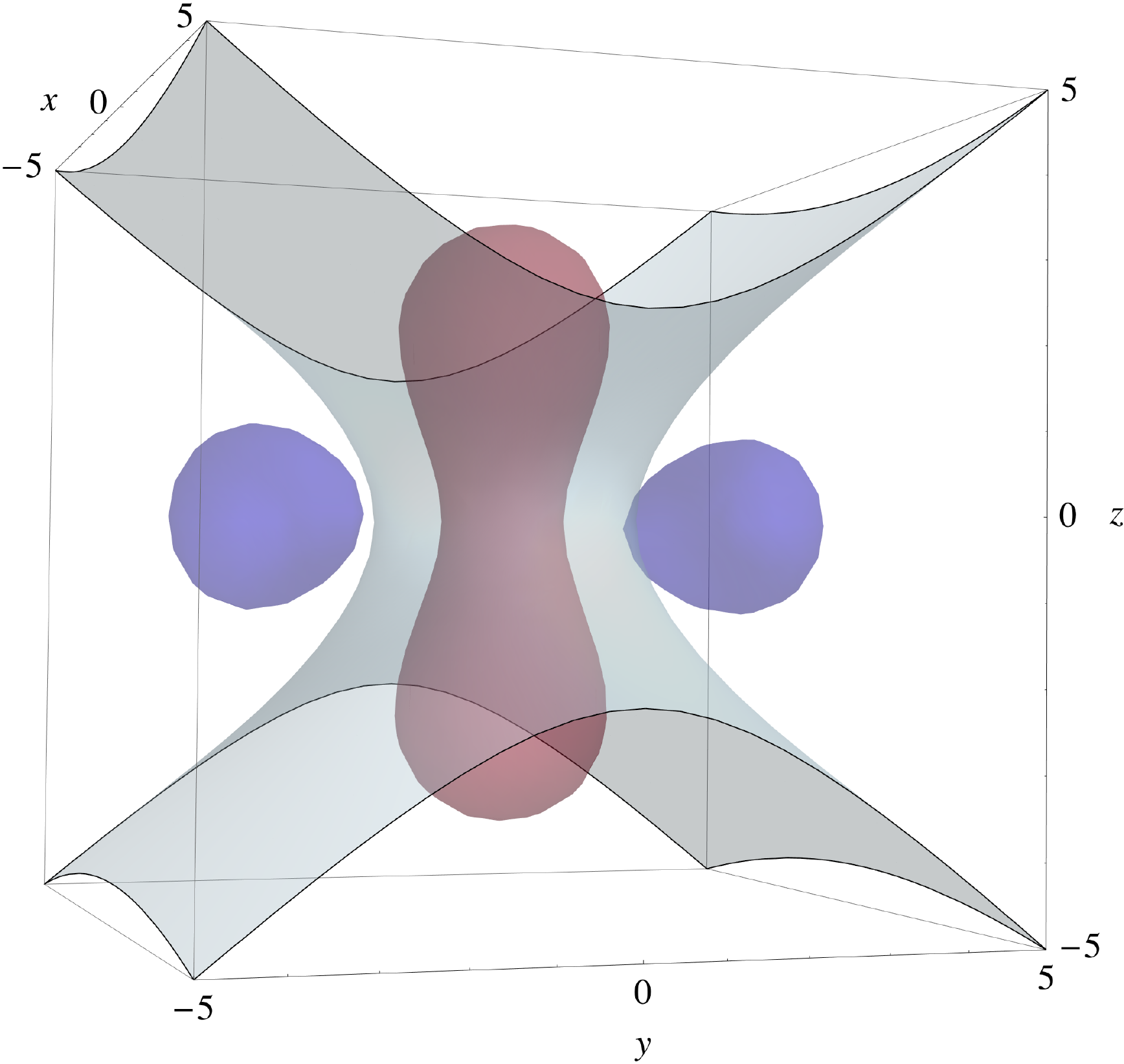}\quad
\includegraphics[width=0.9\columnwidth]{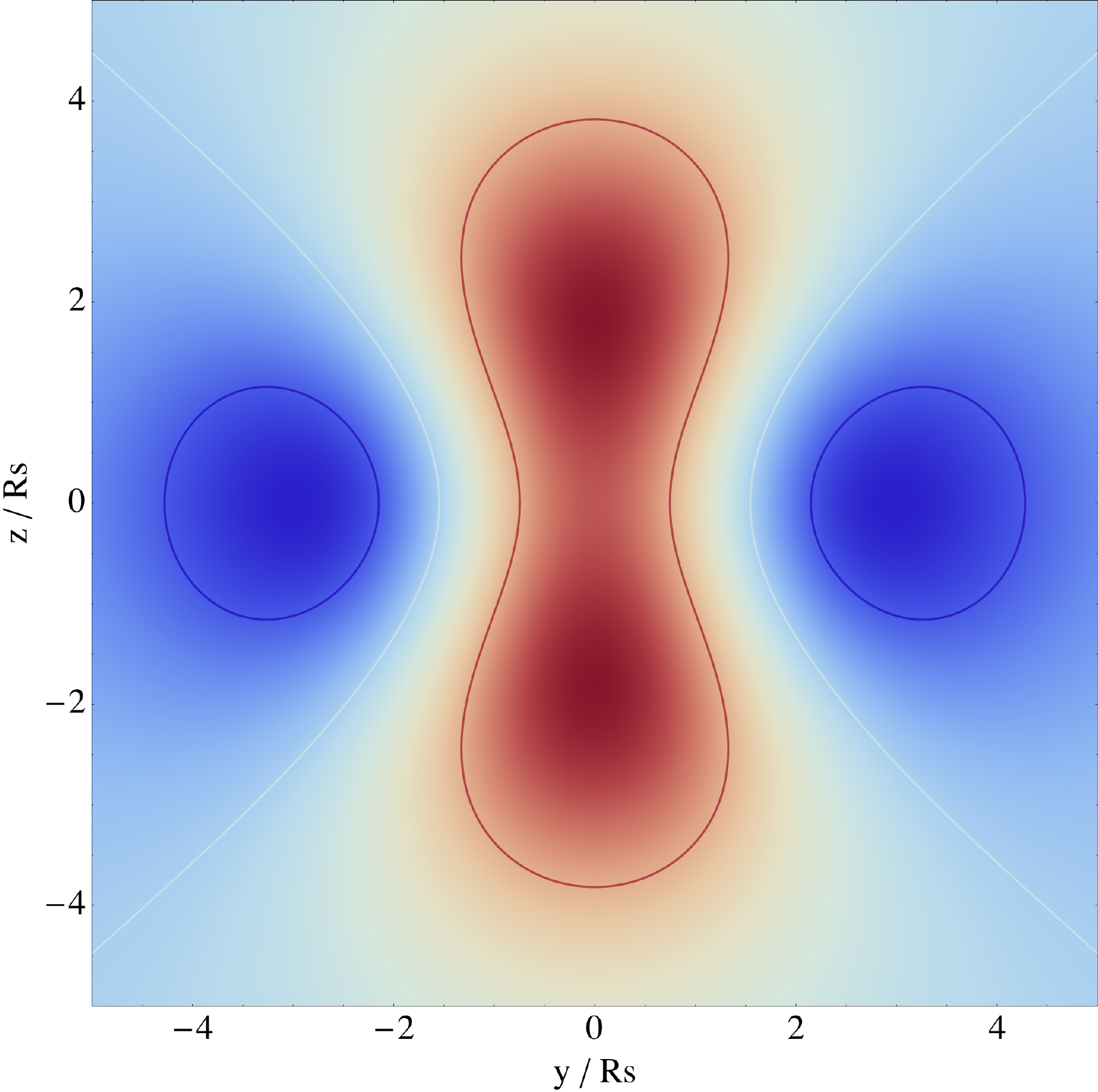}
\caption{Mean density around a filament saddle point of height $\nu=1.25\gamma$, $\lambda_{1}=0.31,\; \lambda_{2}=-0.56,\; \lambda_{3}=-1$ for a power-law 3D power spectrum with spectral index $n=-2$ i.e $\gamma=\sqrt{3}/3$. The $Ox$, $Oy$, and $Oz$ axes are in units of the smoothing length and z is the direction of the filament while the wall is in the plane Oxz. Contours represent the isocontours of the density $\delta=-0.3,0,0.5$ from blue to red. A 3D view is displayed on the left panel and a cut in the plane Oyz is shown on the right panel.
}
\label{fig:3Ddensity}
\end{center}
\end{figure*}
%
Figure~\ref{fig:3Ddensity} displays the mean density field around a typical filament-type saddle point. The elongation of the filament along the Oz axis together with the flattening of the filament in the plane of the wall (Oxz) are clearly-visible on this figure.

   \subsubsection{Mean spin field around a critical point}
   \label{sec:3Dspin}
 As in two dimensions, the  expected spin  can also be computed. 
 In three dimensions, the spin, $\mathbf s$, is a  vector  which components are  given by
\begin{equation}
{s}_i = \sum_{j,k,l}\epsilon_{ijk}  x_{kl}\phi_{lj} \,, \label{eq:defL3D}
\end{equation}
with 
$\epsilon$ the rank 3 Levi-Civita tensor.
 It is found to be orthogonal to the separation and can be written as the sum of two terms
 \begin{equation}
 \mathbf{s}(\mathbf{r}| {\rm crit},I_{1},\kappa_{1},\kappa_{2},\nu)=-15
(\mathbf s^{(1)}
+\mathbf s^{(2)})\cdot(\hat {\mathbf r}^{\rm T}\!\!\cdot \epsilon\cdot \overline {\mathbf{ H}}\cdot \hat {\mathbf{r} })
\,, \label{eq:defL3Dsol}
\end{equation}
where 
  $\mathbf s^{(1)}$ is a scalar operator that depends on the height $\nu$ and trace of the Hessian $I_{1}$ 
     \begin{align}	
 \mathbf{ s}^{(1)}=&\left(  \frac{\nu}{1-\gamma^{2}}  \left[
  (\xi_{\phi \phi}^{\Delta +}+\gamma \xi_{\phi x}^{\Delta +})\xi_{x x}^{\times\times}
  -(\xi_{\phi x}^{\Delta +}+\gamma \xi_{x x}^{\Delta +})\xi_{\phi x}^{\times\times}
  \right]\right.\nonumber
  \\
 & \hskip -1cm+\!\!\! \left.
 \frac{I_{1}}{1-\gamma^{2}}
  \left[
 (\xi_{\phi x}^{\Delta +}+\gamma \xi_{\phi \phi}^{\Delta +})\xi_{x x}^{\times\times}
-
  (\xi_{x x}^{\Delta +}+\gamma \xi_{\phi x}^{\Delta +})\xi_{\phi x}^{\times\times}
    \right]\,\right) \mathbb{I}_{3}\,,\nonumber
     \end{align}
     and $\mathrm s^{(2)}$ a combination of a matricial and a scalar operator that depends on the detraced part of the Hessian
          \begin{multline}	
\mathbf{s}^{(2)}=
  -\frac{5}{8}
   \left[
 2((\xi_{\phi x}^{\Delta +}-\xi_{\phi x}^{\Delta \Delta})\xi_{x x}^{\times\times}
  -( \xi_{x x}^{\Delta +}-\xi_{x x}^{\Delta \Delta})\xi_{\phi x}^{\times\times}) 
  \overline{\mathbf{ H}}
 \right.\nonumber
  \\
+  \left.
\Big( (7\xi_{x x}^{\Delta \Delta}+5 \xi_{x x}^{\Delta +})\xi_{\phi x}^{\times\times}-(7\xi_{\phi x}^{\Delta \Delta}+5\xi_{\phi x}^{\Delta +})\xi_{x x}^{\times\times}
  \Big)
(\hat {\mathbf r}^{\rm T}\cdot  \overline{\mathbf{ H}}\cdot \hat {\mathbf{r} })\mathbb{I}_{3}
    \right]\,
   \nonumber
     \end{multline}
with    $\mathbb{I}_{3}$ the identity matrix,  operating on the vector
     \begin{align}
     \hat {\mathbf r}^{\rm T}\!\!\cdot \epsilon\cdot \overline{\mathbf{ H}}\cdot \hat {\mathbf{r} }=   \sum_{ikl} \hat {\mathbf r}^{}_{i}\epsilon_{ijk}\overline{\mathbf{ H}}_{kl}\hat {\mathbf{r} }_{l}\,.
     \end{align}
     Note that the dependence with the distance $r$ is encoded in the two-point correlation functions, $\xi$,
      while the geometry of the critical point  is encoded in the terms corresponding to the peak height, trace and detraced part of the Hessian and the orientation of the separation is in $\mathbf{\hat r}$. 
     Equation~(\ref{eq:defL3Dsol}) is also remarkably simple: 
     as expected the symmetry of the model induces zero spin along the principal directions of the Hessian (where $\hat {\mathbf r}^{\rm T}\!\!\cdot \epsilon\cdot \overline{\mathbf{ H}}\cdot \hat {\mathbf{r} }=0$) and a point reflection symmetry ($\mathbf{\hat r}\rightarrow-\mathbf{\hat r} $).
Note that the correlation functions, $\xi$ can be evaluated for  arbitrary power spectra (such as power laws, see Appendix~\ref{sec:xi},
or  $\Lambda$CDM, see Appendix~\ref{sec:xi-LCDM}), hence Equation~(\ref{eq:defL3Dsol}) is completely
general.

For scale-invariant density power spectra with index $n$ ($n-4$ for the {\sl potential}), $\mathbf{s}$ 
can be computed explicitly. 
At small separation, 
the term proportional to $ \hat {\mathbf r}^{\rm T}\cdot \overline{\mathbf{ H}}\cdot \hat {\mathbf{r} }$ goes like $r^{4}$
and is thus negligible compare to the rest (that scales like $r^{2}$). The spin coordinates in the frame of the Hessian are therefore quadrupolar
 \begin{equation}
\mathbf{s} \propto  (f(\lambda_{i},\nu,n)yz,g(\lambda_{i},\nu,n)xz,h(\lambda_{i},\nu,n)xy) \label{eq:L-asymptote}\,.
 \end{equation}

Figure~
\ref{fig-momentum3D} illustrate the mean spin geometry around a typical saddle point. All the symmetry properties (antisymmetry, octopole, ...) described in this section are clearly seen on this figure. In the plane of the saddle point, spins are aligned with the filament direction. When moving towards the nodes, the spins become more and more perpendicular (and more and more along $\mathbf{e}_\phi$).

\begin{figure}
\begin{center}
\includegraphics[width=0.95\columnwidth]{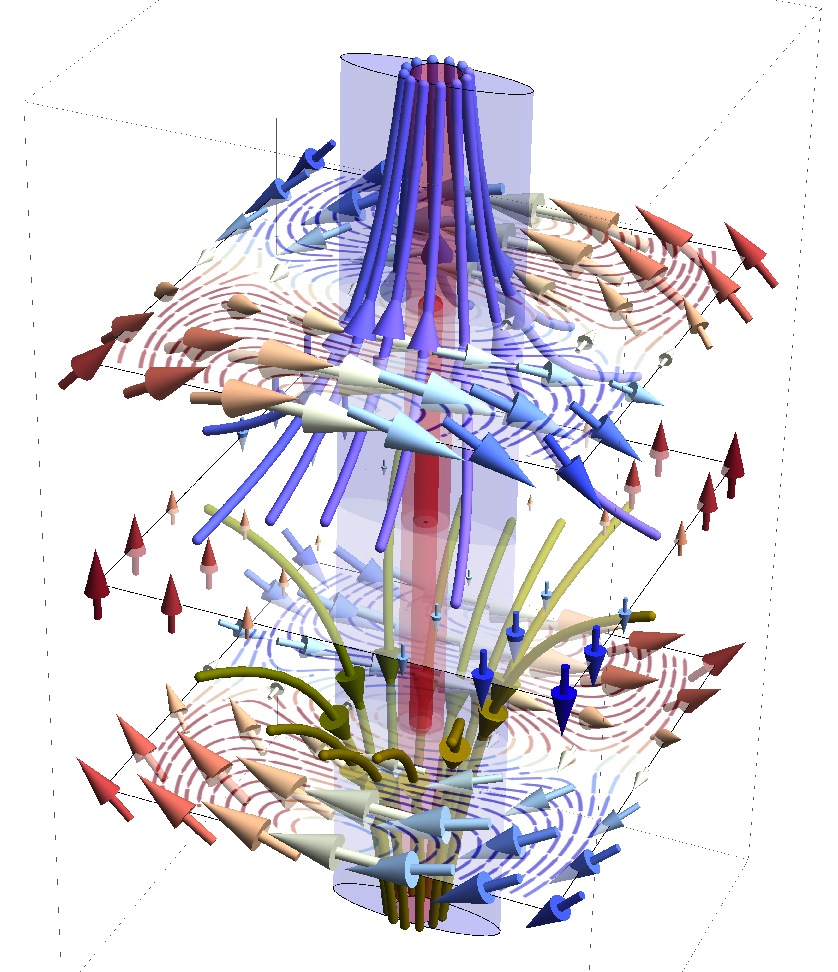}
\caption{\small{
The velocity and Spin flow near a vertical filament (in red) embedded in 
a (purple) wall  {\green for $(x,y,z)\in[-2R_{s},2R_{s}]^{3}$}. The purple and green flow lines trace the (Lagrangian) 3D velocities
(upwards and downwards respectively).
The red and blue arrows show
the spin 3D distribution, while 
the three horizontal cross sections show spin flow lines in the corresponding plane.
Note that the  spin is along $\mathbf e_z$ in the mid plane and along $\mathbf e_\phi$ away from it,
and that it rotates in opposite direction above and below the mid-plane.
See also the interactive version at \url{http://www.iap.fr/users/pichon/AM-near-saddle.html}
}}
\label{fig-momentum3D}
\end{center}
\end{figure}

\subsubsection{Cosmic variance on 3D spin }
\label{sec:variance3D}
It is of interest to also study the variance of the spin alignment $\sigma(\mathbf{r}| {\rm ext},I_{1},\kappa_{1},\kappa_{2},\nu)$ defined as
\begin{equation}
\sigma=\sqrt{\left\langle \cos^{2} \theta\right\rangle-\left\langle\cos\theta\right\rangle^{2}}\,,
\end{equation}
where $\cos\theta=\mathbf{s}\cdot\mathbf{e_{z}}/||\mathbf{s}||$.
It requires the numerical evaluation of a 12D integral. 
In contrast, the mean of the spin $\mathbf{s}$ (as computed in Section~\ref{sec:3Dspin}) or its square $\mathbf{s}^{2}$ can be analytically computed. We therefore propose to approximate the dispersion of the spin alignment with the following related estimator
\begin{equation}
\label{eq:sigma-3D}
\tilde\sigma=\sqrt{\frac{\left\langle s_{z}^{2}\right\rangle-\left\langle s_{z}\right\rangle^{2}}{\left\langle\mathbf{s}\cdot\mathbf{s}\right\rangle}}\,,
\end{equation}
where $s_{z}$ is the component of the spin along the z-axis i.e along the filament direction.
For the sake of readability, we do not write down the result of the integration here but display in Figure~\ref{fig:sigma-3D} the map of the alignment dispersion $\tilde \sigma$ around a typical saddle point. This standard deviation is roughly constant around $\approx 0.6$ and decreases to $\approx0.3$ in the close vicinity of the saddle point.
Note that the  spin direction is again  best defined along its minor axis. This would be the best place to measure 
spin alignments in observations.

\begin{figure*}
\begin{center}
\includegraphics[width=\columnwidth]{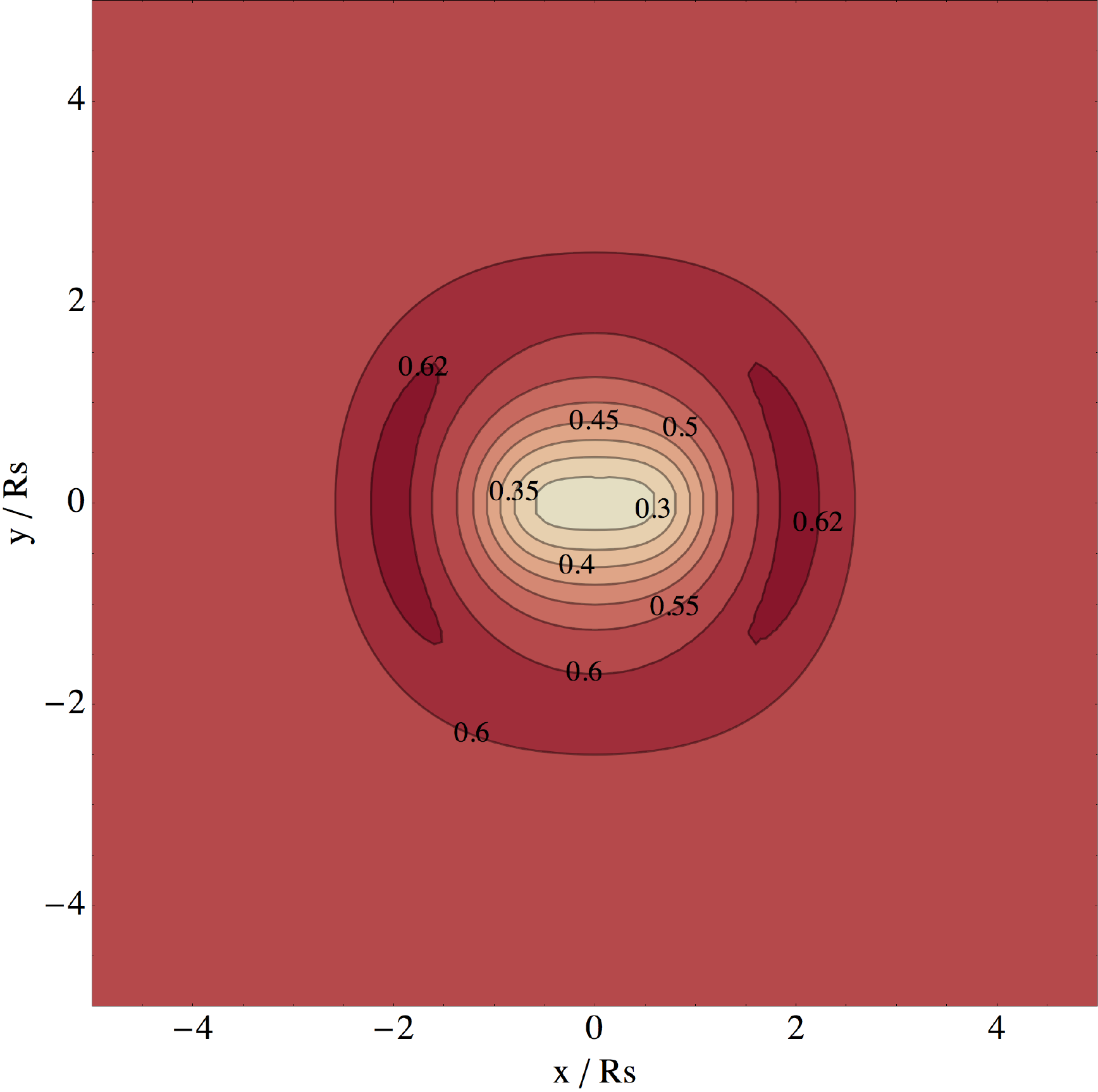}
\includegraphics[width=\columnwidth]{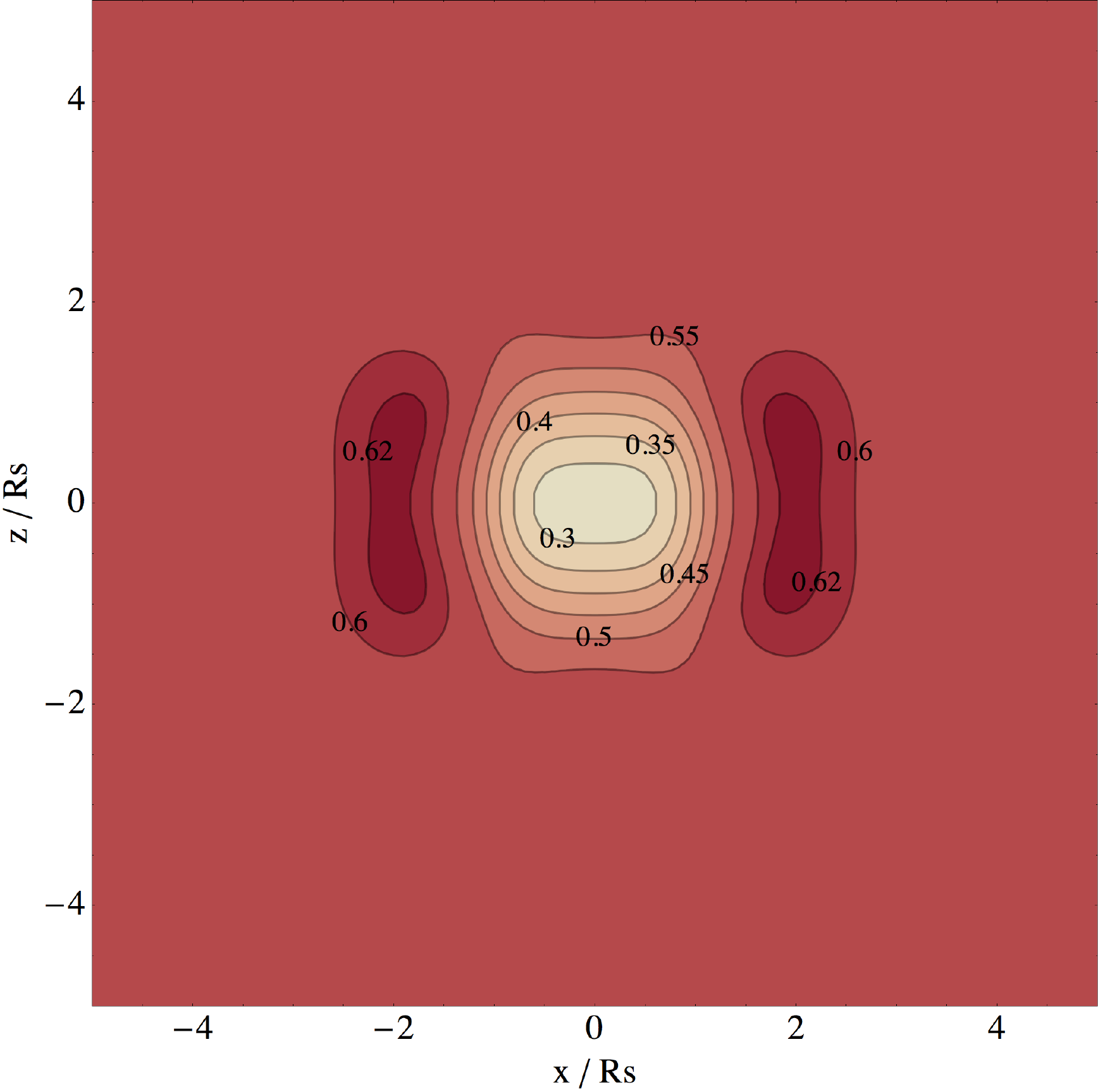}
\caption{Alignment dispersion $\tilde \sigma$ (defined in Equation~\ref{eq:sigma-3D}) around a typical filament-type saddle point of height $\nu=1.25\gamma$, $\lambda_{1}=0.31,\; \lambda_{2}=-0.56,\; \lambda_{3}=-1$ for a power-law 3D power spectrum with spectral index $n=-2$. The left panel displays a cut along the  plane, $Oxy$, of the saddle point and right panel along the plane, $Oxz$, of the wall.
The uncertainty on spin direction is smallest near the saddle.}
\label{fig:sigma-3D}
\end{center}
\end{figure*}

\subsection{Mean saddle point geometry}

Here we want to compute the mean values of $\nu$, $\lambda_{1}<\lambda_{2}<0<\lambda_{3}$ of a typical saddle-point of filament type.
Let us start from the so-called Doroshkevich formula for the PDF of these variables:
\begin{multline}
{\cal P}(\nu,\lambda_{i})=
\!
\frac{135 \left( 5 /{2\pi} \right)^{3/2}
}{4\sqrt{1-\gamma^{2}}}
\!
\exp 
\!\left[-\frac 1 2 \zeta^{2}-3 I_{1}^{2}+\frac {15} 2 I_{2}\right]
\nonumber\\
\times (\lambda_{3}-\lambda_{1})(\lambda_{3}-\lambda_{2})(\lambda_{2}-\lambda_{1})\,,
\end{multline} 
where
$\zeta=(\nu+\gamma I_{1})/\sqrt{1-\gamma^{2}}$,
$I_{1}=\lambda_{1}+\lambda_{2}+\lambda_{3}$,
$I_{2}=\lambda_{1}\lambda_{2}+\lambda_{2}\lambda_{3}+\lambda_{1}\lambda_{3}$
and $I_{3}=\lambda_{1}\lambda_{2}\lambda_{3}$.
Subject to a saddle-point constraint, this PDF becomes
\begin{equation}
{\cal P}(\nu,\lambda_{i}|\textrm{sad})=
 \frac{540  \sqrt{5 \pi } {\cal P}(\nu,\lambda_{i})
}{29 \sqrt{2}+12 \sqrt{3}} I_{3}\Theta(\lambda_{3})\Theta(-\lambda_{2})\,,
\end{equation} 
after imposing the condition of saddle point $|\det \partial_{i}\partial_{j}\delta|\delta_{D}({\mathbf{\nabla}\delta})\Theta(\lambda_{3})\Theta(-\lambda_{2})$ for which as the gradient is decoupled from the density and the Hessian, only the condition on the sign of the eigenvalues and the determinant contribute.
From this PDF, it is straightforward to compute the expected value of the density and the eigenvalues at a saddle point position: $\left\langle\nu\right\rangle\approx 0.76 \gamma$, $\left\langle\lambda_{1}\right\rangle\approx-0.87$, $\left\langle\lambda_{2}\right\rangle\approx -0.40$ and $\left\langle\lambda_{3}\right\rangle\approx 0.51$. However, this saddle point does not belong to the skeleton of the density field but to its inter-skeleton (see \cite{pogo09}). We thus want to impose an additional constraint which is $\lambda_{2}+\lambda_{3}<0$. Let us call those saddle points ``skeleton saddles''. The PDF at those points becomes
\begin{equation}
\hskip -0.33cm
{\cal P}(\nu,\lambda_{i}|\textrm{skl})=
 \frac{26460 \sqrt{5 \pi } {\cal P}(\nu,\lambda_{i}) I_{3} \Theta(\lambda_{3})
}{1421 \sqrt{2}-735 \sqrt{3}+66 \sqrt{42}}\Theta(-\lambda_{2}-\lambda_{3})\,.\label{eq:defskelsaddle}
\end{equation} 
The expected value of the density and the eigenvalues at a skeleton saddle position now becomes $\left\langle\nu\right\rangle\approx 1.25 \gamma$, $\left\langle\lambda_{1}\right\rangle\approx-1.0$, $\left\langle\lambda_{2}\right\rangle\approx -0.56$ and $\left\langle\lambda_{3}\right\rangle\approx 0.31$.

\subsection{Spin flip : from spatial to mass transition}
%
The geometry of the spin distribution near a typical skeleton saddle point (as defined by equation~(\ref{eq:defskelsaddle})) allows us to compute the mean alignment angle between the spin and the filament (see Section~\ref{sec:alignment} below).
In turn, the shape of the density profile in the vicinity of the same critical point,
together with an extension of the Press-Schechter theory involving a filament background split,
allows us to estimate the ``typical'' mass of the dark matter haloes forming in any spatial position around the saddle point  (Section~\ref{sec:massgrad} below).
The alignment-angle map and the typical-mass map will together yield a prediction for the transition mass.
\subsubsection{Spin flip along filaments}
\label{sec:alignment}
Section ~\ref{sec:3DS} showed that the mean spin flips from alignment in the plane of the saddle point to orthogonality when going towards the nodes. This can be quantified by measuring the curvilinear coordinate along the filament at which the spin flips. 

Let us consider the mean {\sl modulus} of the projection of the  spin along the $\mathbf{e_{z}}$ and $\mathbf{e_{\phi}}$ axes within a plane of height $z$ 
\begin{equation}
{\hat{ \bar s}_{z/\phi}}(z)=\int {\rm d} x {\rm d} y \left| \bar s_{z/\phi}(\mathbf{r})\right|/||\mathbf{s}(\mathbf{r})|| \,. \label{eq:midplaneavg}
\end{equation}
Figure~\ref{fig-ztransition} displays ${\hat{ \bar s}_{z/\phi}}(z)$ as a function of $z$ along the 
filament.
Let us define $\hat \theta$ the flip angle so that 
\begin{equation}
\cos \hat \theta(z) =\frac{{\hat{ \bar s}_{z}}(z)}{\sqrt{{\hat{ \bar s}_{z}}(z)^2+{\hat{ \bar s}_{\phi}(z)}^2}}=\frac{1}{\sqrt{2}}\,.
\label{eq:deforientation-profile}
\end{equation}
In Figure~\ref{fig-ztransition}, this flip angle is found to occurs around $z=1.5 R_{s}$ which is very close to the $r_{\star}$ measured in two dimensions (see Section~\ref{sec:Transition}).

\begin{figure}
\begin{center}
\includegraphics[width=0.95\columnwidth]{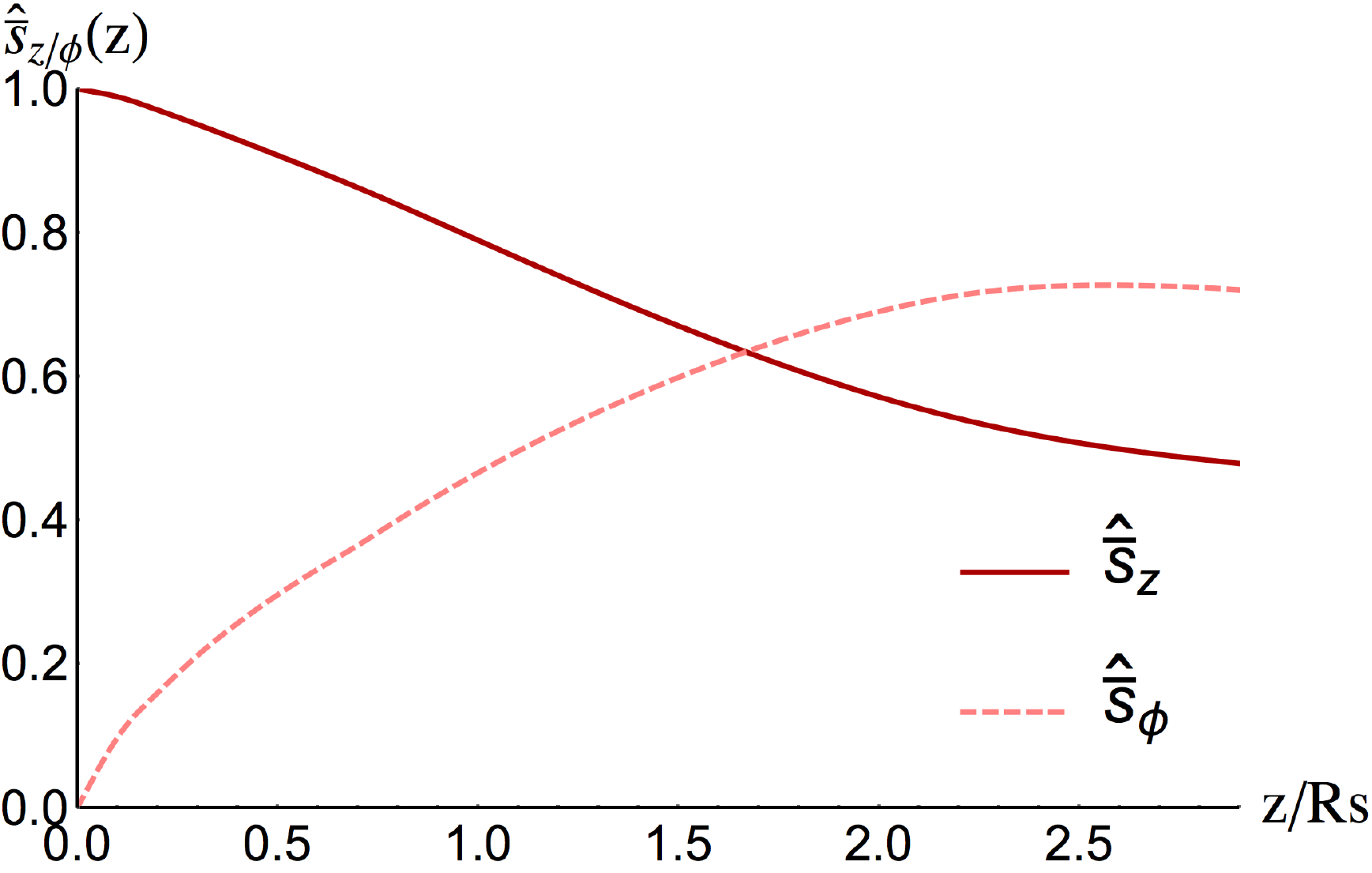}
\caption{
measure the height  $z$ corresponding to the transition from aligned to perpendicular to the filaments. The amplitude in both direction is averaged by plane following equation~(\ref{eq:midplaneavg}). 
The transition curvilinear coordinate is  $z_{\rm tr}=1.5 R_{s}$.
}
\label{fig-ztransition}
\end{center}
\end{figure}

Alternatively, one can also compute at each position the mean alignement with the filament direction $\mathbf{e_{z}}$
\begin{equation}
\cos \theta (\mathbf{r})=\frac{ \mathbf{s}(\mathbf{r}| {\rm crit})\cdot \mathbf{e_{z}}}{||\mathbf{s}(\mathbf{r}| {\rm crit})||}\,.
\label{eq:deforientation-profile}
\end{equation}
The result is shown on the right panel of  Figure~\ref{fig-mass-angle-profile}. Spins tend to align with the filament (region in red) in the plane of the saddle point
and becomes perpendicular to it when moving towards the nodes (region in blue). This is a transition in 
Lagrangian space. Section~\ref{sec:massgrad} shows how to convert it into a transition in mass.

\begin{figure*} 
\begin{center}
\includegraphics[width=0.95\columnwidth]{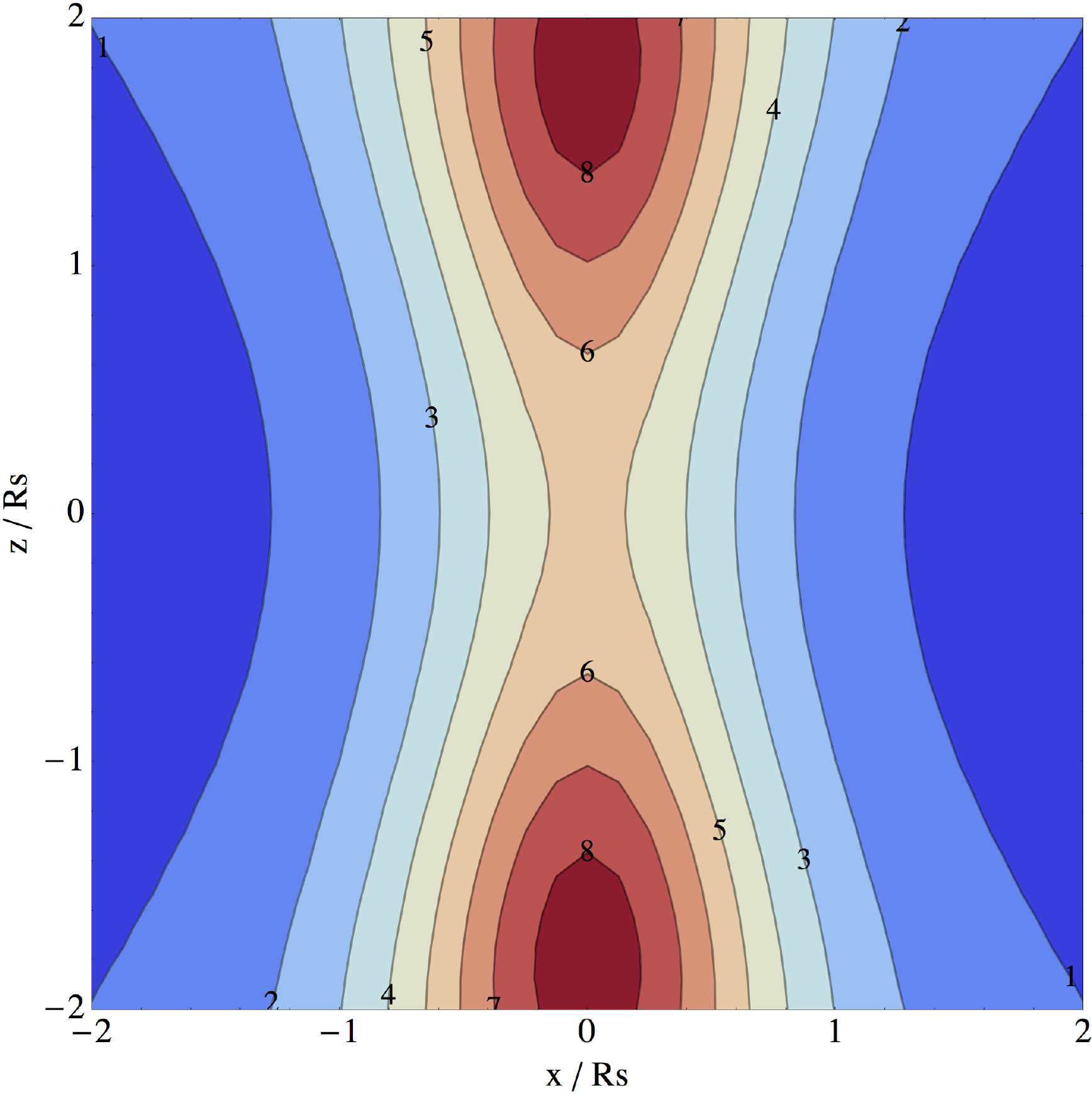}\hspace{1cm}
\includegraphics[width=0.95\columnwidth]{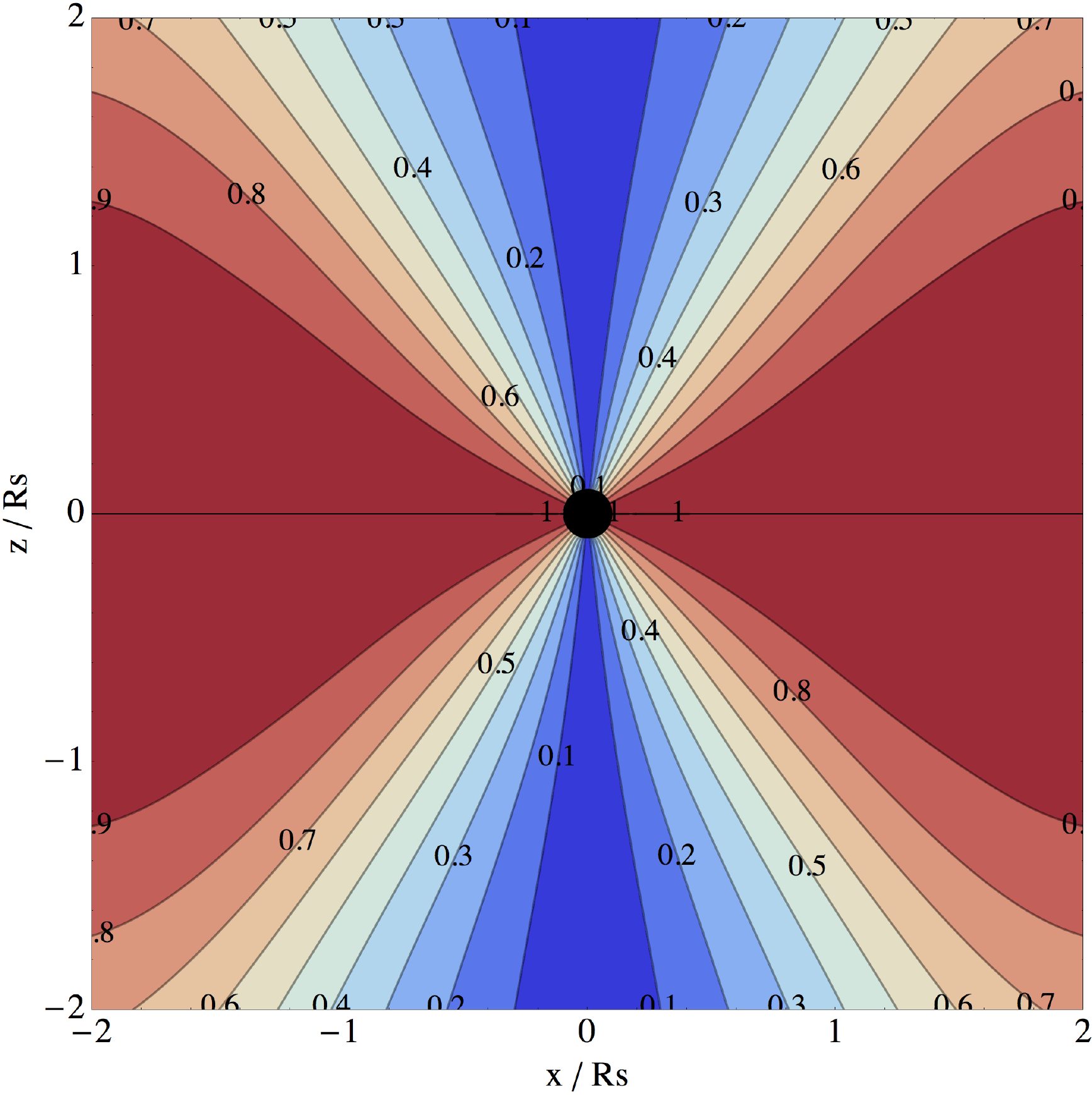}
\caption{Left: cross section of  $M_p(r,z)$ {\green (in units of $10^{12}M_{\odot}$)}  along the most likely filament and in the direction $x=y$.
Right: corresponding cross section of $\langle\cos \hat \theta\,\rangle(r,z)$, the normalised component of the spin aligned with the filament. The black dot represents the position of the saddle point.
The mass of haloes increases towards the nodes, while the spin flips.
 }
\label{fig-mass-angle-profile}
\end{center}
\end{figure*}

\subsubsection{ Halo mass gradient along filament}
\label{sec:massgrad}

The local mass distribution of haloes is expected to vary along the large-scale 
filament due to  changes in the underlying long-wave density. In the linear regime,
the typical {\green overdensity} near the end points (nodes) of the filament, where it joins
the protocluster regions, may exceed the typical{ \green overdensity} near the saddle point
by a factor of two \citep{Pogosyanetal1998}. During epochs 
before the whole filamentary structure has collapsed, this
leads to a shift in the hierarchy of the forming haloes towards larger masses
near the filament end points (the clusters) relative to the filament middle point (the saddle). 
This can be easily understood using the formalism of barrier crossing
\citep{PH1990,Bondetal1991,Paranjapeetal2012,MussoSheth2012},
which associates the density of objects of a given mass to
the statistics for the
random walk of halo density  as the field is  
smoothed with decreasing filter sizes. Specifically these authors predict the first upcrossing 
probability for 
the critical threshold  at
the filter scale corresponding to the mass of interest.
The precise outcome of the formalism depends on the spectral properties of the field
and the form of the smoothing filter, however it is clear that, in general,
decreasing the barrier threshold increases the probability that such first
upcrossing will happen at large smoothings, i.e large mass. 
A larger fraction
of the Lagrangian space will then belong to large-mass haloes, at the expense
of the low-mass ones.

Following the presentation of \cite{Paranjapeetal2012}  of the  
Peacock-Heavens \citep{PH1990}
approximation --that was found to fit numerical simulations rather well,
the number density of dark haloes in the interval $[M,M+dM]$ is
\begin{equation}
\frac{d n(M)}{d M} d M = \frac{\rho}{M}
f(\sigma^2,\delta_c) d \ln \sigma^2\,,
\end{equation}
where $f(\sigma^2,\delta_c) $ is given by the function
\begin{align} \label{eq:deff}
f(\sigma^2,\delta_c) &= \exp\left( \frac{1}{\Gamma} \int_0^{\sigma^2}
\frac{d s^\prime}{s^\prime} \ln p(s^\prime,\delta_c) \right) \times  \\
&\times \left( - \sigma^2 \frac{d p(\sigma^2,\delta_c)}{d \sigma^2}
- \frac{1}{\Gamma} p(\sigma^2,\delta_c) \ln p(\sigma^2,\delta_c)\right)\nonumber\,.
\end{align}
Here $\sigma^2$ is the variance of the density fluctuations smoothed 
at the scale corresponding to $M$ and 
$p(\sigma^2,\delta_c)\equiv {1}/{2} \left(1+
\mathrm{erf}(\delta_c/\sqrt{2}\sigma) \right) $ is the probability of
a Gaussian process with variance $\sigma^2$ to yield value below
some critical threshold $\delta_c$. In Equation~(\ref{eq:deff}),
$\Gamma$ is the parameter dependent on the  filtering scale and, to less extend the underlying power spectrum,  
that specifies how correlated
 the density values at the same  point when smoothed at different scales are.
For Gaussian filter, the value $\Gamma \approx 4$ is advocated.

The overall mass distribution of haloes is well described by the choice
$\delta_c={3}/{5}\left({3\pi}/{2}\right)^{2/3}=1.681$,
motivated by the so-called spherical collapse model.
When haloes form on top of  a large-scale structure background, however,
the long-wave over-density $\overline \delta(z)$ adds to  the over-density in the proto-halo
peaks.  The effect on halo mass distribution, in this  so-called peak-background split
approach, can be approximated as a {\sl shifted} threshold 
$\delta_c(z) = 1.681 - \overline \delta(z)$ for halo formation.
In Figure~\ref{fig:PH}, we show that, as expected, the result of this long wavelength mode is a
shift of the halo mass distribution towards larger masses.
\begin{figure}
\includegraphics[width=0.5\textwidth]{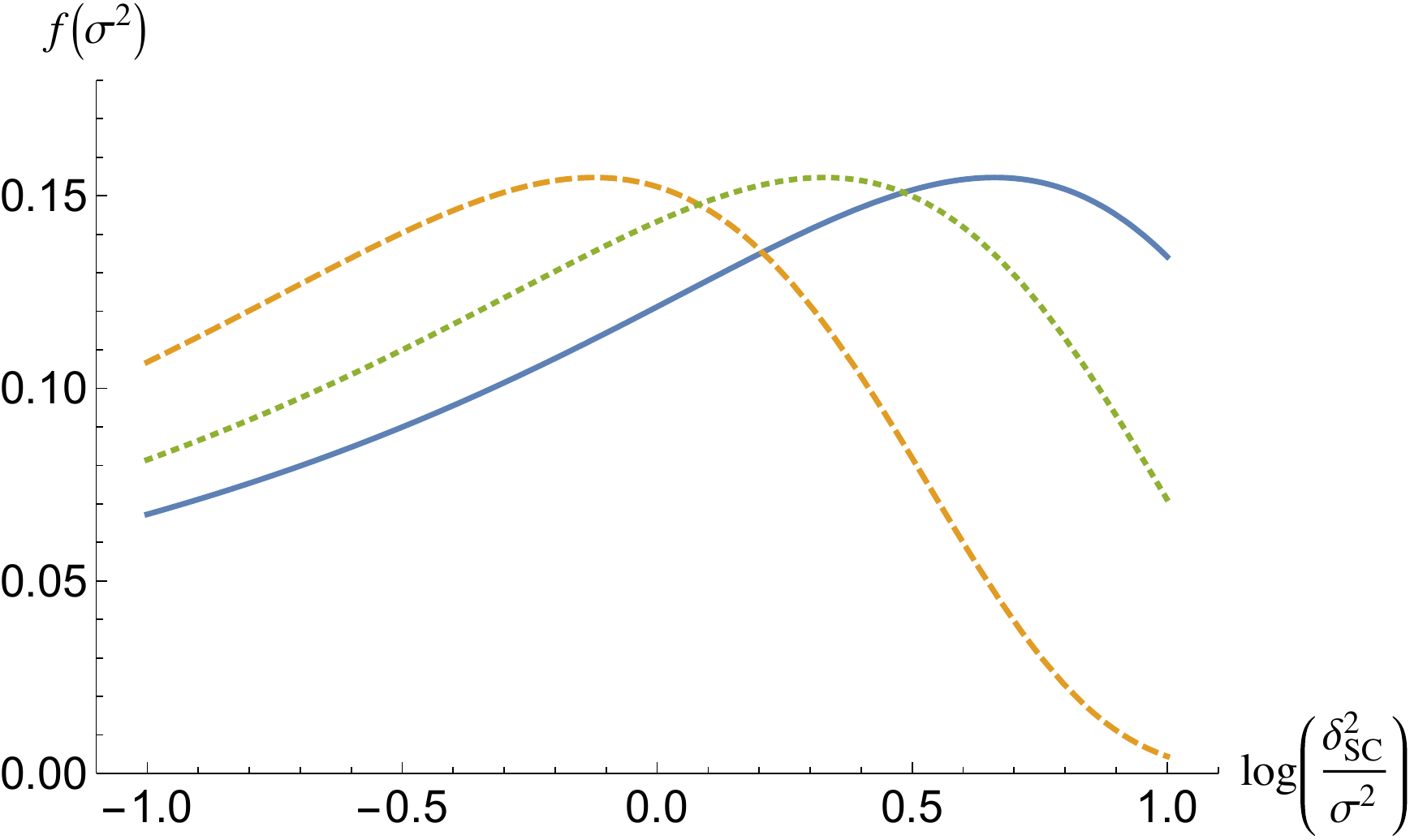}
\caption{Mass distribution for three values of density threshold,
$\delta_c=1.681,1,0.681$ from left {\green(yellow dashed line) to right (blue solid line)}. The  displayed function $f(\sigma^{2})$ is defined in Equation~\ref{eq:deff}.
}
\label{fig:PH}
\end{figure}
This shift can be characterized by the dependence on the threshold of
$M_*(\delta_c)$, defined as $\sigma_*(M_*) = \delta_c$, or of the mass
$M_p(\delta_c)$ that corresponds to the peak of $f(\sigma^2,\delta_c)$,
i.e. the variance $\sigma^2_p(z)$ defined by
\begin{equation}
\sigma^2_p(z)\equiv \underset{\displaystyle\sigma^2} {\mathrm{argmax}}\Big( f(\sigma^2,\delta_c(z)) \Big)  \,.\label{eq:defsigmap}
\end{equation}
\begin{figure}
\includegraphics[width=0.5\textwidth]{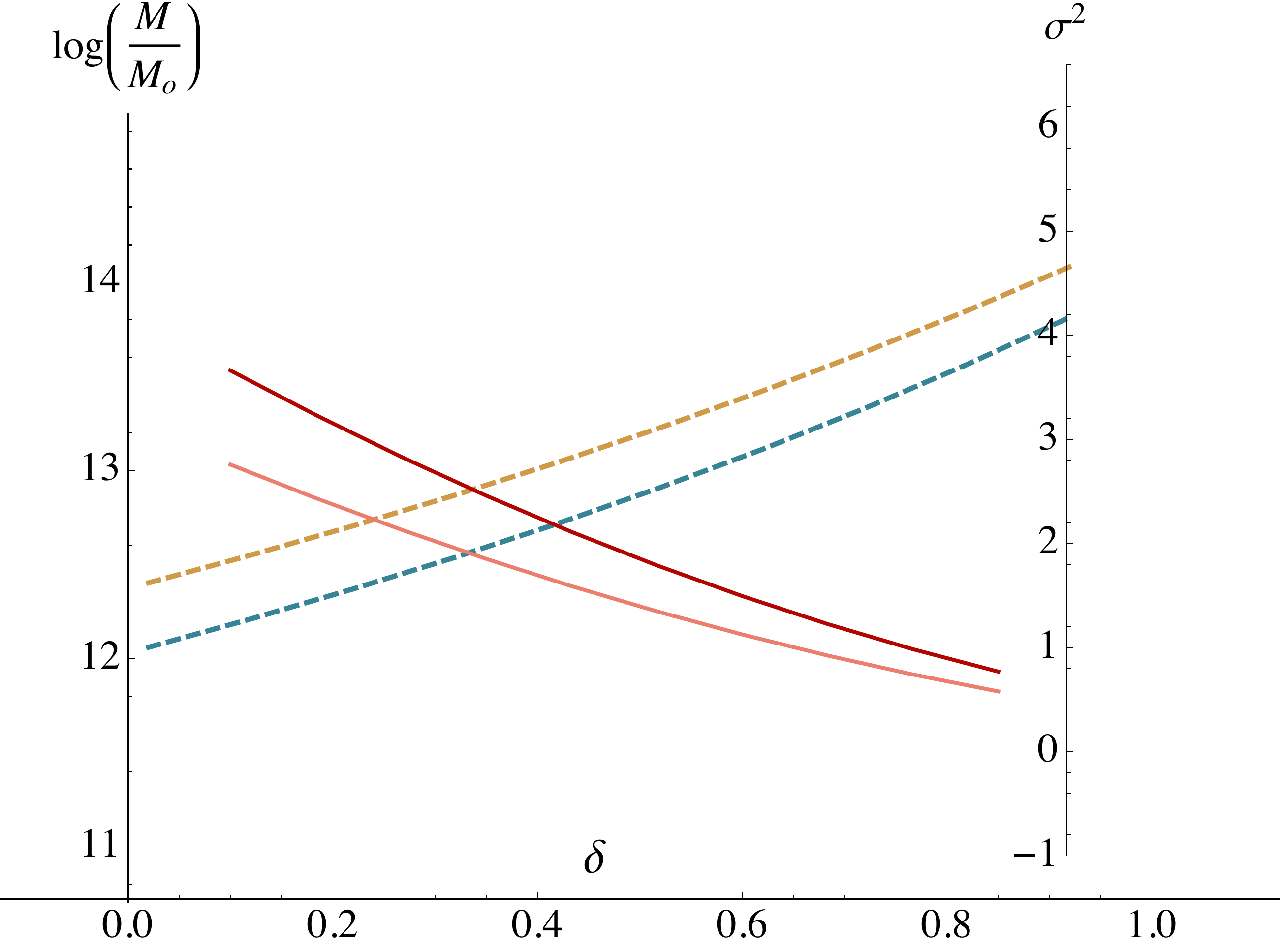}
\caption{Characteristic variances (plain) and $M_*$ (yellow dashed) and $M_p$ halo masses (blue dashed) as functions of the
large-scale density in the peak-background split picture. 
}
\label{fig:M*delta}
\end{figure}
Figure~\ref{fig:M*delta}, right axis, shows these two
characteristic variances as functions of the threshold, $\delta_c$.

The link to cosmology is established by relating the variance $\sigma^2$ 
to the mass of the objects. If a background field is absent,
the variance is just
the integral of the power spectrum $P(k)$ smoothed over a sphere of mass $M$
\begin{equation}
\sigma^2=\sigma^2(M,Z) \equiv D^2(Z) {\int k^2 dk P(k) W_{\mathrm{TH}}^2
\left( \left( {3 M}/{4 \pi \bar\rho}\right)^{1/3} \right) }\nonumber
\label{eq:sigmaM}
\end{equation}
where $D(Z)$ is the linear growing mode of perturbations as a function of redshift $Z$ and $W_\mathrm{TH}$
is the top-hat filter. 
However, when large-scale structures are considered as
fixed background,  
the variance of the relevant 
small-scale density fluctuations
that are responsible for object formation is reduced, approximately as
\begin{equation}
\sigma^2 \approx \sigma^2(M) - \sigma^2(M_{\rm LSS})\,,
\label{eq:sigmaMcorrected}
\end{equation}
where $\sigma^2(M_{\rm LSS})$, given as well as $\sigma^2(M)$ by
Equation~(\ref{eq:sigmaM}), is the unconstrained variance
at the scales at which we have defined the background large-scale density.
This correction is negligible when there is distinct scale separation between
non-linear forming objects and the large-scale density, i.e 
$\bar \delta(x) \ll 1.681$ but becomes important, truncating
the mass hierarchy at $M_{\rm LSS}$, whenever large-scale structures are themselves
non-linear.

On Figure~\ref{fig:M*delta},  left axis,
 the variances are converted  into  masses, $M_\star$ and $M_p$ 
according to
equation~(\ref{eq:sigmaMcorrected}). We choose here $\sigma_8=0.8$, $Z=0$, we
define the mass in a $8 h^{-1}$ Mpc
comoving sphere for the best-fit cosmological mass density
and we approximate the spectrum with a power-law of index $n=-2$,
which allows to solve Equation~(\ref{eq:sigmaMcorrected}) explicitly, giving 
the
$M(\sigma)$ relation as
\begin{equation}
M(\sigma,Z) = 2.6 \times 10^{14} M_\odot 
\left(\frac{\sigma^2 + \sigma^2(M_{\rm LSS})}{\sigma_8^2 D(Z)^2} \right)^{-\frac{3}{n+3}}\,. \label{eq:defMsigmaZ}
\end{equation}
We consider filaments to be defined with $R=5 h^{-1}$Mpc Gaussian smoothing,
which gives $\sigma^2({M_{\rm LSS}}) \approx 0.66$.
The evolution of $M_p(\bar\delta,Z)$ follows from putting equation~(\ref{eq:defsigmap}) into equation~(\ref{eq:defMsigmaZ}).

\subsubsection{Spin orientation versus mass}

From the above described  $M_{p}$-$\delta$ relation, one can attribute a mass to each position depending on the value of the mean density at that location.
The result is illustrated in Figure~\ref{fig-mass-angle-profile} where the left and right panels display respectively the mass map and the spin alignment map around a typical saddle point.
Eliminating the spatial position, $\mathbf{r}$, between these two maps yields $\langle\, \cos \theta\,\rangle$ as a function of  $M_{p}$ as shown on
Figure~\ref{fig:align-mass}. The transition mass, ${\cal M}_{\rm tr}$ for spin flip ($\langle\, \cos \theta\,\rangle=0.5$) is found to be of the order of $4\, 10^{12} M_\odot$,
assuming a smoothing scale of 5 Mpc$/h$, as used in \cite{codisetal12}.
This mass is in qualitative agreement with the transition mass found in that paper, all the more so as the redshift evolution of this transition mass will also be consistent (scaling as the mass of non-linearity).

It is quite striking that the geometry of the saddle point alone allows us to predict this mass. 
The two main ingredients for success are the point reflection symmetry of the spin distribution near the 
most likely filament-like saddle point on the one hand, and the peak background split mass distribution gradient along the 
filament towards the nodes of the cosmic web on the other hand.

\begin{figure}
\includegraphics[width=0.5\textwidth]{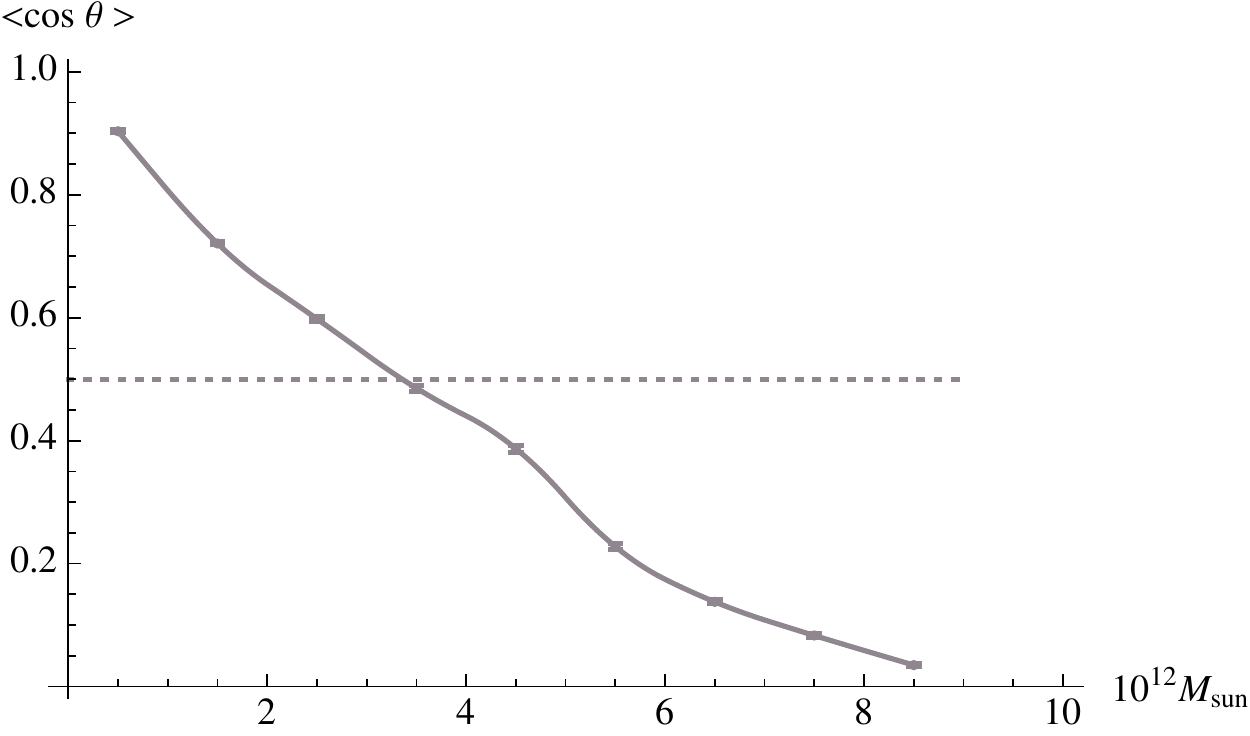}
\caption{Mean alignment as a function of mass for a smoothing scale  for filaments of 5 Mpc$/h$.  Error bars represent the error on the mean cosine in each bin of mass for the region $[-2 R_{s},2R_{s}]\times[-2R_{s},2R_{s}]\times[-2R_{s},2R_{s}]$ around a typical filament saddle point.
The  flip transition mass corresponds to $4\, 10^{12} M_\odot$.
}
\label{fig:align-mass}
\end{figure}

\section{Statistics }
\label{sec:statistics}

Up to now, we have considered the neighbourhood of a  {\sl given  unique}
typical saddle point as a proxy for the behaviour within a 
Gaussian random field (GRF).
In view of our finding let us now first analyze the statistics of alignment 
for GRF,
 and then for fields corresponding to their simulated cosmic evolution down to redshift zero.

\subsection{Validation on GRF }
\label{sec:simulation}

Let us consider the following experiment. Let us generate 2D or 3D 
realizations of GRF smoothed on two successive 
scales, $L_h$ and $L_s\gg L_h$. In the first maps, let us build a catalogue of positions, $\mathbf r_h$ and heights, $\nu_h$
corresponding to ``small-scale'' peaks. 
 From the second maps, let us identify the loci, $\mathbf r_s$ of the corresponding ``large-scale'' peaks (in 2D) and (filament type) saddles (in 3D), 
 and build the corresponding fields $\mathbf s(\mathbf r)$ (via {\tt fft} using equation~(\ref{eq:defL3D})).
 This field allows us to assign a spin to each `halo' at position $\mathbf r_h$ and a closest saddle, $\mathbf r_s$. 
 Given the relative position $\mathbf r_h- \mathbf r_s$ as measured in the frame defined by the Hessian at $\mathbf r_s$,  
we may project the direction of the spins, $\hat{\mathbf{s}}\equiv \mathbf s/s$ of all `haloes' in the vicinity along the corresponding
local cylindrical coordinate $(\mathbf e_R,\mathbf e_\phi,\mathbf e_z)$. We may then compute  
the one point statistics of 
$\mu_z\equiv \hat{\mathbf{s}}\cdot \mathbf e_z$ {\sl per octant}. 

\subsubsection{2D GRF fields spin flip }
\label{sec:def2DGRF}

In two dimensions, the expectation is that the spin should be aligned
or anti-aligned with $e_z$ depending on each quadrant.

Let us first start with a set of 25  2D $2048^2$ maps from a power-spectrum with 
$n=-1/2$.
The map is first smoothed with Gaussian filter of width $L_h=4$  pixels, and the positions of the peaks are identified. It is then smoothed 
again over $L_s=24$ pixels, exponentiated {\green (in order to mimick the almost log-normal statistics of the  evolved cosmic density field)}, and the corresponding Hessian and tidal fields are computed, together with the momentum map,
which is thresholded above $1/30^{\rm th}$ of its highest value (see Figure~\ref{fig:2D-spin-GRF}). The peaks of this second map are identified as
'saddles' for contrasts higher than 2.5.   
Figure~\ref{fig:2D-spin-GRF-2} shows that the average spin of `haloes' in each quadrant is flipping from one quadrant to the next, with a statistically 
significant non zero mean value in each quadrant.

\begin{figure}
\begin{center}
\includegraphics[width=0.95\columnwidth]{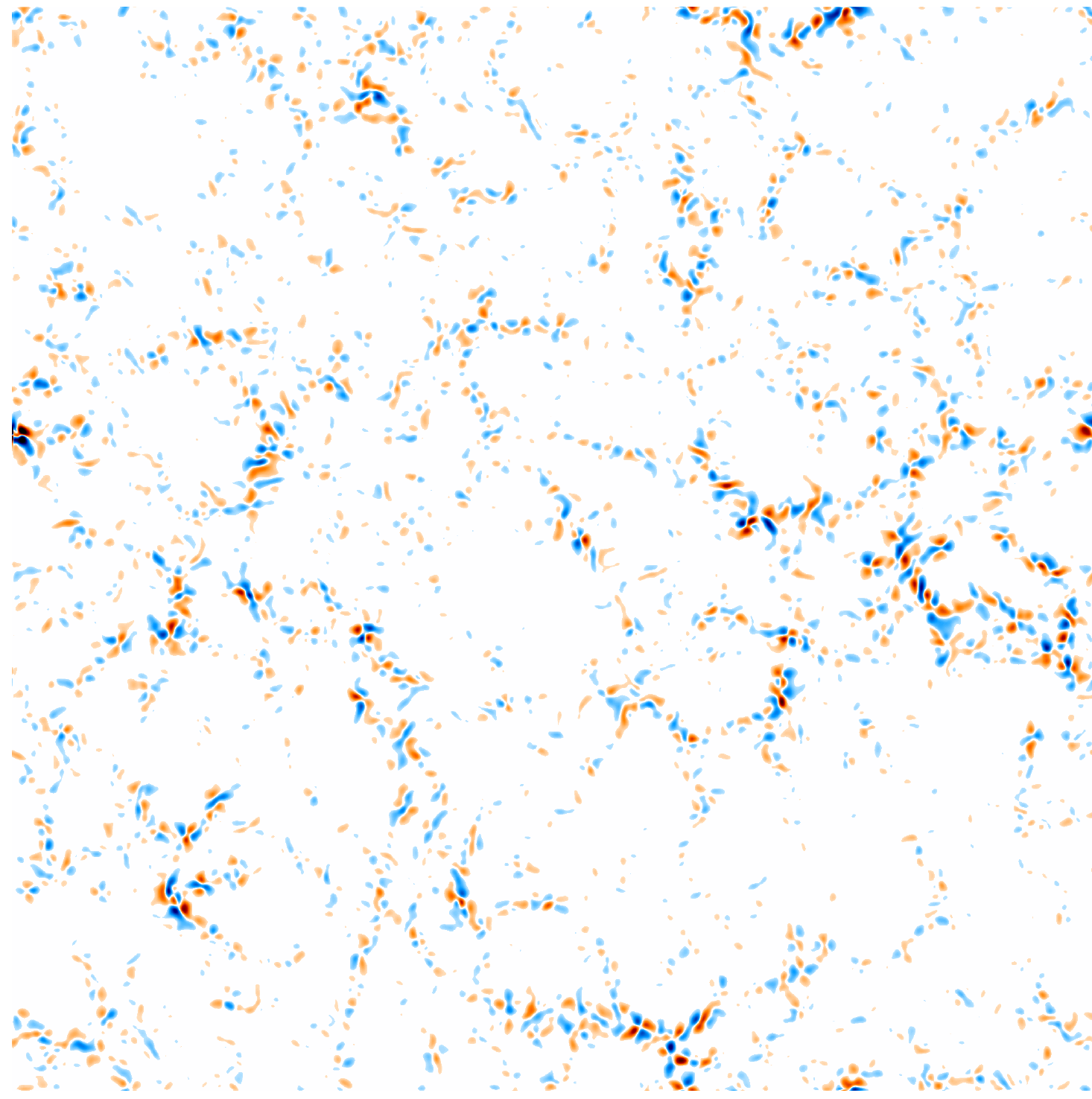}
\includegraphics[width=0.475\columnwidth]{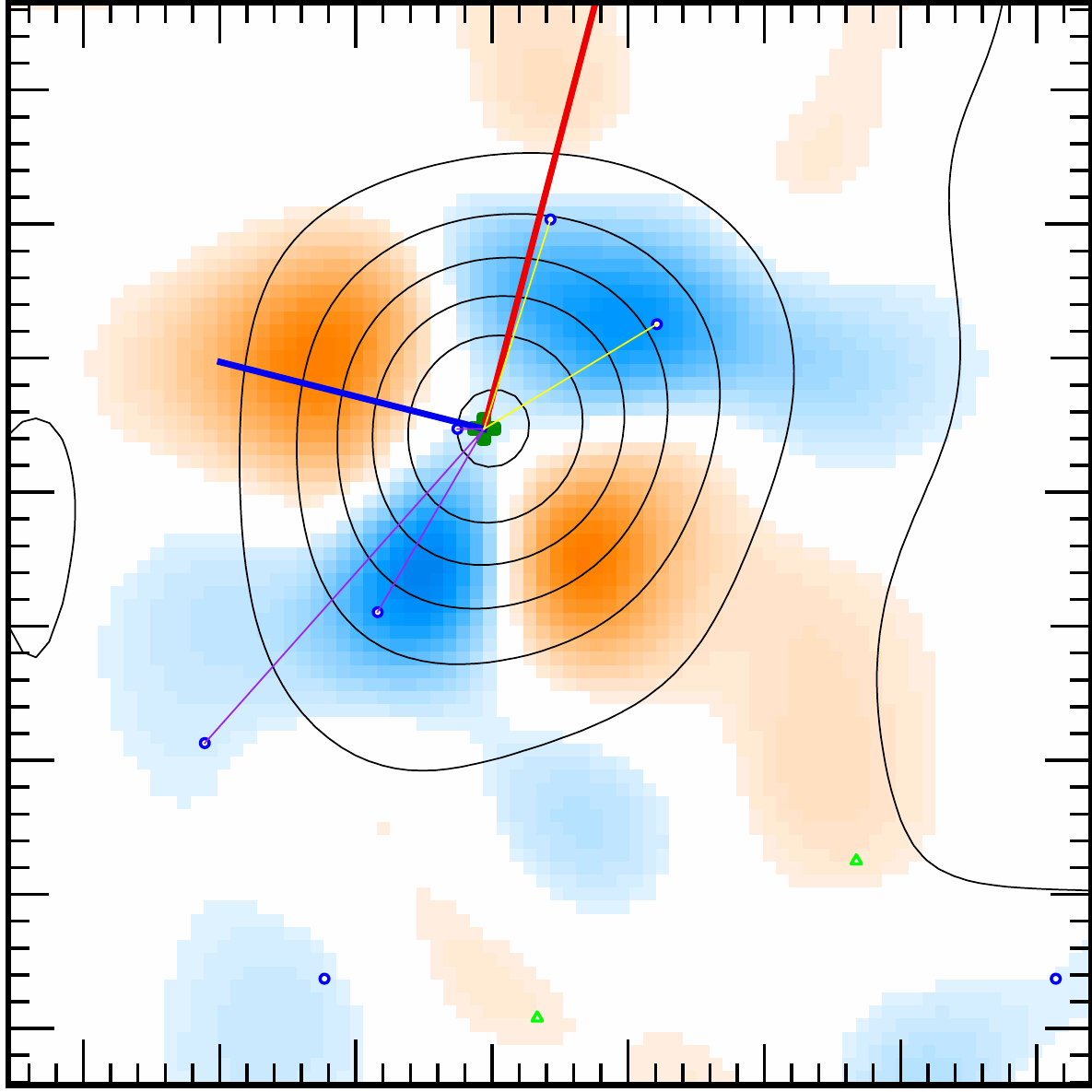}
\includegraphics[width=0.475\columnwidth]{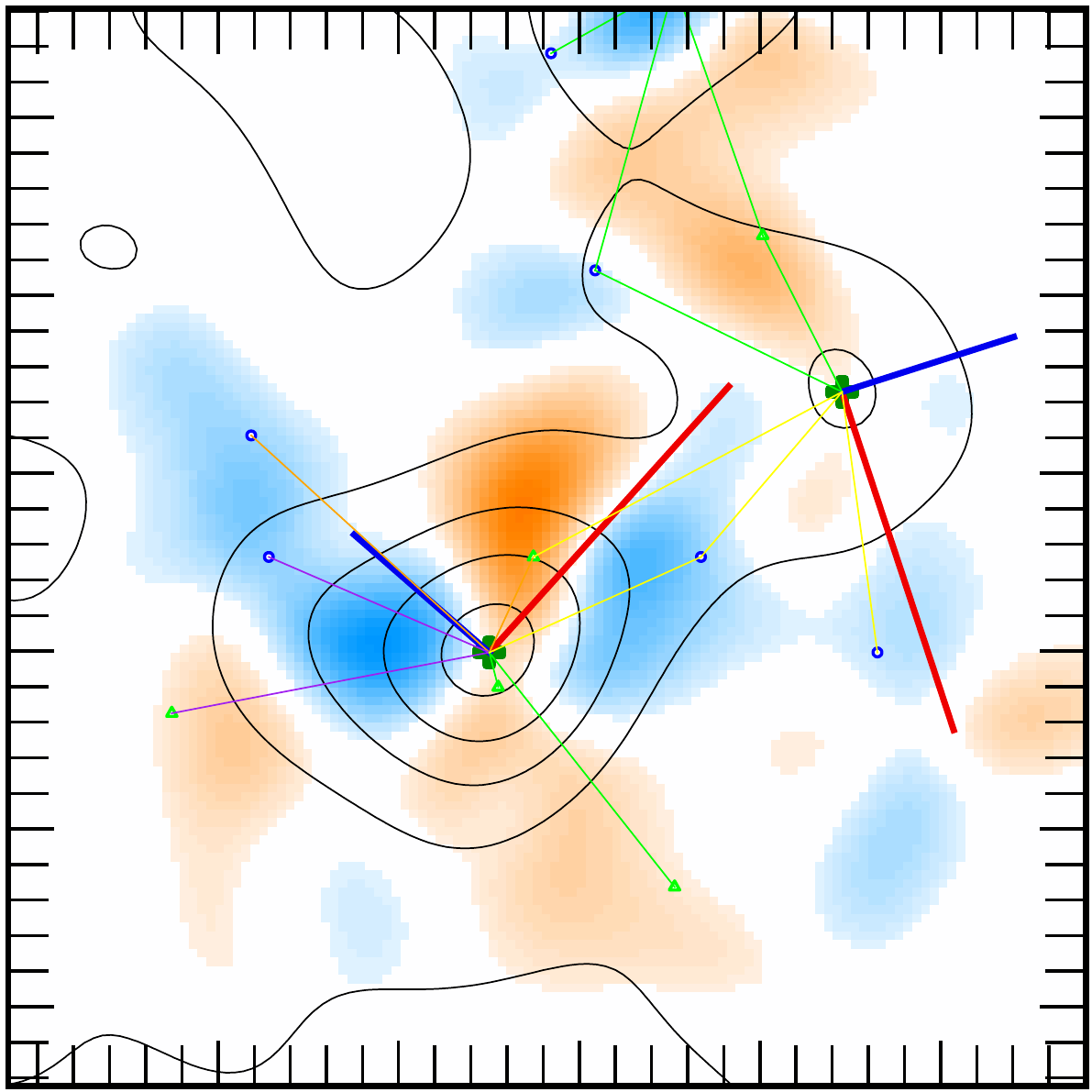}
\caption{
{\sl Top: } example of spin map (colour coded by sign) generated following the prescription of Section~\ref{sec:def2DGRF}.
{\sl Bottom: } the local frame (in red, long axis and blue) around a couple of `saddle's.
The black contours correspond to the density.}
\label{fig:2D-spin-GRF}
\end{center}
\end{figure}

\begin{figure}
\begin{center}
\includegraphics[width=0.75\columnwidth]{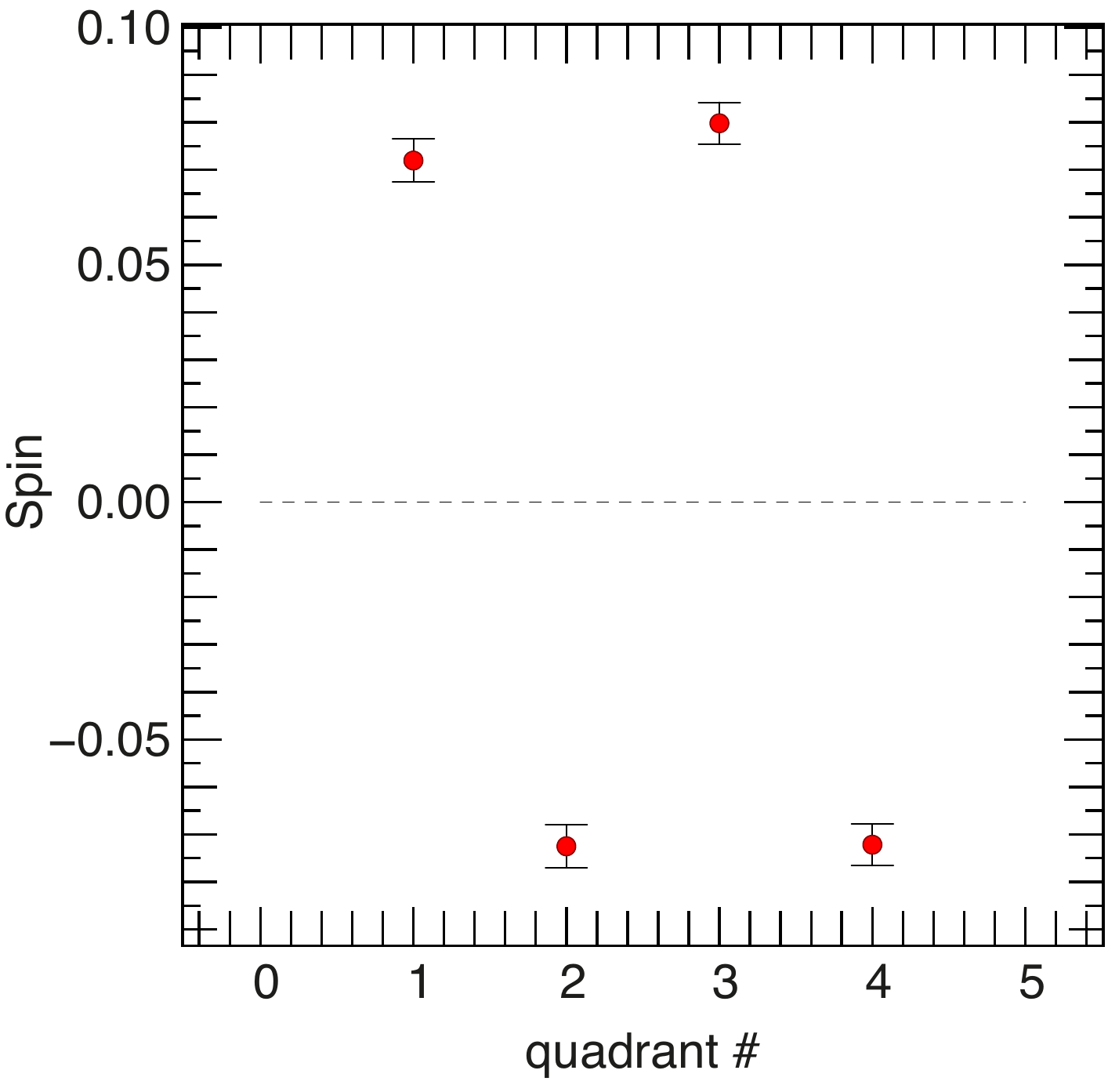}
\caption{\small{
 Alignment of `spin' along $\mathbf e_z$ in two dimensions as a function of quadrant rank, clockwise.
As expected, from one quadrant to the next, the spin is on average  unambiguously  flipping sign. }}
\label{fig:2D-spin-GRF-2}
\end{center}
\end{figure}

\subsubsection{3D GRF fields spin flip }

Let us similarly consider a set of 
20  three dimensionnal $256^3$ cubes from a power-spectrum with $n=-2$.
The cube is first smoothed with Gaussian filter of width $L_h=4$  pixels, and the positions of the peaks are identified. It is then smoothed 
again over $L_s=24$ pixels, exponentiated, and the corresponding Hessian and tidal fields are computed, together with the spin field,
which is thresholded above $1/30^{\rm th}$ of its highest value. The saddle of this second cube are identified as  for contrasts higher than 1.  Only peaks closer than one smoothing length from the large-scale saddles are kept. The angle between their spin and the filament axis is computed and stored depending on the octant they belong to. In this section, the octants are numbered from 1 to 8 depending on the separation from the peak to the saddle $\mathbf{r}=(x,y,z)$:  $x,y,z>0$ (\#1), $x<0\;\&\; y,z>0$ (\#2), $x,y<0\;\&\; z>0$ (\#3), $y<0\;\& \;x,z>0$ (\#4), $z<0\;\&\;x,y>0$ (\#5), $x,z<0\;\&\; y>0$ (\#6), $x,y,z<0$ (\#7) and $y,z<0\;\&\; x>0$ (\#8).
Fig.~\ref{fig:3D-spin-GRF} shows that, as expected, the component of the spin aligned with the filament axis is flipping sign from one octant to the other.

\begin{figure}
\begin{center}
\includegraphics[width=0.75\columnwidth]{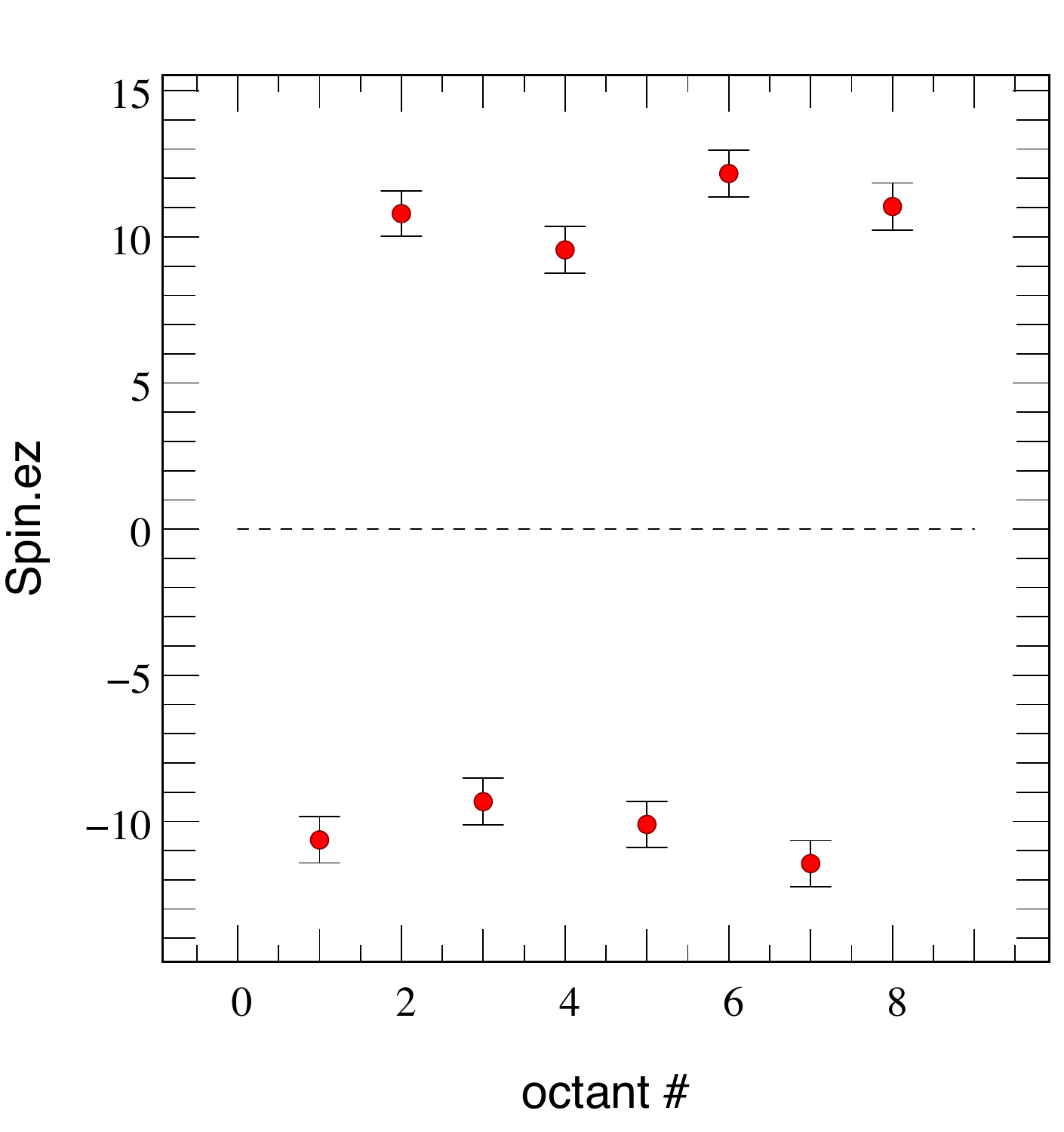}
\caption{
Alignment of the spin along the filamentary direction depending on the considered octant. As predicted by the theory, the z-component of the spin is flipping sign from one octant to the other.}
\label{fig:3D-spin-GRF}
\end{center}
\end{figure}

\subsection{Validation on dark matter simulations at $z=0$}
\label{sec:simulation}

Let us now identify the Eulerian implication at redshift zero of the above sketched Lagrangian theory.
For this we must rely on N-body simulations.
Hence we now make use  of the $43$ million dark matter haloes detected at redshift zero in the Horizon 4$\pi$ N-body simulation \citep{teyssieretal09}  to test some of the outcomes of the Anisotropic Tidal Torque Theory presented in this paper. 
This simulation contains $4096^3$ DM particles distributed in a 2 $h^{-1}$Gpc periodic box and is characterized by the following $\Lambda$CDM cosmology: $\Omega_{\rm m}=0.24 $, $\Omega_{\Lambda}=0.76$, $n=0.958$, $H_0=73 $ km$\cdot s^{-1} \cdot $Mpc$^{-1}$ and $\sigma _8=0.77$ within one standard deviation of WMAP3 results \citep{Spergeletal03}. 
The initial conditions were evolved non-linearly down to redshift zero using the adaptive mesh refinement code RAMSES \citep{teyssier02}, on a $4096^3$ grid. The motion of the particles was followed with a multi grid Particle-Mesh Poisson solver using a Cloud-In-Cell interpolation algorithm to assign these particles to the grid (the refinement strategy of  40 particles as a threshold for refinement allowed us to reach a constant physical resolution of 10 kpc, see the above mentioned two references).

The Friend-of-Friend Algorithm \citep{huchra82} was used  over $18^3$ overlapping subsets of the simulation with a linking length of  0.2 times the mean interparticular distance  to define dark matter haloes. 
In the present work, we only consider haloes with more than 40 particles (the particle mass being $7.7\times 10^{9}M_{\odot}$). The mass dynamical range of this simulation spans about 5 decades.

The filament's direction is then defined via the global skeleton algorithm introduced by \citet{sousbie09} and based on Morse theory. It defines the skeleton as the set of critical lines joining the maxima of the density field through saddle points following the gradient. 
In practice \citet{sousbie09} 
 define the peak and void patches of the density field as the set of points converging to a specific local maximum/minimum while following the field lines in the direction/opposite direction of the gradient. The skeleton is then the set of intersection of the void patches i.e. the subset of critical lines connecting the saddle points and the local maxima of a density field and parallel to the gradient of the field.
In practice, the $\sim$70 billion particles of the Horizon-4$\pi$ were sampled on a $2048^3$ cartesian grid and 
{\green{ the  density field was smoothed  using {\tt mpsmooth} \citep{prunetetal08} over  a scale of 5 $h^{-1}$Mpc corresponding to a mass of 
$1.9 \times 10^{14} M_\odot$. }}This cube was then divided into  $6^3$ overlapping sub-cubes and  the skeleton was computed for each of these sub-cubes. It was then reconnected across
the entire simulation volume to produce a catalog of segments which locally defines the direction of the filaments.

Figure~\ref{fig:3D-spin-hz4pi} demonstrates that the spins of  the 43 million  dark haloes of the simulation obey the expected 
mass dependent flip predicted by the theory presented in Section~\ref{sec:3D}. On top of the  alignment with the filament direction found e.g in \cite{codisetal12}, haloes are shown to 
have a spin increasingly perpendicular to $\mathbf{e}_\phi$ at low-mass (red) and up to the critical mass ($\simeq 10^{12}M_{\odot}$), while high-mass haloes have a spin parallel to the $\mathbf{e}_\phi$ direction. The transition from alignment to orthogonality occurs around $M_{\rm tr}\simeq 5\cdot 10^{12}M_{\odot}$. 

\begin{figure}
\begin{center}
\includegraphics[width=0.95\columnwidth]{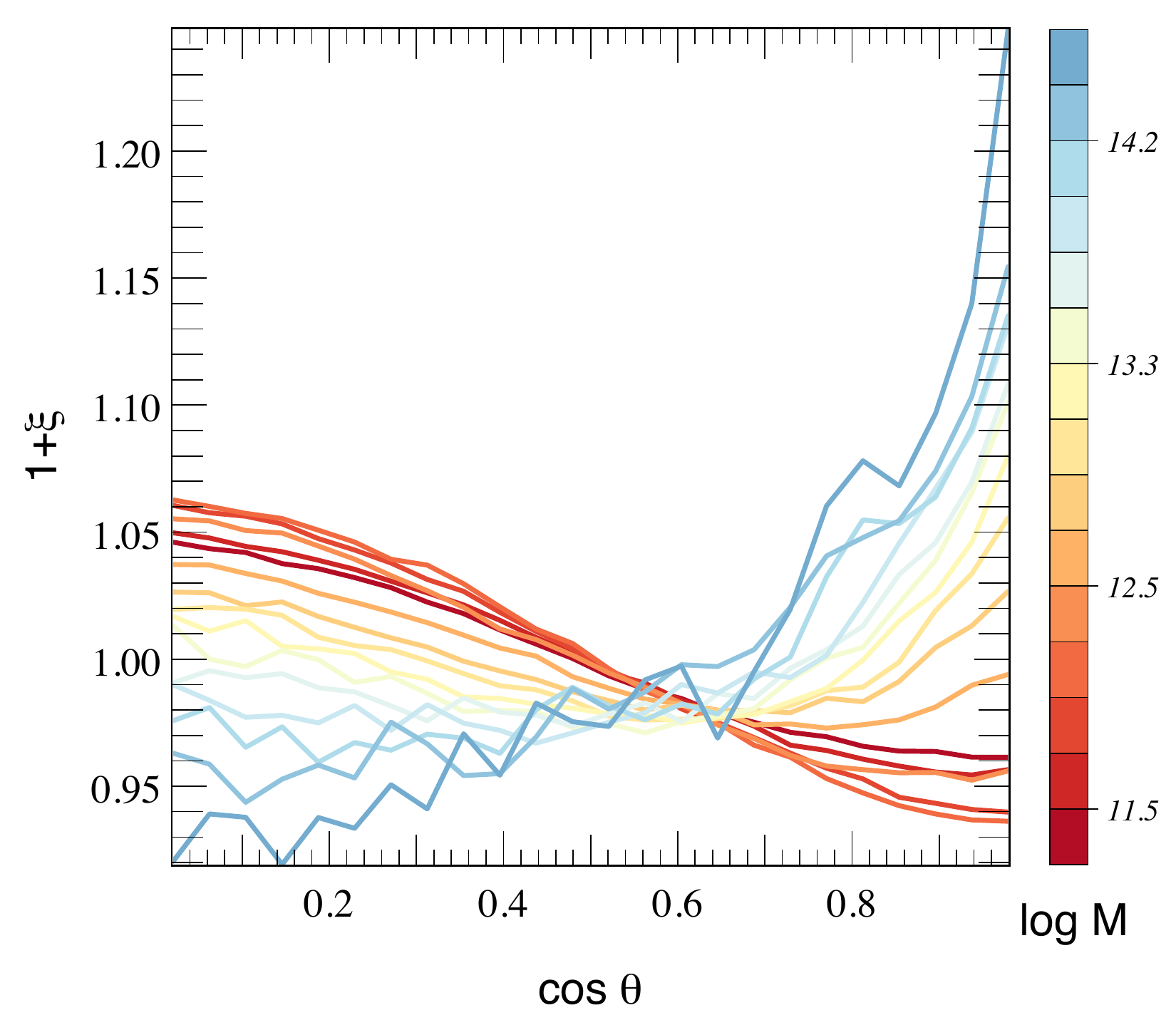}
\caption{
 Alignment of `spin' along $\mathbf{e}_\phi$  in the Horizon-4$\pi$  simulation. The normalised histogram of the cosine of the angle between the spins and the closest filament's direction is displayed. Deviations from the $\xi=0$ uniform distribution are detected and depends on the dark matter halo mass.
Haloes have a spin aligned with the $e_{\phi}$ direction on average at low-mass (red) and perpendicular to it at larger mass (blue).
}
\label{fig:3D-spin-hz4pi}
\end{center}
\end{figure}

Figure~\ref{fig:3D-spin-hz4pi-distancetosaddle} shows that the spins tend to be more aligned with the filament axis when getting closer to the saddle point. The alignment decreases from $\cos \theta=0.511$ at $r\simeq 20$ Mpc/h to $\cos \theta=0.506$ at $r < 1$ Mpc/h. This qualitative trend is in full agreement with the anisotropic tidal torque theory picture presented in Section~\ref{sec:3D} for which on average, spins are aligned with the filament axis in the plane of the saddle point and become misaligned when going away from this saddle point.

\begin{figure}
\begin{center}
\includegraphics[width=0.95\columnwidth]{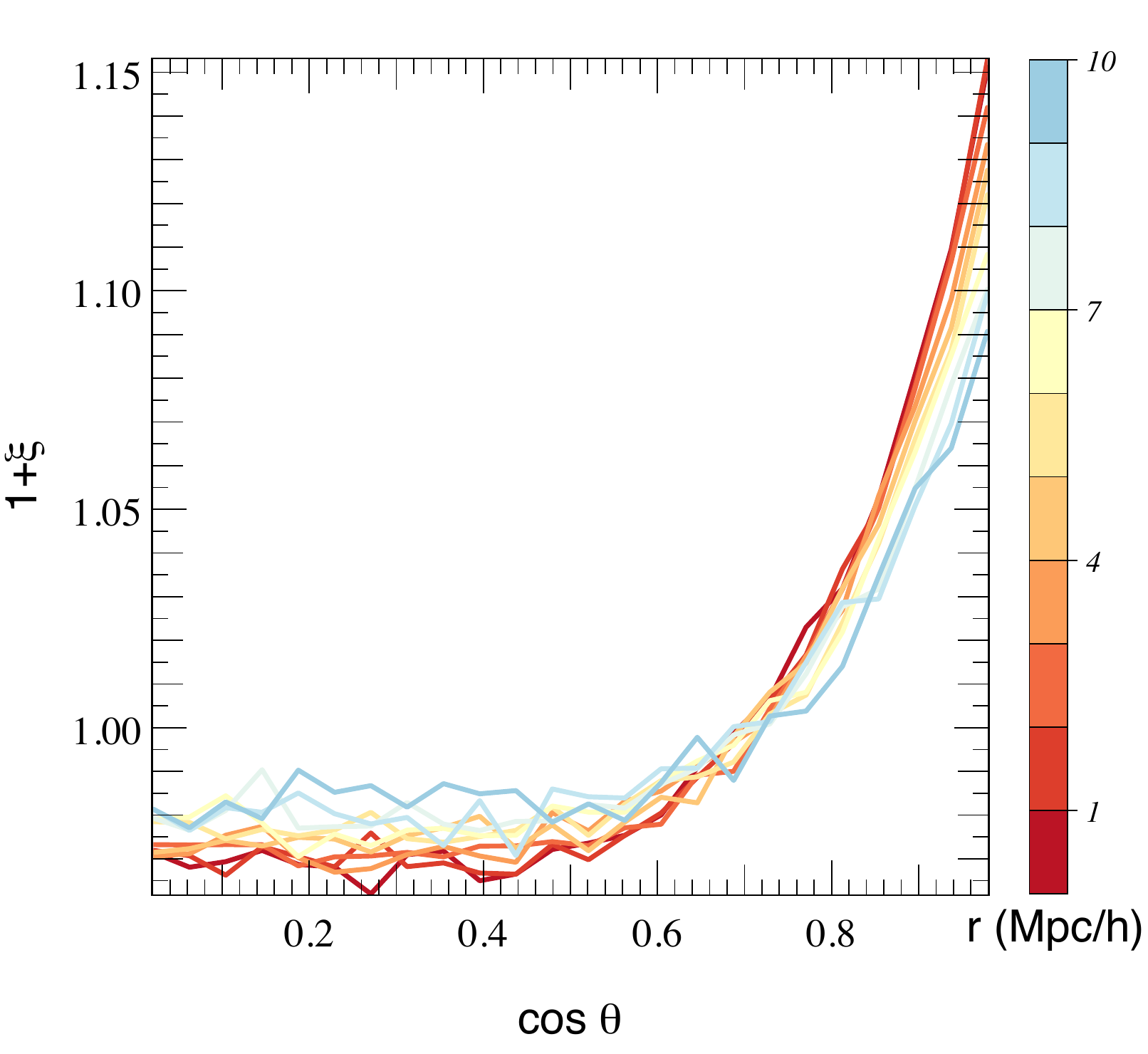}
\caption{
 Alignment of the spins and the filaments  in the Horizon-4$\pi$  simulation as a function of the distance to the closest saddle point (from red -- 0Mpc/h -- to blue --10 Mpc/h --).
The alignment decreases with the distance to the saddle point as predicted by the anisotropic tidal torque theory model.}
\label{fig:3D-spin-hz4pi-distancetosaddle}
\end{center}
\end{figure}

Figure~\ref{fig:mass-distancetosaddle} displays the occupancy of haloes along the filaments. It appears that the higher the mass, the more concentrated they are far from the saddles. This is in good agreement with the halo mass gradient along the filaments described in Section~\ref{sec:massgrad}.

\begin{figure}
\begin{center}
\includegraphics[width=0.95\columnwidth]{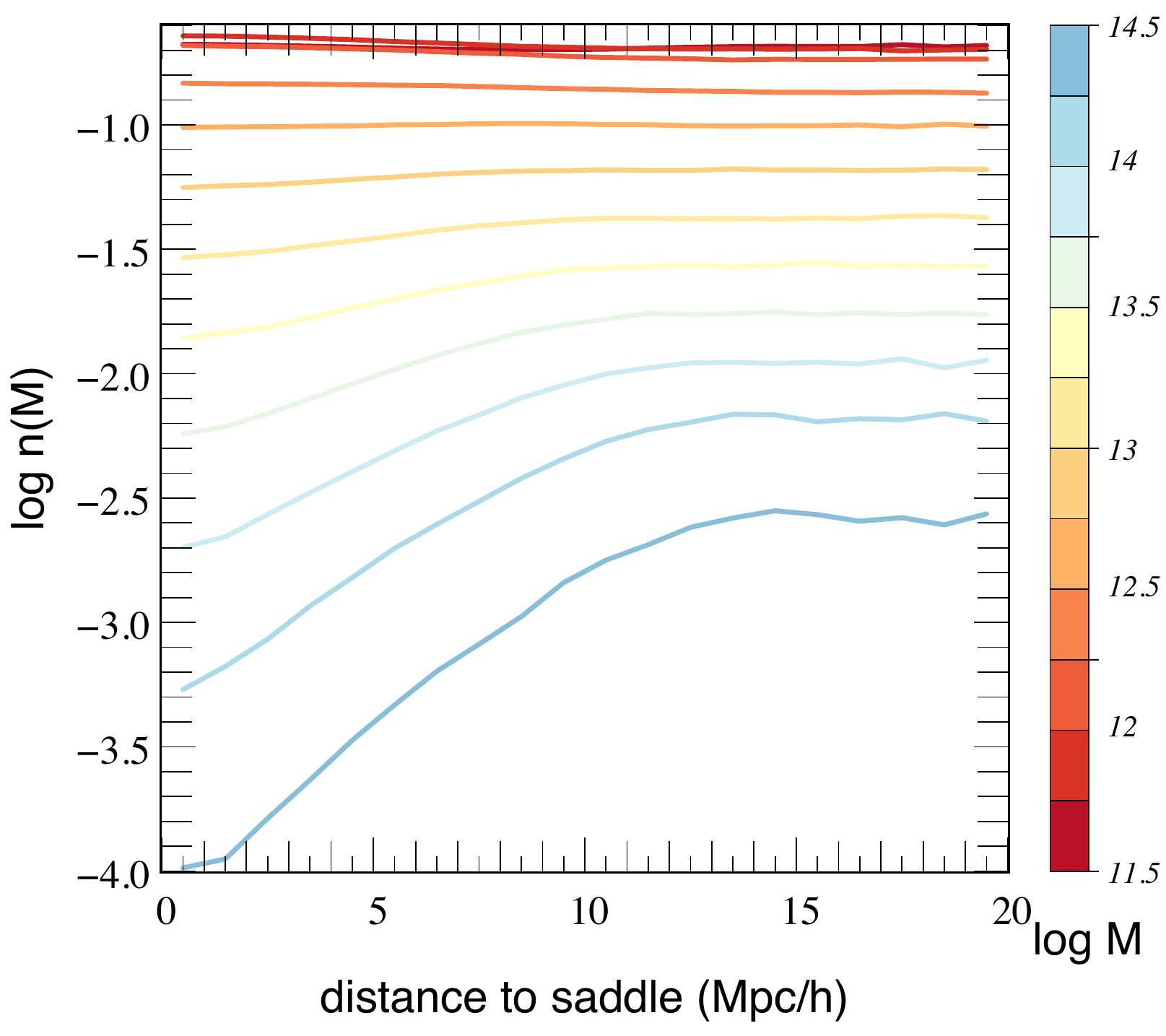}
\caption{
(Log-) Fraction of haloes of different mass (from red to blue in solar mass units) as a function of the distance to the saddle point in the Horizon-4$\pi$ simulation. Low-mass haloes (red) lie almost uniformly along the filaments (with a small concentration --not clearly seen in logarithmic units-- around the saddles due to consumption when going towards the nodes) while high-mass haloes (blue) are more concentrated far from the saddles.
\label{fig:mass-distancetosaddle}}
\end{center}
\end{figure}

Overall, the above GRF experiments as well as the 
 re-analysis of the  Horizon 4$\pi$ N-body simulation seem consistent with the prediction of the theory 
presented in Section~\ref{sec:2D} and \ref{sec:3D}. While the former demonstrates that interferences from 
neighbouring saddles do not wash out the tide correlations, the latter suggests that 
on the scales probed by this experiment, this Eulerian  measure still captures features of the underlying Lagrangian 
theory.

\section{ Conclusions and perspectives}
\label{sec:conclusion}

TTT was revisited while focussing on an anisotropic peak background split
in the vicinity of a saddle point.  Such critical point captures as a point process the 
geometry of a typical filament embedded in a given wall \citep{Pogosyanetal1998}. 
The induced  mis-alignment between the tidal tensor and the hessian of the density   
simply  explains  the surrounding  transverse and 
longitudinal point reflection-symmetric geometry of the spin distribution near filaments.

This geometry of the spin field predicts in particular that less massive galaxies have their spin parallel to the filament,
while more massive ones have their spin in the azimuthal direction.
The corresponding  transition mass  follows from this 
geometry together with  its  scaling with the mass of non linearity, in good agreement with measurements in simulations.

The main findings of this paper are:
i) galaxies form near filaments embedded in walls, and flow towards the nodes: this anisotropic environment produces the long wave modes 
on top of which galactic haloes pass the turnaround threshold;
ii) a typical filament is elongated and flattened: as a point process,  it is therefore best characterized by its triaxial saddle points;
iii) the spin geometry is octupolar in the vicinity of the saddle point, displaying a point reflection symmetry;
iv) the mean spin field is parallel to the filament axis in the plane of the saddle point and becomes azimuthal away from it;
v) the constrained tidal torque theory presented in this paper allows to accurately predict the transition mass of the spin-filament alignment measured in simulations.
vi) this theory seems consistent  with both GRF experiments and results from N-body simulations.
vii) a dual theory describes spin alignments in voids (see Appendix~\ref{sec:voids}).

\subsection{Discussion}

One of the striking
 features of this anisotropic extension of TTT is 
  the  induced quadrupolar point-symmetric flattened geometry of the spin distribution
  near a saddle point,
   which effectively scales down by one order of magnitude
the transition mass away from the mass of non-linearity, in  agreement with 
the measured scaling.
The qualitative analysis derived from first principles in the vicinity of a given 
saddle point seems to hold when considering  realizations of  GRF, once proper account of
the induced geometry near such points is taken care of.
In effect, we have shown that  the geometry of the saddle point provides a natural `metric' 
(the local frame as defined by the hessian at that saddle point) relative to
 which we can study the dynamical evolution of dark haloes along filaments.
It should allow us to study how galactic feeding \citep[via helicoidal cold flow, see][]{dubois14} should vary with curvilinear coordinate along the filament.
It was indeed found in that paper using hydrodynamical simulations
that such flows were reaching  galaxies in the so-called circum-galactic medium with velocities roughly  parallel the polar axis. 
Taken at face value, such findings suggest that the flow feeding galaxies has significant helicity during that phase.

Another striking feature of this {\sl Lagrangian} framework is that it captures 
naturally the arguably 
non-linear {\sl Eulerian} process of spin flip via mergers.
Recently, \cite{laigle2014} showed 
 that angular momentum generation of haloes could  be captured  in Eulerian space via the secondary advection of vorticity which the formation of the filament generates,
 whereas we show in this paper that it may also be described 
in Lagrangian space via the analysis of the anisotropic tides generated by the filament to be.
 No description is more fundamental than the other but are the two (Eulerian versus Lagrangian) sides of the same coin. The mapping between the two descriptions requires
 a reversible time integrator, such as the Zel'dovich approximation, which clearly limits
 its temporal validity to weakly non-linear scales. 
 Our proxy for the spin, equation~(\ref{eq:spin-ATTT}), is an approximation which seems to quantitatively capture the relevant physics.
 It is remarkable that such an (admittedly approximate) straightforward extension of TTT captures what seems to be the driving process of  spin orientation
 acquisition and its initial evolution.
It is also sticking that very  simple closed form for the spin orientation distribution in the vicinity of the saddle point
are available for this proxy.

 Our theory here has focussed on a two-scale process. Given the characteristics  of $\Lambda$CDM hierarchical clustering 
one can anticipate that this process occurs on several nested scales at various epochs - and arguably  on various scales at the same epoch. {The scenario we propose for the origin of this signal is, like the signal itself, relative to the linear scale involved in defining the filaments and as such, multi-scale. It will hold as long as filaments are well defined in order to drive the local cosmic flow.} In other words,
one expects smaller-scale filaments are themselves embedded in larger-scale walls. The induced multi-scale anisotropic flow transpires in the scaling of the transition mass with smoothing, as discussed in \cite{codisetal12}.

Of course, we have here completely ignored  the effect of feedback, which will play some -- yet undefined -- role 
 in redistributing the cosmic pristine gas falling onto forming galaxies.
Another issue would be to estimate for how long  this entanglement between the large-scale dynamics and the kinematic properties of high redshift  pervades, given the disruptions induced by feedback. What will be the effect of AGN feedback \citep{duboisetal2013,prietoetal2014} on tidally biased secondary infall?
 \cite{ocvirketal08} have also shown that at lower redshift, the so-called hot mode of accretion will kick in; 
how will  hot flows  wash out/disintegrate  these ribbons?  Given that they locally reflect the large-scale geometry, will the  gas  continue to flow-in  along preferred 
directions \citep[as does the dark matter, see e.g.][]{aubertetal04}, or does the hot phase erase any  anisotropy?
Will the above-mentioned smaller-scale non-linear dynamics eventually wash out any such trace?

\subsection{Perspectives}

 One possibly significant shortcoming of the analysis is the proxy involved in using the hessian of the density instead of  the inertia tensor (though  see  Section~\ref{sec:multiscale}). 
 This is critical in  order to retain a point process for the induced spin, but is achieved at the expense of having an adequate 
 estimate for the {\sl amplitude} of the spin, which is unfortunate because from the point of view of morphology, the dividing line 
 between spirals and ellipticals is likely to be spin amplitude.
Let us nonetheless  assume that e.g. match to simulations or Ansatz such as those described in \cite{Schafer2012} 
will yield access to reasonable fit to spin amplitude and 
discuss briefly implications  to galaxy formation within its cosmic web.

\subsubsection{Epoch of  maximal spin advection?}

The inspection of hydrodynamical simulations \citep[e.g.][using tracer particles]{codisetal12}  shows that
  ribbon-like caustics  feed the central galaxy along its spin axis 
from both poles. The gas flowing roughly parallel to the spin axis of the disc along both directions will typically impact the disc's circum-galactic medium and shock once more (as it did when it first reached the wall, and then the filaments, forming those above mentioned ribbons), radiating away its vertical momentum \citep[see][] {tillsonetal2012}. 
 These ribbons are generated via the same winding/folding process as the protogalaxy, and represent the dominant source of secondary filamentary infall
which feeds the newly formed galaxy with gas of well-aligned angular momentum.

Having computed the most likely spin (direction) as a function of position, it is therefore of interest to measure its  covariant polar flux through a drifting 
forming galaxy.

From our knowledge of the spin distribution within the neighbourhood of a given saddle we may then compute the 
 rate of advected spin within some galactic volume ${\cal V}=S \Delta z$; it reads 
\begin{align}
{ \mathbf{ \dot s}}&= \int  {\rm d}^2 \mathbf{S}\cdot  \mathbf{v} \otimes\rho \mathbf{s}=\int_{\cal V} {\rm d}^3 \mathbf{r}\,\,\nabla\cdot \left(\rho \mathbf{v}\otimes \mathbf{s}\right) \nonumber
\,, \\
 & \approx  {S}\left[\rho \mathbf{v}\otimes \mathbf{s}\right]^+_- \approx  {S \Delta z }\,\frac{\partial}{\partial z} \left( \rho {v}_z  {s}_z\right)\,,  \label{eq:flux}
\end{align}
where the last equality assumes that the advection is quasi-polar, and that the spin is mostly aligned with 
the filament. In equation~(\ref{eq:flux}) $\mathbf{v}$ is the gradient of the potential.
Let us identify the curvilinear coordinate, $z_{\rm up}$, for which this flux is maximal:
\begin{align}
z_{\rm up} = \underset{\displaystyle z} {\mathrm{argmax}} \frac{\partial}{\partial z} \left( \rho {v}_z  {s}_z\right)=
\Big\{ z \,\Big|{\displaystyle \frac{\partial^2}{\partial z^2}\left( \rho {v}_z  {s}_z\right)=0} \Big\}\,.
\end{align}
The coordinate $z_{\rm up}(\nu,\kappa_1,\kappa_2)$ characterizes the most active regions in the cosmic web for galactic spin up. Focussing on the most likely saddle,  
the argument sketched in Section~\ref{sec:massgrad} allows us to assign a redshift-dependent spin-up mass,
${\cal M}_{\rm up}(Z)$,
 via equations~(\ref{eq:defsigmap}) and~(\ref{eq:defMsigmaZ}). 
There could be an observational signature, e.g. in terms of the cosmic evolution of the  SFR, as 
 maximum spin-up corresponds to efficient pristine  cold and dense gas accretion, which in turn induces 
 consistent and steady star formation.

\subsubsection{Morphological type versus loci on web?}

The magnitude of the spin of galaxies could be taken as a proxy for morphological type.
Indeed, \cite{welkeretal14,welkeretal15} have shown in cosmological hydrodynamical simulations
that spin direction and galactic sizes where sensitive to the anisotropic environment. 
It is shown in particular that the magnitude of the spin of simulated galaxies increases steadily and aligns itself
preferentially with the nearest filament
when no significant merger occurs,  
in agreement with the first phase of the  above described spin-up \citep[see also][]{pichonetal11}. During that phase, the fraction of larger spirals should increase. 
In contrast, following Figure~\ref{fig:mass-distancetosaddle}, if we account for the fact that galactic morphology 
-- the fraction of ellipticals, correlates with dark halo mass, it  should then  increase with distance to saddle.

In order to tackle such process theoretically, it 
 would therefore  be worthwhile to revisit \cite{Quinnetal1992} in the context of this constrained theory of tidal torques and quantify how the dynamics of concentric shells are differentially 
biased by the tides of a saddle point. This would allow us to describe the whole timeline of anisotropic secondary infall.

\subsubsection{Implication for weak lensing?}

Weak lensing attempts to probe the statistics of the cosmic web   between background galaxies -- which shape is assumed to 
be uncorrelated -- and the observer, while assuming that observed shape statistics reflects the deflection of light going through the intervening 
web. In view of Figure~\ref{fig-momentum3D}, if we take as a proxy  spin alignment for shape alignment, we can in principle 
compute the expectation of 
$\xi(\Delta \mathbf r)\equiv\langle \mathbf s(\mathbf r)\cdot \mathbf s(\mathbf r')| \,{\rm skl}\,\rangle$ as a function of $\Delta \mathbf r=\mathbf r - \mathbf r' $. Calling $\delta \mathbf s =  \mathbf s  -\langle \mathbf s(\mathbf r) | \,{\rm skl}\,\rangle$, we have  $\xi(\Delta \mathbf r)=
\langle \mathbf s(\mathbf r)| \,{\rm skl} \rangle \cdot \langle \mathbf s(\mathbf r')| \,{\rm skl}\,\rangle + 
\langle \delta \mathbf s(\mathbf r)\cdot \delta \mathbf  s(\mathbf r')| \,{\rm skl}\,\rangle + 2
\langle \delta \mathbf s(\mathbf r)\cdot  \mathbf  s(\mathbf r')| \,{\rm skl}\,\rangle
$. Let us just focus here on the first term, $\langle \mathbf s(\mathbf r)| \,{\rm skl} \rangle \cdot \langle \mathbf s(\mathbf r')| \,{\rm skl}\,\rangle$.
Given equation~(\ref{eq:defL3Dsol}),  we can compute it and find that  it will typically be non-zero and vary significantly
depending on both the magnitude and the orientation of $\Delta \mathbf r$. E.g. if $\Delta \mathbf r$ is  off axis along the filament, but 
if the pair is close to the saddle and $|\Delta \mathbf r|$ is small it will be positive (spins will align as they 
are both within  coherent region of the saddle's tides), while if $|\Delta \mathbf r|$ is somewhat larger
it will vanish  (spins will be perpendicular). Conversely, if $\Delta \mathbf r$ is transverse to the filament  and $|\Delta \mathbf r|$ is small,
it will be positive, but if $|\Delta \mathbf r|$ is of the order of  the size of one octant it will typically vanish again.
The formalism presented in Section~\ref{sec:shape2D} can clearly be extended (while considering the joint three points statistics) to predict exactly all terms involved in  $\xi(\Delta \mathbf r)$ and 
quantify within this framework the effect of intrinsic alignments on the spin-spin two point correlation.
This is will be to topic of future work (Codis et al, in prep.).

\section*{Acknowledgments}
This work is partially supported by the Spin(e) grants ANR-13-BS05-0005  (\url{http://cosmicorigin.org}) of the French {\sl Agence Nationale de la Recherche}
and by the ILP LABEX (under reference ANR-10-LABX-63 and ANR-11-IDEX-0004-02).
CP thanks D.~Lynden-bell for suggesting to tackle  this problem and Churchill college for hospitality while this work was completed.
Many thanks to J.~Devriendt, A.~Slyz, J.~Binney, Y.~Dubois, V.~Desjacques, C.~Laigle  and S.~Prunet for discussions about tidal torque theory,
and to  our collaborators of the Horizon project (\url{http://projet-horizon.fr}) for helping us produce the Horizon-4$\pi$ simulation,
F.~Bouchet for allowing us to use the {\tt magique3} supercomputer during commissioning, and to S. Rouberol for making it possible,
and running the {\tt horizon} cluster for us.
SC and CP thanks Lena for her hospitality when this work was initiated, and Eric for his help with some figures.

\bibliographystyle{mnras}
\bibliography{author}

\appendix
\section{A multi-scale theory}
\label{sec:A1}
The proxy we take for the spin direction 
\begin{equation}
\label{eq:proxyAppendix}
s_{i}=  \sum_{j,k,l}\epsilon_{ijk} H_{jl}T_{lk}\,,
\end{equation}
is, as mentioned in the main text, a (quadratic) approximation. First, because Equation~(\ref{eq:TTT}) is only valid in the linear regime \citep{porcianietal02}, but  possibly more importantly  because we take the Hessian as a proxy for the inertia tensor.
In practice, recall  that this approximation seems nevertheless to capture the essence of the processes at work in aligning spins with the large-scale structure given its ability to explain observed alignments through the comparison with simulations presented in Section~\ref{sec:simulation}
together with the measured mass transition. This suggests experimentally  that it   is indeed reasonable.
Notwithstanding, while $H_{ij}$ and $I_{ij}$ locally share the same eigenframe, their amplitudes are different, leading to a different weighting of field configurations when computing ensemble averages such as in equation~(\ref{eq:defL2D}).
It is therefore important to investigate this possible  shortcoming further in this Appendix.

\subsection{More realistic spin proxies and peak}
\label{sec:multiscale}
 For this purpose, one can  in principle i) impose an additional peak constraint at the location where the spin is computed in order to impose that a proto-halo will form there, and ii)  use more realistic spin proxies. 
 
 The peak  constraint will typically be at a smaller scale than the filament's constraint, which  requires building a two-scale theory 
 and therefore  increases significantly  the complexity of the formalism.
An additional difficulty with ii)  is  that standard local proxies for the  inertia tensor are highly non-linear and therefore require high-dimensional {\sl numerical} integrations that are fairly difficult to implement in practice.
For instance, considering the proxy that  \cite{Schafer2012}  use to locally approximates the inertia tensor, we have 
\begin{equation}
\label{Iij}
I_{ij}=\frac M 5  \left(
\begin{array}{ccc}
A_{y}^{2}+A_{z}^{2}&0&0\\
0&A_{z}^{2}+A_{x}^{2}&0\\
0&0&A_{x}^{2}+A_{y}^{2}
\end{array}
\right)\,,
\end{equation}
(in the frame of the Hessian) where the mass is $M=4/3 \pi A_{x}A_{y}A_{z}\rho_{0}a_{0}^{3}$ and the semi-axes of the ellipsoid, $A_i$,  are function of the eigenvalues of the Hessian (negative for a peak),
\begin{equation}
A_{i}=\sqrt{\frac{2\nu\sigma_{2}}{-\lambda_{i}}}\,.
\end{equation}
The traceless part of $I_{ij}$ that is relevant for torques is then
proportional to the traceless part of the inverse Hessian
\begin{equation}
\label{eq:inertia}
\overline{I}_{ij} = \frac{2}{5} \nu \sigma_2 M \overline{H}_{ij}^{-1}\,.
\end{equation}
This  introduces singular factors like $1/\sqrt{\det \mathbf{H}}$ in the expectation for $\langle s \rangle$. 
Such factors make the  numerical evaluation  of equation~(\ref{eq:defL3D}) more challenging
 as discussed in Section~\ref{sec:principe}.  We therefore postpone their evaluation in three dimensions to future work.  
Let us briefly investigate their implementation in two dimensions.
In Section~\ref{sec:2D-Rh} we introduce a multi-scale description while in  Section~\ref{sec:2D-pkpk}
 we take into account the proxy given by equation~(\ref{Iij}) and we add an explicit peak condition.
\begin{figure}
\begin{center}
\includegraphics[width=0.95\columnwidth]{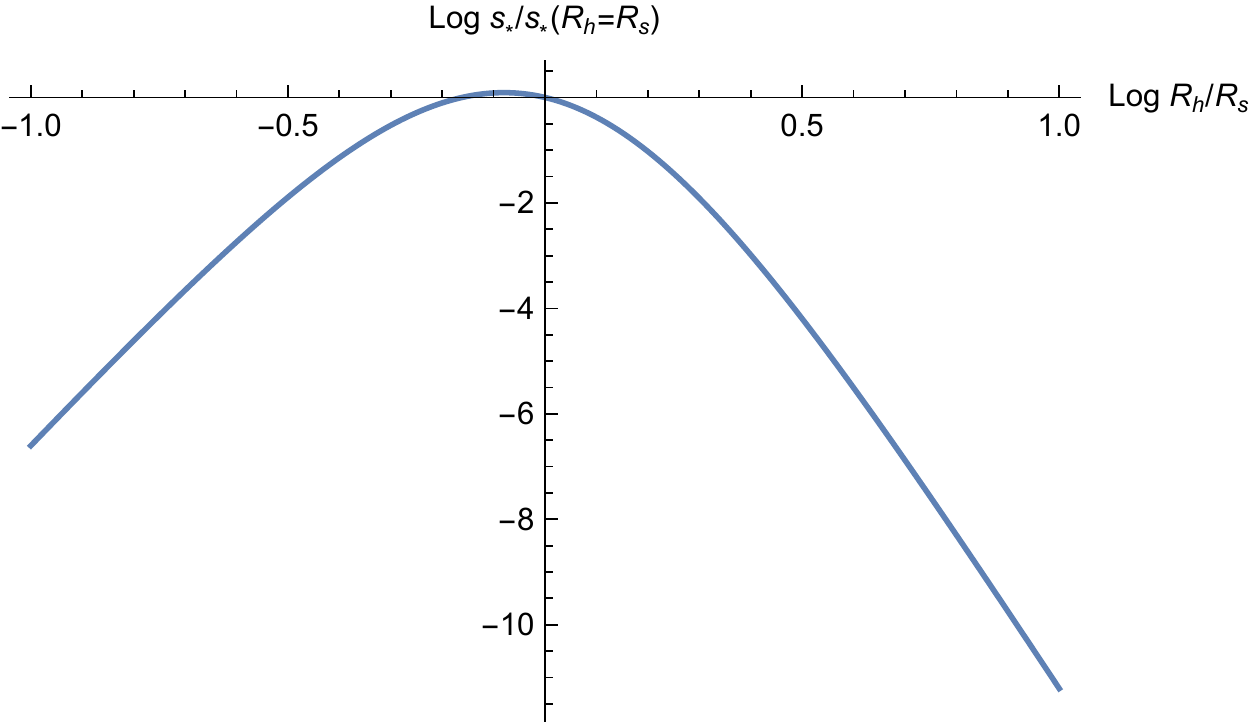}
\caption{
Maximum of spin as a function of the halo's scale $R_{h}$. 
\label{fig:L-Rh}}
\end{center}
\end{figure}

\subsection{A 2D multiscale analysis}
\label{sec:2D-Rh}
The evaluation of $r_{\star}$, see equation~(\ref{eq:defarea}), requires to look for the scale that maximizes the spin amplitude. An improvement  to the main text's approach is to investigate this issue in a two-scale theory where the scale of the central peak is larger than the scale of the halo. In this section, we propose to describe how to implement this multi-scale theory and we show
that this more accurate multi-scale estimate for $r_{\star}$  yields, following the main text, a very similar value  for the critical mass.

More specifically, let us smooth the fields in $\mathbf{r}_{Y}$ (the location of the spin) on $R_{h}$, an additional parameter, characterizing the halo's size and the fields in $\mathbf{r}_{X}$ on a different scale $R_{s}>R_{h}$ in order to impose a large-scale filament (a peak in 2D). If the one-point covariance matrix $\mathbf{C}_{0}$, does not change for a power law power spectrum, the two-point covariance matrix $\mathbf{C}_{\gamma}$ does. In particular, all correlation functions $\xi$ are now function of $r$, $R_{s}$ and $R_{h}$. For instance for a  power-law density power spectrum, $P(k)\propto k^{n}$,
\begin{equation}
\xi_{\phi\phi}^{\Delta\Delta}(r)=\left(\frac{R_{s}^{2}+R_{h}^{2}}{2R_{s}R_{h}}\right)^{-\frac{n+2}{2}}\left[{\cal F}_1^2-\frac{1}{8} (n+2)\frac{ r^2}{R_{s}^{2}+R_{h}^{2}} {\cal F}_2^3\right]\nonumber
\end{equation}
where ${\cal F}_i^j= {}_1F_1\left({n}/{2}+i;j;-{r^2}/{2(R_{s}^{2}+R_{h}^{2})}\right)$.
The mean spin in $\mathbf{r}_{Y}$ is then given by the expectation of $\tilde s_{z}= \sum_{i,j,k} \epsilon_{ij3}  \phi_{ik}   x_{jk}$ given a peak on scale $R_{s}$ in $\mathbf{r}_{X}$ as was  computed in the main text.
Compared to the main text,
the only difference here is that we now also take into account the two-scale process through the two smoothing scales, $R_{s}$ and $R_{h}$. The maximum spin magnitude as a function of the scale $R_{s}/R_{h}$ is then computed and displayed in Figure~\ref{fig:L-Rh}. It appears that the spin magnitude is non-monotonic,
peaking at $R_{h}=0.8 R_{s}$ which is very close to the value of $r_{\star}\approx 0.7 R_{s}$ (when top hat smoothing are taken for both lengths).

 Note that as we are computing here only the component of the spin along the filament, our magnitude plot does not contain
the mass prefactor, and our proxy for the moment of inertia, $\mathbf{I}\approx \mathbf{H}$, reflects only
its orientation, but not its magnitude. We are able to argue, however, that 
Figure~\ref{fig:L-Rh} displays a maximum  spin alignment 
with the filament's direction for some critical value of $R_h$ -- as it shows first an increase and then a 
fall in the  z-component of the spin that is due to Hessian-tidal shear alignment.

\subsection{A 2D multiscale analysis with peak constraint}
\label{sec:2D-pkpk}

Adding a peak constraint at the location of the spin and taking into account the proxy given by equation~(\ref{Iij}) is more difficult as it requires a numerical integration to account for the sign constraints  on the eigenvalues. Notwithstanding this shortcoming, 
we will show now that in two dimensions,  the adjunction of a peak constraint preserves both the qualitative picture (same geometry with four quadrants of opposite spin direction) as well as  the typical scale for $r_{\star}$. 

Figure~\ref{fig:test-proxy-2D} indeed shows the numerical integration of the mean spin when i) the inertia tensor is approximated by equation~(\ref{eq:inertia}) --where the mass is fixed by the smoothing length $R_{h}$--, ii) the scale of the central peak $R_{s}$ is different from the scale of the halo $R_{h}$ (here we take $R_{h}/R_{s}=1/10$), iii) there is a peak constraint at the location where the spin is computed with height $\nu=5/2$, negative eigenvalues and zero gradient.
In short, the mean spin is  now computed as
\begin{equation}
\left\langle \tilde {\mathbf{s}} | \textrm{pk,pk}\right\rangle=\frac{\left\langle \tilde {\mathbf{s}}\det [\mathbf{H}] \,\Theta(-\lambda_{i})\delta_{\rm D}(x-\nu)\delta_{\rm D}(x_{i}) |\textrm{pk}\right\rangle}{\left\langle\det[\mathbf{H}]\, \Theta(-\lambda_{i})\delta_{\rm D}(x-\nu)\delta_{\rm D} (x_{i})|\textrm{pk}\right\rangle}\,,
\end{equation}
where $\tilde s_{i}$  is defined as $\tilde s_{i}=  \sum_{j,k,l}\epsilon_{ijk} H^{-1}_{jl}T_{lk}$ while the expectations $\left\langle\cdot|\textrm{pk} \right\rangle$ are defined as  conditional expectation to a central peak of geometry $\nu=1$, $\lambda_{1}=-1$, $\lambda_{2}=-2$.
The mean spin map is then obtained by numerical integration. Figure~\ref{fig:test-proxy-2D}  clearly shows that the four quadrants of opposite spin direction, as well as  the size of these quadrants are preserved. This test strongly suggests  that in two dimensions, improvements beyond the $\mathbf{I}\approx \mathbf{H}$ approximation do not change the global picture described  in the main text.
%
\begin{figure}
\begin{center}
\center\includegraphics[width=0.95\columnwidth]{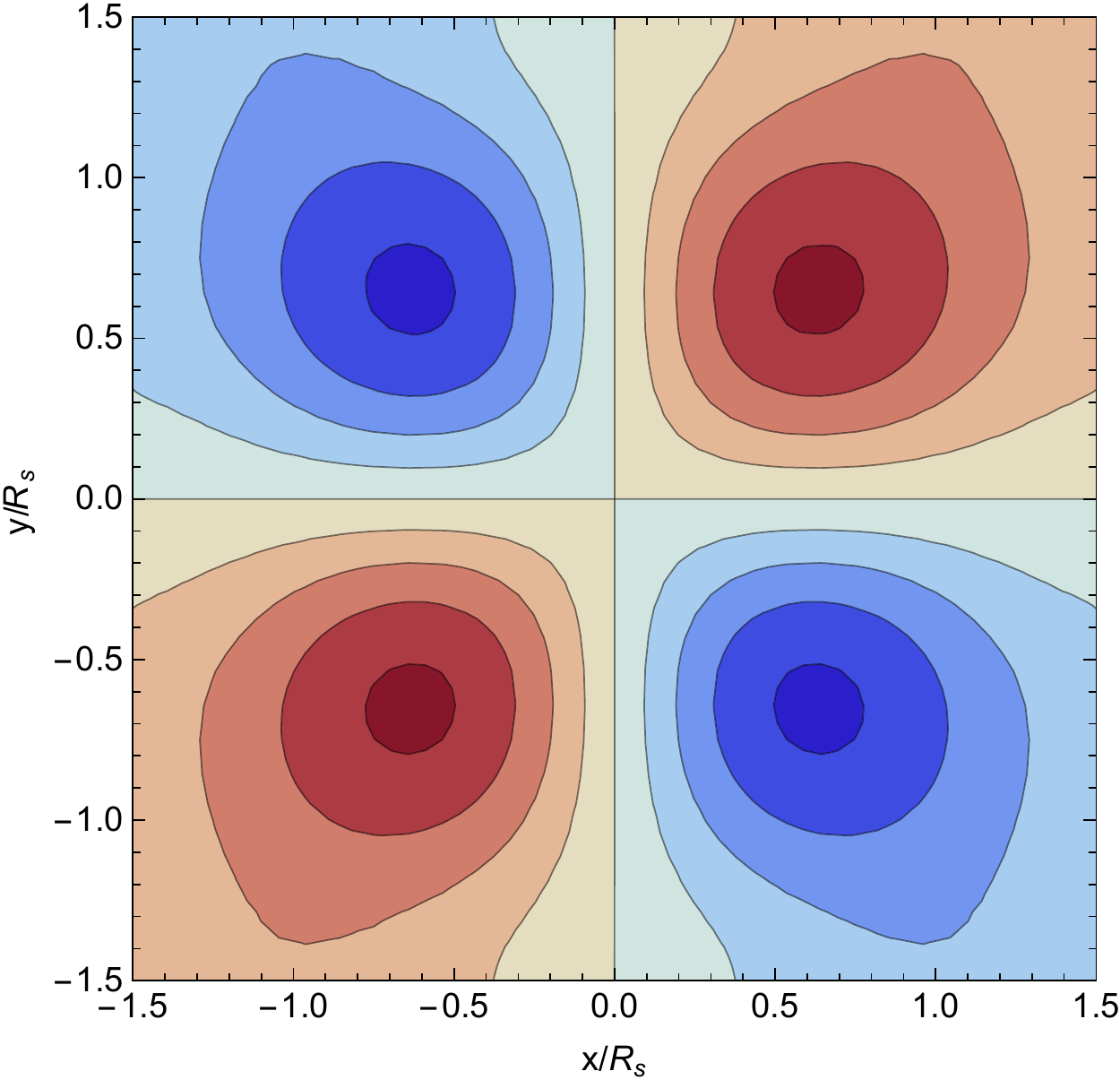}
\caption{
Mean spin computed via numerical integrations when i) a central peak with geometry given by $\nu=1$, $\lambda_{1}=-1$, $\lambda_{2}=-2$ is set, ii) the inertia tensor is approximated by equation~(\ref{Iij}), iii) the scale of the central peak $R_{s}$ is different from the scale of the halo $R_{h}$ (here we take $R_{h}/R_{s}=1/10$), iv) there is a peak constraint at the location where the spin is computed with height $\nu=5/2$, negative eigenvalues and zero gradient. The density power spectrum is a power-law here with spectral index $n=1/2$. 
\label{fig:test-proxy-2D}}
\end{center}
\end{figure}

Further developments, beyond the scope of this paper, could be to carry out the same analysis in three dimensions,  also adding a peak constraint at the location where the spin is computed in order to impose the existence of a proto-halo and use equation~(\ref{Iij}) to define its inertia tensor. While the two-scale analysis is  straightforward enough to implement, the adjunction of a peak constraint in three dimensions is much more tricky and requires in particular the computation of high-dimension numerical integrals (the results will not be analytic anymore) that are left for future investigations.

\section{Dual void theory}
\label{sec:voids}
 \begin{figure*}
\begin{center}
\includegraphics[width=0.85\columnwidth]{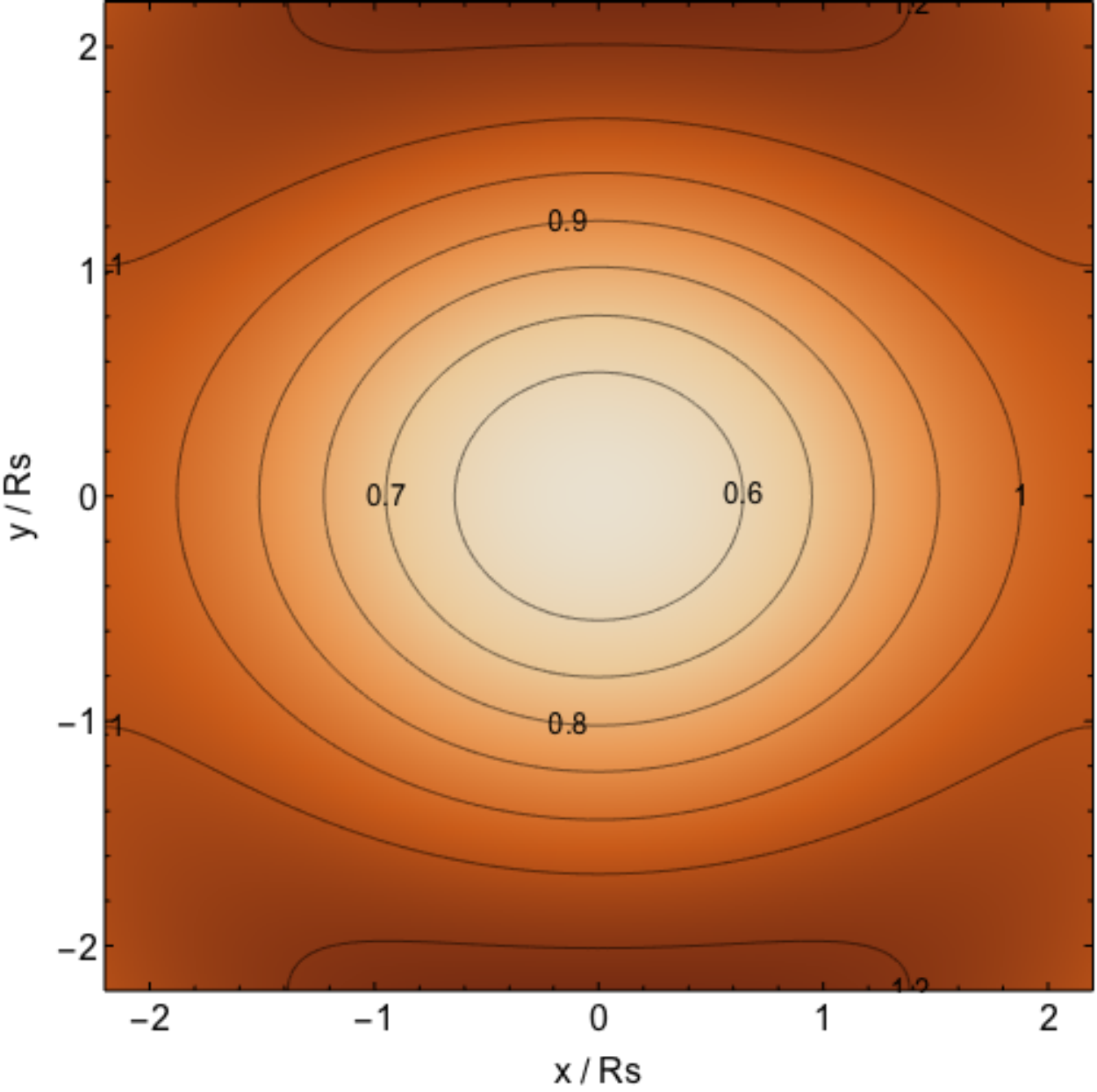}\quad
\includegraphics[width=0.85\columnwidth]{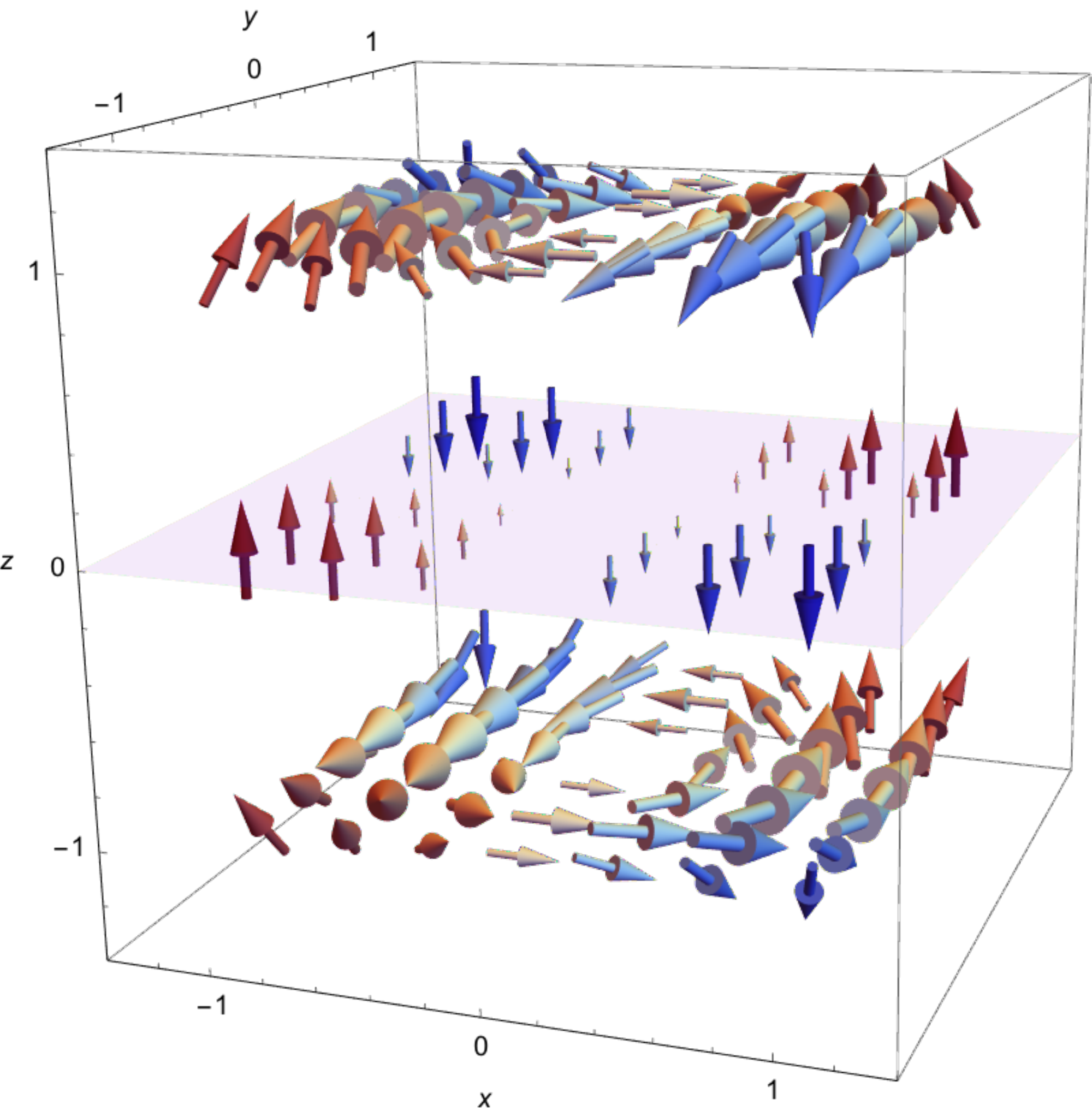}\\
\includegraphics[width=0.85\columnwidth]{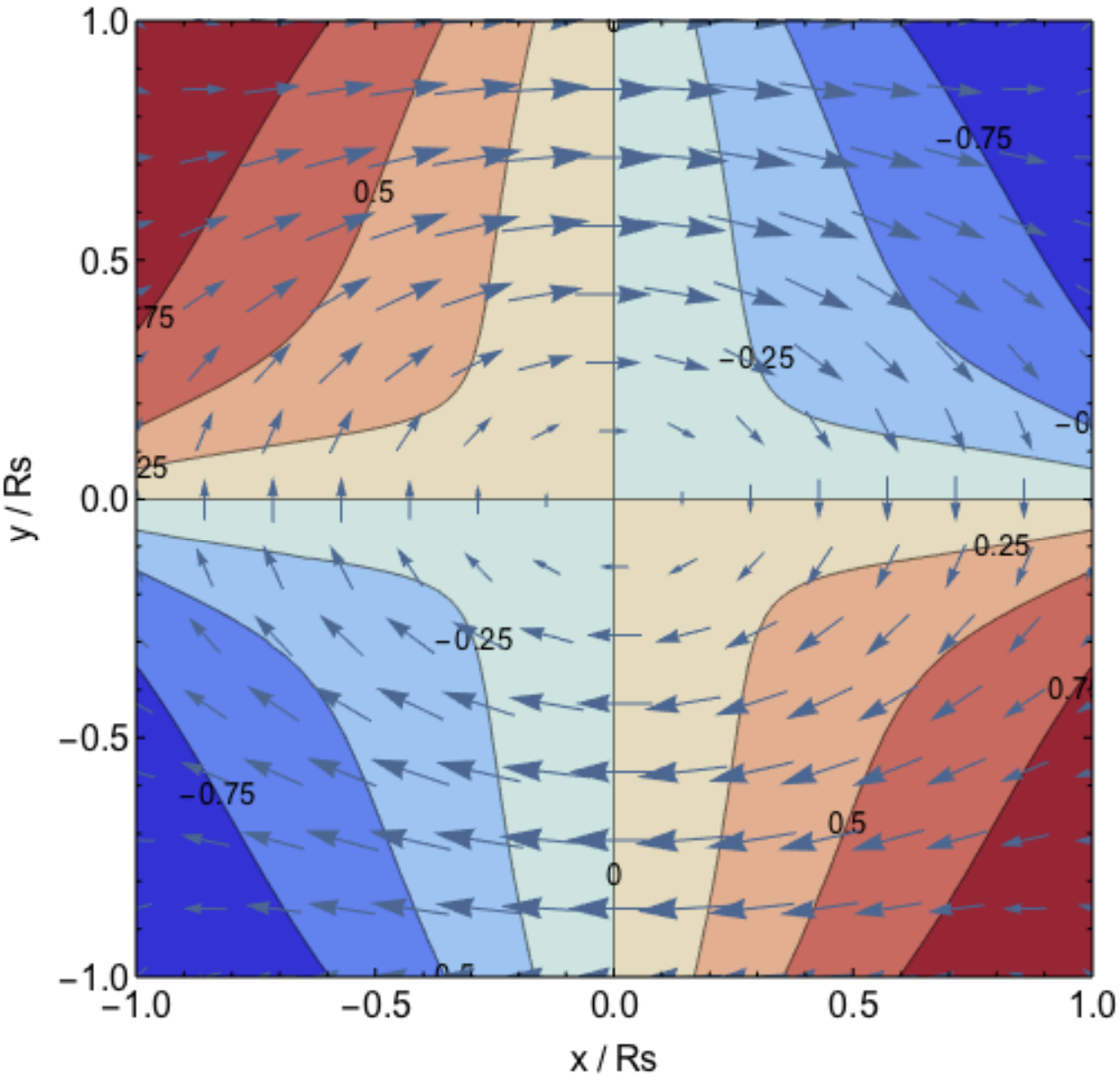}\quad
\includegraphics[width=0.85\columnwidth]{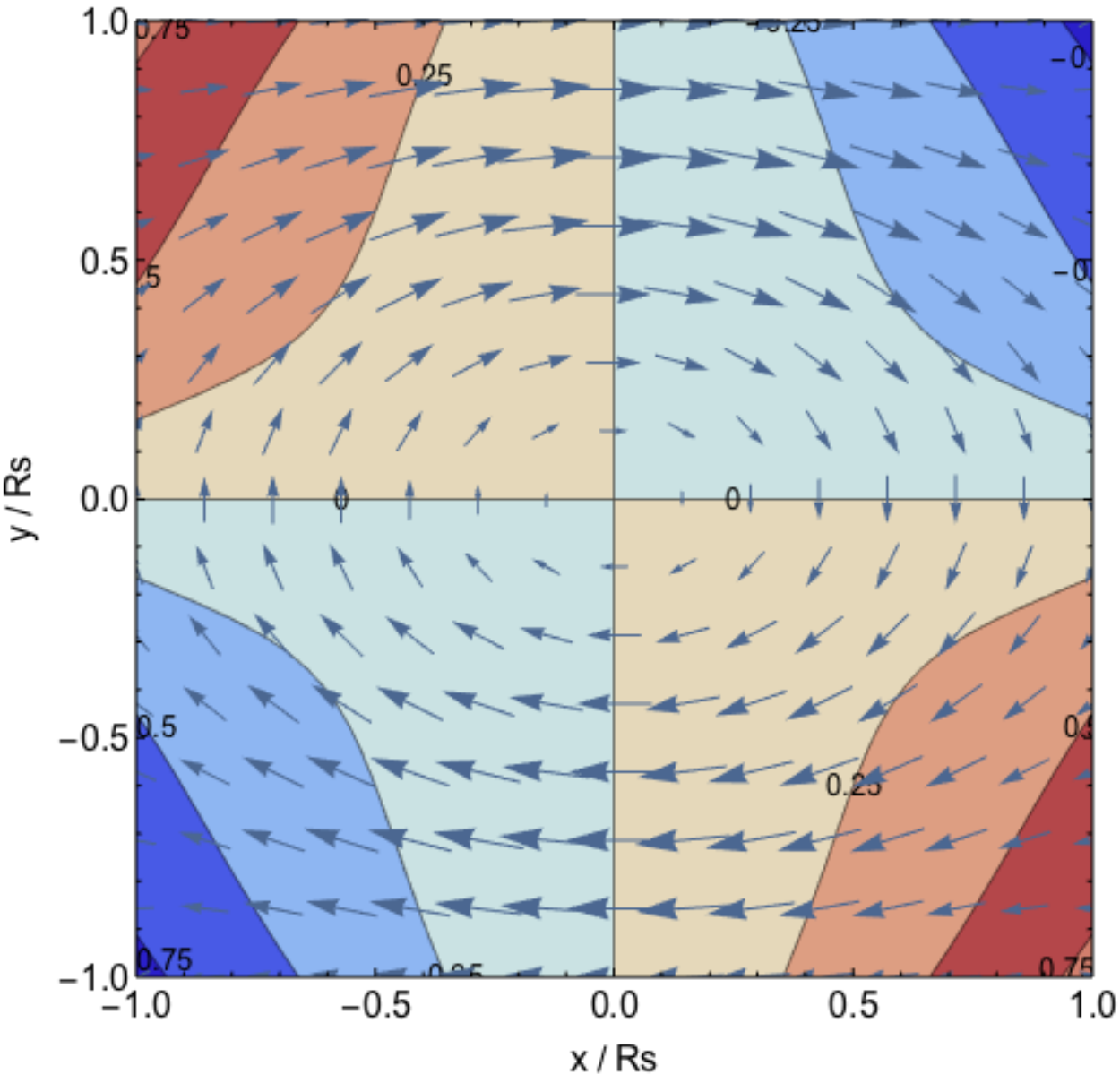}
\caption{Top left-hand panel: Mean density in the plane of the wall $Oxy$ and centered on a wall-type filament with geometry $\nu=0.5$, $\lambda_{1}=0.8$ along the x-axis, $\lambda_{2}=0.6$ along the y-axis and $\lambda_{3}=-0.5$ along the z-axis for a power-law density power spectrum with spectral index $n=-2$. Contours are displayed from $\delta=0.6$ to 1.2 as labeled. The filaments are clearly seen around $y=\pm 2 R_{s}$. Top right-hand panel: Mean spin colour-coded by its projection along the normal to the wall. The spins are aligned with the normal in the plane of the wall and perpendicular to it when going outside the plane of the wall. Bottom panels: mean spin vectors at $z=0.5 R_{s}$ (left) and $z=1 R_{s}$ (right). Contours represent the orientation of the spin with regards to the normal to the wall from -1 (anti-aligned, red) to +1 (aligned, blue) through 0 (perpendicular).
}
\label{fig:voids}
\end{center}
\end{figure*}

The theory presented in Section~\ref{sec:3DS} is algebraic. Effectively no assumption has been made about the signs 
of the eigenvalues of the saddle we are considering.
It is therefore also perfectly valid in the neighbourhood of a wall-type saddle in order to describe the spin alignments 
of dark haloes in that vicinity.  At a qualitative level, Figure~\ref{principe} applies up to a sign: voids and wall saddles repel.
It follows that the spins should rotate around the wall-saddle to void axis and become parallel near the wall with a point symmetric
 change of polarity.   This is indeed what equation~(\ref{eq:defL3Dsol})  predicts and is shown on Figure~\ref{fig:voids}.

The statistical significance of these alignments is likely to be reduced as there are much fewer galaxies in voids and near wall 
saddles. 

\section{Technical complements}\label{sec:technical}
 \subsection{Codes for density and spin in 2/3D}
\label{sec:code}
The expression for the 2 and 3D spin statistics (mean and variance) for scale invariant power spectra  are
available 
both as a mathematica package 
(\url{http://www.iap.fr/users/pichon/spin/code/ATTT.m}), 
and a mathematica notebook
(\url{http://www.iap.fr/users/pichon/spin/code/ATTT-package.nb}).
The following functions are provided:
{\tt  $\delta$2D, spin2D, var2D,	$\delta$3D, spin3D,} 
which correspond resp. to the 2D density, spin, its variance, and in 3D
the density and the spin for scale invariant power spectra of index $n$ as a function of position $r,\theta,(\phi)$
and the geometry of the peak (resp. saddle) 
$\nu,\lambda_1, \lambda_{2},({\lambda_3})$.
Compiled versions are also provided.

 \subsection{Correlation functions for power-law spectra }
 \label{sec:xi}
\label{sec:A3}
  \begin{figure}
\begin{center}
\includegraphics[width=0.95\columnwidth]{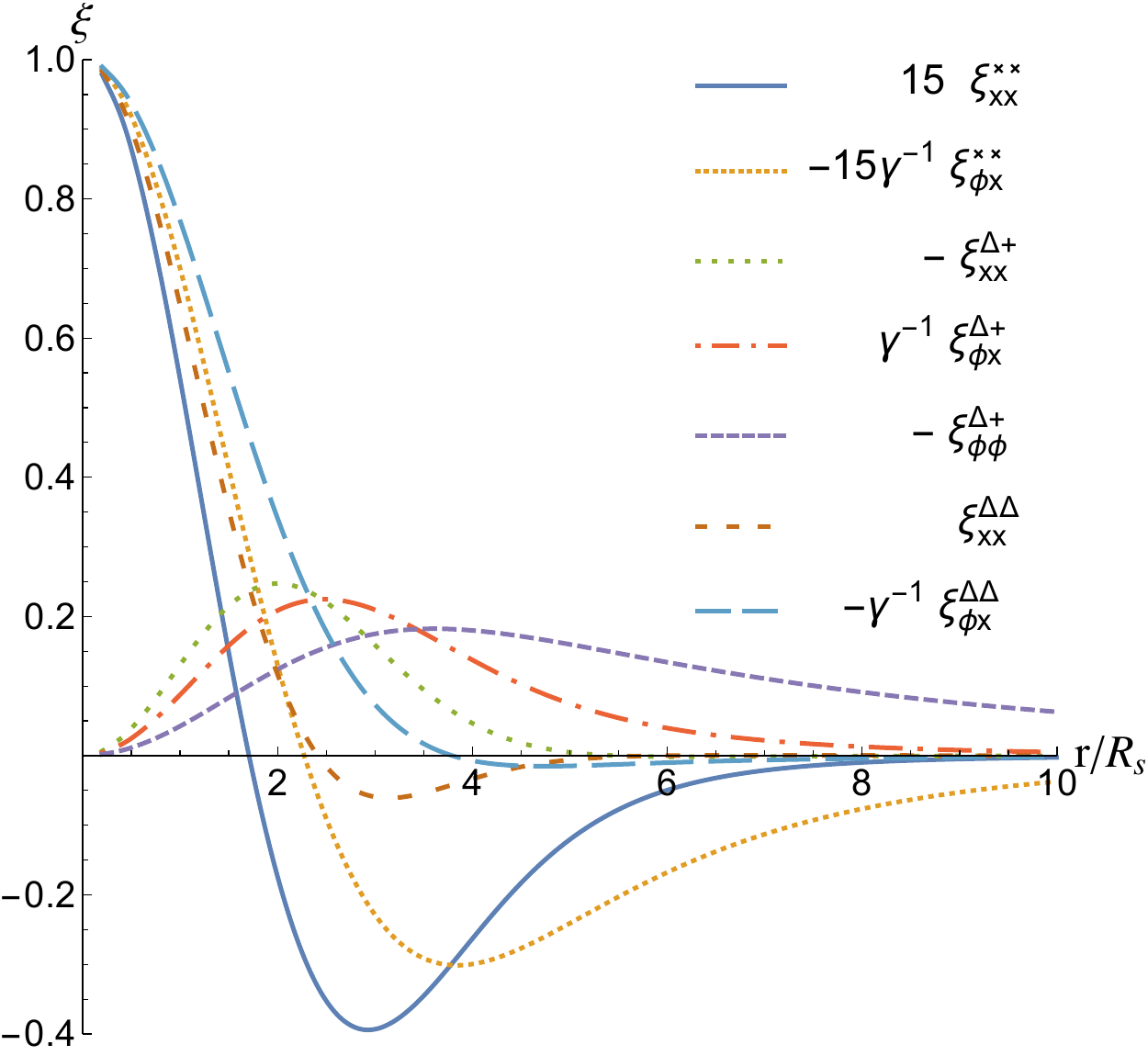}
\caption{Two-point correlation functions as a function of the separation $r$ in units of the smoothing length $R_{s}=5$Mpc$/h$ for a $\Lambda$CDM power spectrum. 
}
\label{fig-xi-LCDM}
\end{center}
\end{figure}
 
 The 2D correlation functions defined in Equation~(\ref{eq:xi})  and  (\ref{eq:defpowespectra})
 can be analytically obtained for density power-law power spectrum $P_{k}(k)\propto k^{n}$ for which the scale parameter is $\gamma=\sqrt{(n+2)/(n+4)}$:
 \begin{align}
 \xi_{\phi\phi}^{\Delta\Delta}(r)\!&={\cal F}_1^2-\frac{1}{16} (n+2) r^2 {\cal F}_2^3, \,\,\,\nonumber\\
 \xi_{\phi x}^{\Delta\Delta}(r)\!&= \gamma \left(\frac{(n+4)}{16} r^2 {\cal F}_3^3- {\cal F}_2^2\right), \nonumber\\
  \xi_{x x}^{\Delta\Delta}(r)\!&={\cal F}_3^2-\frac{1}{16} (n+6) r^2 {\cal F}_4^3, \,\,\,\nonumber\\
\xi_{\phi\phi}^{\Delta +}(r)\!&=-\frac{1}{32} (n+2) r^2 {\cal F}_2^3    ,\,\,\, \nonumber\\
\xi_{\phi x}^{\Delta +}(r)\!&= \frac{\gamma}{32} {(n+4)} r^2 {\cal F}_3^3    ,\,\,\, \nonumber\\
\xi_{x x}^{\Delta +}(r)\!&= -\frac{1}{32} (n+6) r^2 {\cal F}_4^3     ,\nonumber
\end{align}
\begin{align}
\xi_{\phi\phi}^{\times\times}(r)\!&= \frac{1}{8} \left(4 {\cal F}_1^2-3 {\cal F}_1^3\right),\,\,\, \nonumber\\
\xi_{\phi x}^{\times\times}(r)\!&= -\frac{\gamma}{8} \left(4 {\cal F}_2^2-3 {\cal F}_2^3\right)    \nonumber\\
 \,\,\, \xi_{x x}^{\times\times}(r)\!&=   \frac{1}{8} \left(4 {\cal F}_3^2-3 {\cal F}_3^3\right) ,
 \end{align}
  where $r$ is in units of the smoothing length and ${\cal F}_i^j= {}_1F_1\left({n}/{2}+i;j;-{r^2}/{4}\right)$, with $_1F_1$ the Hypergeometric functions of the first kind.
Some of those correlation functions are plotted in Figure~\ref{fig-xi-2D}.

 The 3D correlation functions can similarly be obtained for power-law power spectrum $P_{k}(k)\propto k^{n}$(some of those correlations are plotted in Figure~\ref{fig-xi}) for which the scale parameter is $\gamma=\sqrt{(n+3)/(n+5)}$ and we define ${\cal G}_i^j= _1F_1\left({n+i}/{2}; j/2;-{r^2}/{4}\right) $:
 
\begin{table*}
 \begin{align}
 \xi_{\phi\phi}^{\Delta\Delta}(r)\!&=\frac{-32 (n-1) {\cal G}_1^3-\left((2-4 n) r^2+r^4-32\right) {\cal G}_{-1}^1+(n-2)
   \left(2 n r^2-r^4+32\right){\cal G}_{-1}^3}{2 \left(n^2-1\right) r^2},\nonumber \\
 \xi_{\phi x}^{\Delta\Delta}(r)\!&= \frac{\Gamma \left(\frac{n+1}{2}\right) \left(32 (n+1){\cal G}_3^3+\left(-2 (2 n+3) r^2+r^4-32\right){\cal G}_1^1+n \left(-2 (n+2) r^2+r^4-32\right){\cal G}_1^3\right)}{8 r^2
   \sqrt{\Gamma \left(\frac{n+3}{2}\right)} \sqrt{\Gamma \left(\frac{n+7}{2}\right)}},\nonumber \\
  \xi_{x x}^{\Delta\Delta}(r)\!&=\frac{-32 (n+3){\cal G}_5^3+\left(2 (2 n+7) r^2-r^4+32\right){\cal G}_3^1+(n+2)
   \left(2 (n+4) r^2-r^4+32\right) {\cal G}_3^3}{2 (n+3) (n+5) r^2},\nonumber \\
\xi_{\phi\phi}^{\Delta +}(r)\!&=\frac{\left(4 n \left(r^2+3\right)-\left(r^2+8\right) r^2-40\right) {\cal G}_{-1}^1+16 (n-1) {\cal G}_1^3+(n-2) \left(2 (n-3) r^2-r^4-28\right) {\cal G}_{-1}^3}{2
   \left(n^2-1\right) r^2}      ,\nonumber
\\
 \xi_{\phi x}^{\Delta +}(r)\!&= \frac{\Gamma \left(\frac{n+1}{2}\right) \left(-16 (n+1) {\cal G}_3^3+\left(-4 n \left(r^2+3\right)+r^4+16\right) {\cal G}_1^1+n \left(-2 (n-1) r^2+r^4+28\right) {\cal G}_1^3\right)}{8 r^2
   \sqrt{\Gamma \left(\frac{n+3}{2}\right)} \sqrt{\Gamma \left(\frac{n+7}{2}\right)}}    ,\nonumber\\
\xi_{x x}^{\Delta +}(r)\!&= \frac{16 (n+3) {\cal G}_5^3+\left(4 (n+2) r^2+12 n-r^4+8\right){\cal G}_3^1+(n+2)
   \left(2 (n+1) r^2-r^4-28\right) {\cal G}_3^3}{2 (n+3) (n+5) r^2}     ,\nonumber\\
\xi_{\phi\phi}^{\times\times}(r)\!&= \frac{\left((n-2) r^2 \left(r^2+10\right)-48\right) {\cal G}_{-1}^3+\left(-2 (n-6) r^2+r^4+48\right){\cal G}_{-1}^1}{\left(n^2-1\right) r^4}     ,\nonumber\\
\xi_{\phi x}^{\times\times}(r)\!&= -\frac{\Gamma \left(\frac{n+1}{2}\right) \left(\left(n r^2 \left(r^2+10\right)-48\right) {\cal G}_1^3+\left(-2 (n-4)
   r^2+r^4+48\right) {\cal G}_1^1\right)}{4 r^4 \sqrt{\Gamma \left(\frac{n+3}{2}\right)} \sqrt{\Gamma \left(\frac{n+7}{2}\right)}}   ,\nonumber\\
\xi_{x x}^{\times\times}(r)\!&=   \frac{\left((n+2) r^2 \left(r^2+10\right)-48\right) {\cal G}_3^3+\left(-2 (n-2) r^2+r^4+48\right) {\cal G}_3^1}{(n+3) (n+5) r^4}   \,.\nonumber
 \end{align}
 \end{table*}

 \subsection{Correlation functions for LCDM spectra }
 \label{sec:xi-LCDM}
 The same $\xi$ correlation functions can also be computed for a $\Lambda$CDM power spectrum using \cite{BBKS} and 
 equation~(\ref{eq:defpowespectra}).
 The corresponding functions are shown on Figure~\ref{fig-xi-LCDM} for a Gaussian smoothing length of $R_{s}=5{\rm Mpc}/h$ and a WMAP-7 cosmology. Note that those correlation functions are quite similar to $n=-2$ power-law power spectrum (see Figure~\ref{fig-xi}).
 Given these correlations, it would be straightforward to compute the corresponding spin.

\end{document}